\documentclass{JFM-FLM_Au}
\usepackage{subfigure}
\usepackage{multirow}
\usepackage{float}
\usepackage{makecell}
\usepackage{upgreek}
\usepackage{graphicx}

\lefttitle{Xiaozhe Xi et al.}
\righttitle{Journal of Fluid Mechanics}

\title{Surrogate-Based Aerodynamic Shape Optimization in Multiscale Flows via the Implicit Unified Gas-Kinetic Scheme}

\author{Xiaozhe Xi\aff{1}, Wenpei Long\aff{1}, Wenzhi Guo\aff{1}, Junzhe Cao\aff{1} \and Kun Xu\aff{1, 2, 3}}

\affiliation{\aff{1}Department of Mathematics, Hong Kong University of Science and Technology, Hong Kong, China
\aff{2}Department of Mechanical and Aerospace Engineering, Hong Kong University of Science and Technology, Hong Kong, China
\aff{3}HKUST Shenzhen Research Institute, Shenzhen, 518057, China}

\corresau{Kun Xu, makxu@ust.hk}

\begin{document}
\maketitle

\begin{abstract}

While hypersonic glide vehicles such as the HTV-2 continue to be a focal point in aerospace research, their aerodynamic characteristics in complex near-space environments are not yet fully understood. Because traditional continuum assumptions fail to accurately capture multiscale flow features across varying rarefied altitudes, this study investigates the aerodynamic shape optimization of an HTV-2-type aircraft across multiple flow regimes.
An automated optimization framework is developed by coupling surrogate-based optimization (SBO) with the implicit unified gas-kinetic scheme (IUGKS). To ensure relevance to practical engineering requirements, both volumetric and center-of-pressure constraints are incorporated into the optimization process. The resulting optimized configurations are subsequently validated through high-fidelity computations, detailed flow-field evaluations, and global sensitivity analyses.
Under volumetric constraints, the optimized lift-to-drag ratio ($L/D$) increases significantly at altitudes ranging from 70 km to 100 km. The optimal aerodynamic strategy is shown to shift with altitude: at 70 km, reducing the windward radius ($R_1$) weakens the oblique shock wave, whereas at highly rarefied altitudes, reducing the leeward radius ($R_3$) enhances the expansion wave. Correspondingly, sensitivity analyses confirm that as flow rarefaction increases, aerodynamic dominance shifts toward $R_3$. Furthermore, reducing the wingtip bluntness ($R_2$)yields consistent aerodynamic benefits across the entire flight envelope, ultimately driving the optimized geometries toward a flatter and more slender profile.

\end{abstract}

\begin{keywords}
Hypersonic Technology Vehicle; Surrogate-based optimization; Global sensitivity analysis; Implicit unified gas-kinetic scheme; Multiple flow regimes;
\end{keywords}


\section{Introduction}\label{Sec: introduction}

Driven by the demand for higher speeds, increased altitudes, and enhanced trajectory control, hypersonic vehicles remain a primary focus in aerospace research. Current technological developments are primarily centered on air-breathing propulsion systems and hypersonic glide vehicles. To achieve global-scale, point-to-point flight, these configurations must balance high volumetric efficiency with superior aerodynamic performance. As an evolution of earlier experimental platforms such as the Common Aero Vehicle, the Hypersonic Technology Vehicle 2 (HTV-2) represents a technical transition from traditional boost-glide trajectories to high lift-to-drag ratio ($L/D$) glide flight~\citep{walker2008darpa, ken2004hypersonic}. The HTV-2 utilizes a lifting-body configuration featuring a sharp leading edge and swept-back trailing-edge flaps, demonstrating a clear design trend toward waverider geometries. This aerodynamic layout is intended to maximize $L/D$ to support long-duration and long-range flight trajectories (up to 16,000 km) within the atmosphere~\citep{walker2005falcon}. However, despite the integration of advanced thermal protection systems—such as a tantalum-based shell with a melting point of 3290K -- flight test anomalies in 2010 and 2011 indicate that aerodynamic stability and unsteady flow mechanisms in complex near-space environments are not yet fully understood.

Existing studies on HTV-2-type aircraft have primarily focused on high-temperature effects on aerodynamic characteristics~\citep{liu2015high}, electromagnetic scattering during thermochemical ablation~\citep{shao2016analysis}, aerothermodynamic optimization~\citep{shi2017aerodynamic}, and near-field sonic phenomena~\citep{zou2024computational}. While these investigations have provided valuable insights into various aspects of the vehicle's aerodynamics, the pronounced multiscale flow effects encountered across different near-space altitudes remain insufficiently characterized.
According to the trajectory profiles reported by~\citet{walker2008darpa}, the typical flight envelope of the HTV-2 spans an altitude range of 50 to 120 km. Within this near-space environment, the flow field around the vehicle exhibits strong multiscale characteristics, extending from the high-density shock compression region at the nose to the highly rarefied expansion region at the tail. For context, the local Knudsen number along the surface of similar hypersonic platforms, such as the X-38, can vary by up to five orders of magnitude~\citep{jiang2019implicit}. Such distinct cross-regime features cannot be accurately captured using traditional continuum assumptions. Instead, resolving the underlying physical processes across all flow regimes necessitates the use of gas-kinetic methods based on the Boltzmann equation.
Therefore, investigating the multiscale aerodynamic mechanisms of HTV-2-type aircraft and optimizing their aerodynamic layouts across multiple flow regimes are essential steps toward understanding their performance evolution and advancing hypersonic glide flight technologies.

The intensification of hypersonic research in rarefied flow regimes has catalyzed a surge in the development of optimization frameworks based on kinetic theory. Beyond early explorations into topology optimization using discrete velocity methods (DVM)\citep{sato2019topology} and adjoint direct simulation Monte Carlo (DSMC) frameworks \citep{caflisch2021adjoint, yang2023adjoint}, research has expanded to encompass cross-regime aerodynamic shape optimization. As multiscale methods become more computationally efficient, the associated reduction in overhead has made aerodynamic shape optimization across multiple flow regimes an active area of research.
Within the realm of deterministic methodologies, recent efforts have advanced adjoint- and surrogate-based optimization utilizing schemes such as DUGKS and GSIS~\citep{yuan2024design, yuan2025adjoint, zhang2025fast, li2025surrogate}. However, because kinetic solvers must discretize both physical and velocity spaces, the resulting prohibitive memory footprints and computational overheads restrict these studies primarily to two-dimensional or simplified geometries. Consequently, the refined optimization of complex 3D configurations remains nearly intractable. Alternatively, stochastic approaches—such as DSMC and the unified stochastic particle (USP) method—have been coupled with machine learning techniques to capture cross-regime aerodynamic effects~\citep{liu2025aerodynamic}. Nevertheless, the inherent statistical noise of these methods severely hinders stable sensitivity analysis during the optimization process. Furthermore, the stringent resolution requirements associated with stochastic methods in the near-continuum regime render them computationally prohibitive for full-envelope design.

In this work, the deterministic implicit unified gas-kinetic scheme (IUGKS) is employed for surrogate-based shape optimization of an HTV-2-type aircraft. By directly modeling the underlying physical laws within a discretized space, the unified gas-kinetic scheme (UGKS)\citep{xu2010unified} overcomes the microscopic scale limitations typical of near-continuum flows. Specifically, its integral solution of the kinetic model is utilized to construct fluxes that couple the free transport and collision effects of the gas distribution function, thereby recovering the gas-kinetic scheme (GKS)\citep{xu2001gas} in the continuum regime. The UGKS has been successfully applied to a diverse range of physical systems, including hypersonic flows~\citep{long2024implicit}, microscale flows~\citep{liuchang2020}, thermal and chemical non-equilibrium flows~\citep{wei2024adaptive, wei2025unified}, radiation transport~\citep{quan2025radiative}, plasma transport~\citep{pu2025ugks}, and multiphase systems~\citep{LIU2019264}. Through a series of numerical advancements~\citep{zhu2016implicit, zhu2019implicit, xu2022ugks}, the integration of implicit iteration techniques and an unstructured velocity space~\citep{wei2024adaptive} has improved computational efficiency by one to three orders of magnitude. This acceleration, combined with the inherent accuracy of the kinetic solver, enables the IUGKS to efficiently handle large-scale non-equilibrium flow simulations, ultimately making the cross-regime shape optimization presented in this study computationally feasible.

The remainder of this paper is organized as follows. Section~\ref{sec: Numerical method} introduces the IUGKS methodology. Next, the geometric model of an HTV-2-type aircraft is detailed in Section~\ref{sec: Geometric model}. Section~\ref{sec: Optimization and sensitivity analysis} describes the surrogate-based optimization (SBO) framework and the Sobol global sensitivity analysis. The optimization results, along with an analysis of the aerodynamic characteristics at various altitudes, are discussed in Section~\ref{sec: Results and analyses}. Finally, Section~\ref{sec: conclusion} summarizes the conclusions of this study.

\section{Numerical method}\label{sec: Numerical method}

To achieve computational efficiency in multiscale flow regimes, the IUGKS is adopted. Under the framework of the finite volume method, the implicit evolution of gas distribution function $f_{i,k}$ at the discrete velocity $\boldsymbol{u}_k$ with the backward Euler method is
\begin{equation}
    f_{i,k}^{n+1} = f_{i,k}^n - \frac{\upDelta t_i}{\upOmega_i} \sum_{j \in N(i)} \mathscr{F}_{ij,k}^{n+1} \mathscr{A}_{ij}
    + \frac{\upDelta t_i}{\tau_i^{n+1}} \left( g_{i,k}^{n+1} - f_{i,k}^{n+1} \right),
    \label{eq: iugks in fvm}
\end{equation}
where $\upOmega_i$ is the cell volume, $\upDelta t_i$ is the local numerical time step, $\mathscr{A}_{ij}$ denotes the area of cell interface $ij$, and $N(i)$ represents the set of neighboring cells. For the collision term, the kinetic BGK-Shakhov model is employed to characterize it,
\begin{equation}
    \frac{\partial f}{\partial t} + \boldsymbol{u} \cdot \frac{\partial f}{\partial \boldsymbol{x}} = \frac{g -f}{\tau},
\end{equation}
where $\tau$ is the relaxation time which is determined by $\tau = \mu /p$ with the macroscopic pressure $p$ and dynamic viscosity $\mu$. And $g$ is defined as follows~\citep{shakhov_generalization_1968} to result in a realistic Prandtl number ($\mathrm{Pr}$),
\begin{equation}
    \begin{aligned}
        &g = g^{eq}\left[1 + (1 - \Pr) (\boldsymbol{u} - \boldsymbol{U}) \cdot q \left( \frac{(\boldsymbol{u} - \boldsymbol{U})^2 + \boldsymbol{\xi}^2}{RT} - 5 \right) / (5pRT) \right],\\
        &g^{eq}(x, t, \boldsymbol{u}) = \rho (\frac{\lambda}{\pi})^{\frac{\mathrm{K}+3}{2}}  e^{- \lambda [(\boldsymbol{u} - \boldsymbol{U})^2 + \boldsymbol{\xi} ^2]}.
    \end{aligned}
    \label{eq: modified equilibrium distribution}
\end{equation}
Here, $g^{eq}$ represents the local equilibrium Maxwellian distribution, $R$ is the gas constant, $\mathrm{K}$ is the internal degree of freedom of gas, and $\lambda$ is related to the temperature $T$ by $\lambda = 1/2RT$. $\boldsymbol{\xi} = (\xi_1, \xi_2, ..., \xi_\mathrm{K})$ is the internal variable.
The relation of macroscopic conservative variables of density, total momentum, and total energy in the control volume $\boldsymbol{W} = (\rho, \rho \boldsymbol{U}, \rho E)^T$ and the distribution function $f$ is
\begin{equation}
    \boldsymbol{W} = \int f \boldsymbol{\psi} \mathrm{d} \boldsymbol{\Xi} ,
\end{equation}
where $\boldsymbol{\psi} = (1, \boldsymbol{u}, \frac{1}{2}(\boldsymbol{u}^2+\boldsymbol{\xi}^2))^T$ and $\mathrm{d} \boldsymbol{\Xi} = \mathrm{d} \boldsymbol{u} \mathrm{d} \boldsymbol{\xi}$.
Therefore, $g_{i,k}^{n+1}$ and $\tau_i^{n+1}$ in Eq.~\ref{eq: iugks in fvm} are calculated from the macroscopic conservative variables $\tilde{\boldsymbol{W}}_i^{n+1}$ given by macroscopic implicit prediction as
\begin{equation}
    \tilde{\boldsymbol{W}}_i^{n+1} = \boldsymbol{W}_i^n - \frac{\upDelta t_i}{\upOmega_i} \sum_{j \in N(i)} \boldsymbol{F}_{ij}^{n+1} \mathscr{A}_{ij}.
    \label{eq: implicit macroscopic conservative variable}
\end{equation}

For the macroscopic implicit prediction part, by subtracting the time-averaged flux $\boldsymbol{F}_{ij}^n$ corresponding to the macroscopic variables at time step $n$ from both sides of the macroscopic implicit prediction formula Eq. ~\ref{eq: implicit macroscopic conservative variable}, we can rewrite it in the $\upDelta$ form
\begin{equation}
    \frac{\upOmega_i}{\upDelta t_i} \upDelta \boldsymbol{W}_i + \sum_{j \in N(i)} \upDelta \boldsymbol{F}_{ij} \mathscr{A}_{ij} = \boldsymbol{R}_i,
    \label{eq: implicit conservative variable in Delta form}
\end{equation}
where $\upDelta \boldsymbol{W}_i = \tilde{\boldsymbol{W}}_i^{n+1} - \boldsymbol{W}_i^{n}$, $\upDelta \boldsymbol{F}_{ij} = \boldsymbol{F}_{ij}^{n+1} - \boldsymbol{F}_{ij}^n$, and $\boldsymbol{R}_i$ is the residual
\begin{equation}
    \boldsymbol{R}_i = -\sum_{j \in N(i)} \boldsymbol{F}_{ij}^n \mathscr{A}_{ij}.
\end{equation}
For the UGKS, the macroscopic flux $\boldsymbol{F}_{ij}^n$ is provided by taking moments of the time-average flux of the discrete gas distribution function,
\begin{equation}
    \boldsymbol{F}_{ij}^n = \sum_k \mathscr{F}_{ij,k}^n \boldsymbol{\psi}_k \mathscr{V}_k.
\end{equation}
where $\mathscr{V}_k$ is the corresponding weight in the discrete velocity space.
The time-averaged flux modeled by the integral solution is expressed as
\begin{equation}
    \mathscr{F}_{ij,k}^n = \boldsymbol{u}_k \cdot \boldsymbol{n}_{ij} \left( L_1^u g_k + L_2^u \boldsymbol{u}_k \cdot \frac{\partial g_k}{\partial \boldsymbol{x}} + L_3^u \frac{\partial g_k}{\partial t} + L_4^u f_k + L_5^u \boldsymbol{u}_k \cdot \frac{\partial f_k}{\partial \boldsymbol{x}} \right),
\end{equation}
where $\boldsymbol{n}_{ij}$ is the unit normal direction of the interface $ij$, and $L_1^u$ to $L_5^u$ are scale-related time coefficients considering the local time step
\begin{equation}
    \begin{aligned}
        L_1^u &= 1 - \frac{\tau}{\upDelta t_{ij}} \left(1 - e^{-\upDelta t_{ij}/\tau}\right), \\
        L_2^u &= -\tau + \frac{2\tau^2}{\upDelta t_{ij}} - e^{-\upDelta t_{ij}/\tau} \left( \frac{2\tau^2}{\upDelta t_{ij}} + \tau \right), \\
        L_3^u &= \frac{1}{2} \upDelta t_{ij} - \tau + \frac{\tau^2}{\upDelta t_{ij}} \left(1 - e^{-\upDelta t_{ij}/\tau}\right),  \\
        L_4^u &= \frac{\tau}{\upDelta t_{ij}} \left(1 - e^{-\upDelta t_{ij}/\tau}\right), \\
        L_5^u &= \tau e^{-\upDelta t_{ij}/\tau} - \frac{\tau^2}{\upDelta t_{ij}} \left(1 - e^{-\upDelta t_{ij}/\tau}\right). \\
    \end{aligned}
\end{equation}

The $\upDelta \boldsymbol{F}_{ij}$ is approximated by the Euler equations-based flux splitting method as
\begin{equation}
    \upDelta \boldsymbol{F}_{ij} = \frac{1}{2} \left[ \upDelta \boldsymbol{T}_i + \upDelta \boldsymbol{T}_j + \Gamma_{ij} (\upDelta \boldsymbol{W}_i - \upDelta \boldsymbol{W}_j) \right],
    \label{eq: delta flux}
\end{equation}
where $\upDelta \boldsymbol{T}_i = \boldsymbol{T}_i^{n+1} - \boldsymbol{T}_i^n$. $\boldsymbol{T}$ represents the Euler flux calculated from macroscopic variables by
\begin{equation}
\boldsymbol{T} = \begin{pmatrix}
\rho U_n \\
\rho \boldsymbol{U} U_n + p \boldsymbol{n}_{ij} \\
(\rho E + p) U_n
\end{pmatrix},
\end{equation}
where $U_n = \boldsymbol{U} \cdot \boldsymbol{n}_{ij}$. $\Gamma_{ij}$ represents the spectral radius of the Euler flux Jacobian matrix, and an additional stabilizing term related to kinetic viscosity is added considering the viscous effects
\begin{equation}
\Gamma_{ij} = |\boldsymbol{U} \cdot \boldsymbol{n}_{ij}| + c + \frac{2\mu}{\rho |\boldsymbol{n}_{ij} \cdot (\boldsymbol{x}_j - \boldsymbol{x}_i)|},
\end{equation}
where $c$ is sound speed. By substituting the expression of $\upDelta \boldsymbol{F}_{ij}$ in Eq.~\ref{eq: delta flux} into Eq.~\ref{eq: implicit conservative variable in Delta form}, the macroscopic linear system is
\begin{equation}
    \left( \frac{\upOmega_i}{\upDelta t_i} + \frac{1}{2} \sum_{j \in N(i)} \Gamma_{ij} \mathscr{A}_{ij} \right) \upDelta \boldsymbol{W}_i
    + \frac{1}{2} \sum_{j \in N(i)} (\upDelta \boldsymbol{T}_j - \Gamma_{ij} \upDelta \boldsymbol{W}_j) \mathscr{A}_{ij}
    = -\sum_{j \in N(i)} \boldsymbol{F}_{ij}^n \mathscr{A}_{ij},
    \label{eq: macro governing eq}
\end{equation}
where $\upDelta \boldsymbol{T}_j$ can be obtained by the direct differential calculation
\begin{equation}
\upDelta \boldsymbol{T}_j = \boldsymbol{T}(\boldsymbol{W}_j^n + \upDelta \boldsymbol{W}_j) - \boldsymbol{T}(\boldsymbol{W}_j^n).
\end{equation}
When solving the macroscopic variables prediction equation, Eq.~\ref{eq: macro governing eq}, on a finite volume mesh, a large sparse matrix is obtained. After a simple renumbering process, the point relaxation scheme~\citep{rogers1995comparison} is adopted. Instead of one forward-backward scanning process of the traditional lower–upper symmetric Gauss-Seidel (LU-SGS) method, the point relaxation scheme adopts multiple iteration steps. For the $\left[ m \right]$ th forward scanning process,
\begin{equation}
\begin{aligned}
&\left( \frac{\upOmega_i}{\upDelta t_i} + \frac{1}{2} \sum_{j \in N(i)} \Gamma_{ij} \mathscr{A}_{ij} \right) \upDelta \boldsymbol{W}_i^*
+ \\
&\frac{1}{2} \sum_{j \in L(i)} \mathscr{A}_{ij} \left[ \boldsymbol{T} (\boldsymbol{W}_j^n + \upDelta \boldsymbol{W}_j^*) - \boldsymbol{T} (\boldsymbol{W}_j^n) - \Gamma_{ij} \upDelta \boldsymbol{W}_j^* \right]
+ \\
&\frac{1}{2} \sum_{j \in U(i)} \mathscr{A}_{ij} \left[ \boldsymbol{T} (\boldsymbol{W}_j^n + \upDelta \boldsymbol{W}_j^{\left[ m-1 \right]}) - \boldsymbol{T} (\boldsymbol{W}_j^n) - \Gamma_{ij} \upDelta \boldsymbol{W}_j^{\left[ m-1 \right]} \right]
= \boldsymbol{R}_i,
\end{aligned}
\end{equation}
and the $\left[ m \right]$-th backward scanning process is
\begin{equation}
\begin{aligned}
&\left( \frac{\upOmega_i}{\upDelta t_i} + \frac{1}{2} \sum_{j \in N(i)} \Gamma_{ij} \mathscr{A}_{ij} \right) \upDelta \boldsymbol{W}_i^{[m]}
+ \\
&\frac{1}{2} \sum_{j \in L(i)} \mathscr{A}_{ij} \left[ \boldsymbol{T} (\boldsymbol{W}_j^n + \upDelta \boldsymbol{W}_j^*) - \boldsymbol{T} (\boldsymbol{W}_j^n) - \Gamma_{ij} \upDelta \boldsymbol{W}_j^* \right]
+ \\
&\frac{1}{2} \sum_{j \in U(i)} \mathscr{A}_{ij} \left[ \boldsymbol{T} (\boldsymbol{W}_j^n + \upDelta \boldsymbol{W}_j^{[m]}) - \boldsymbol{T} (\boldsymbol{W}_j^n) - \Gamma_{ij} \upDelta \boldsymbol{W}_j^{[m]} \right]
= \boldsymbol{R}_i,
\end{aligned}
\end{equation}
where $\boldsymbol{W}_j^{\left[ m-1 \right]}$ is the result of $\left[ m-1 \right]$-th iteration and $\upDelta \boldsymbol{W}_i^*$ is the  result of $\left[ m \right]$-th forward scanning process. $L(i)$ and $U(i)$ are subsets of $N(i)$, which denotes neighboring cells occupying the lower and upper triangular area of the large matrix, respectively.

After the macroscopic implicit prediction, the equilibrium distribution function $\tilde{g}^{n+1}$ and relaxation time $\tilde{\tau}^{n+1}$ at $n+1$ step can be directly calculated from the predicted $\tilde{\boldsymbol{W}}^{n+1}$. By substituting $\tilde{g}^{n+1}$ and $\tilde{\tau}^{n+1}$, and subtracting the time-averaged microscopic flux at $n$ step, the delta-form of the microscopic implicit iteration equation is
\begin{equation}
    \left( \frac{\upOmega_i}{\upDelta t_i} + \frac{\upOmega_i}{\tilde{\tau}_i^{n+1}} \right) \upDelta f_{i,k} + \sum_{j \in N(i)} \upDelta \mathscr{F}_{ij,k} \mathscr{A}_{ij} = r_{i,k},
    \label{eq: implicit governing equation}
\end{equation}
where $r_{i,k}$ is the residual of discretized gas distribution function at $\boldsymbol{u}_k$
\begin{equation}
r_{i,k} = \frac{\upOmega_i}{\tilde{\tau}_i^{n+1}} \left( \tilde{g}_{i,k}^{n+1} - f_{i,k}^n \right) - \sum_{j \in N(i)} \mathscr{F}_{ij,k}^n \mathscr{A}_{ij}.
\end{equation}
For microscopic distribution function flux, a single first-order upwind scheme is adopted
\begin{equation}
    \upDelta \mathscr{F}_{ij,k} = \bar{u}_{ij,k} \left\{ H(\bar{u}_{ij,k}) \upDelta f_{i,k} + \left[1 - H(\bar{u}_{ij,k})\right] \upDelta f_{j,k} \right\}.
    \label{eq: microscopic distribution function flux}
\end{equation}
By substituting the microscopic distribution function flux Eq.~\ref{eq: microscopic distribution function flux} into the implicit governing equation Eq.~\ref{eq: implicit governing equation}, we have
\begin{equation}
    D_{i,k} \upDelta f_{i,k} + \sum_{j \in N(i)} D_{j,k} \upDelta f_{j,k} = r_{i,k},
\end{equation}
where
\begin{equation}
    \begin{aligned}
        D_{i,k} &= \frac{\upOmega_i}{\upDelta t_i} + \frac{\upOmega_i}{\tilde{\tau}_i^{n+1}} + \sum_{j \in N(i)} \bar{u}_{ij,k} \mathscr{A}_{ij} H(\bar{u}_{ij,k}), \\
        D_{j,k} &= \bar{u}_{ij,k} \mathscr{A}_{ij} \left[1 - H(\bar{u}_{ij,k})\right]. \\
    \end{aligned}
\end{equation}
Considering the computation cost of the discrete velocity space operations, the microscopic implicit equation adopts the traditional LU-SGS method to decompose the process into a forward scanning process
\begin{equation}
D_{i,k} \upDelta f_{i,k}^* + \sum_{j \in L(i)} D_{j,k} \upDelta f_{j,k}^* + \sum_{j \in U(i)} D_{j,k} \upDelta f_{j,k}^n = r_{i,k},
\end{equation}
and a backward scanning process
\begin{equation}
D_{i,k} \upDelta f_{i,k}^{n+1} + \sum_{j \in L(i)} D_{j,k} \upDelta f_{j,k}^* + \sum_{j \in U(i)} D_{j,k} \upDelta f_{j,k}^{n+1} = r_{i,k}.
\end{equation}

Further details regarding the algorithm can be found in \citet{long2024implicit}.

\section{Geometric model and CFD results}\label{sec: Geometric model}

\subsection{Baseline configuration and aerodynamic characteristics}
\begin{figure}
	\centering
	\subfigure[Back view]
        {
			\includegraphics[height=0.2 \textwidth]{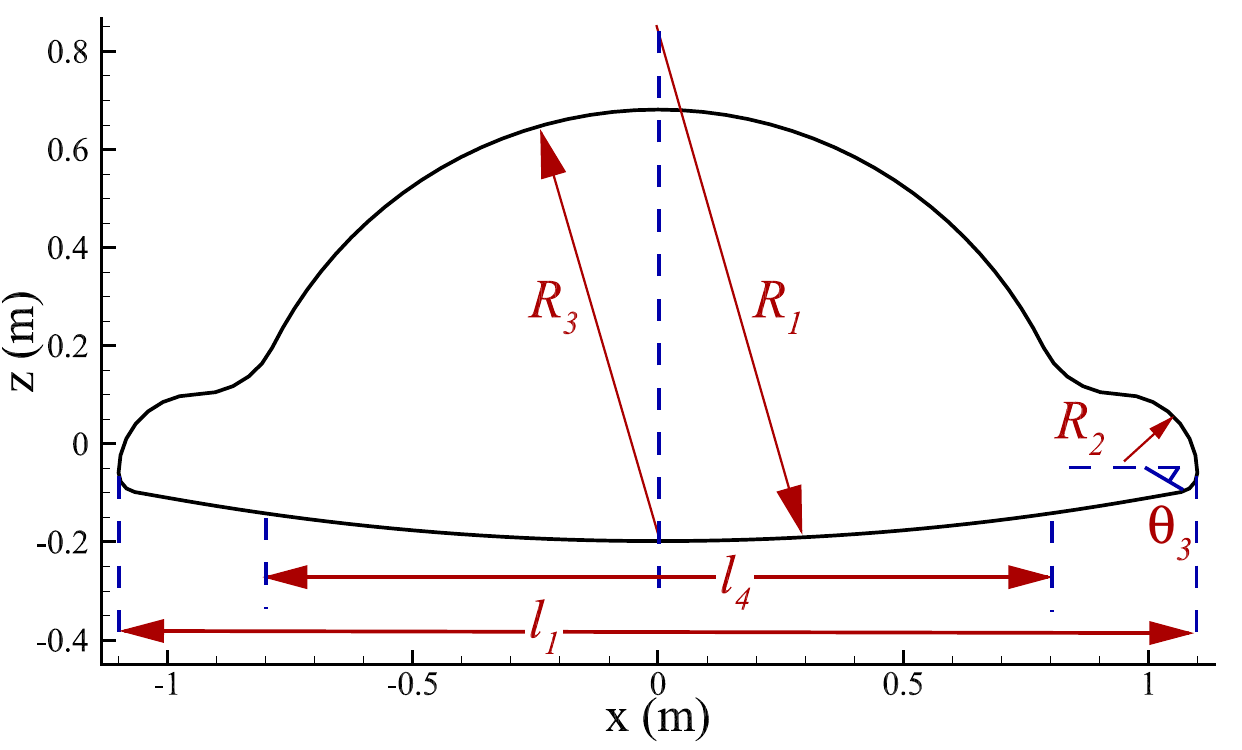}
		}
    \subfigure[Side view]
        {
    		\includegraphics[height=0.2 \textwidth]{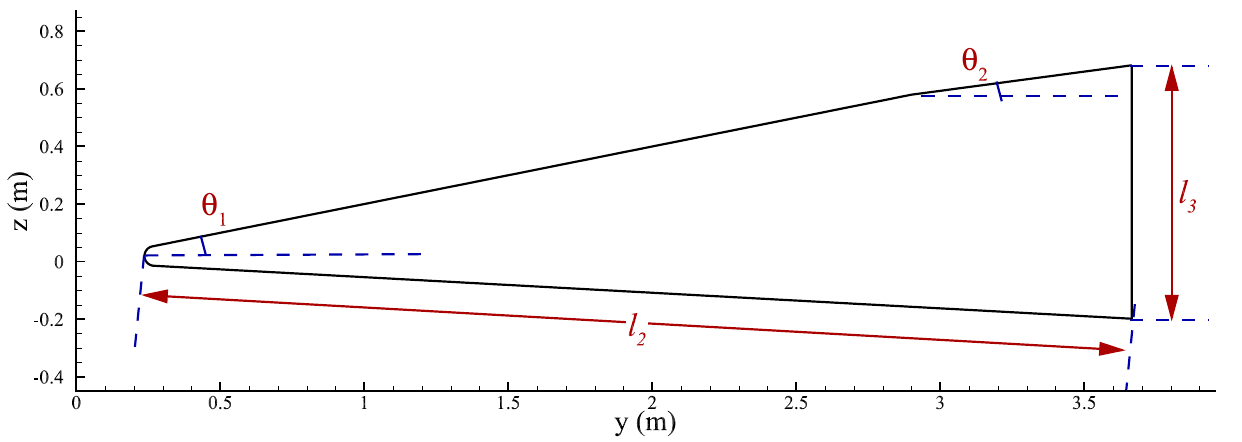}
    	}
    \subfigure[Isometric view]
        {
			\includegraphics[height=0.3333 \textwidth]{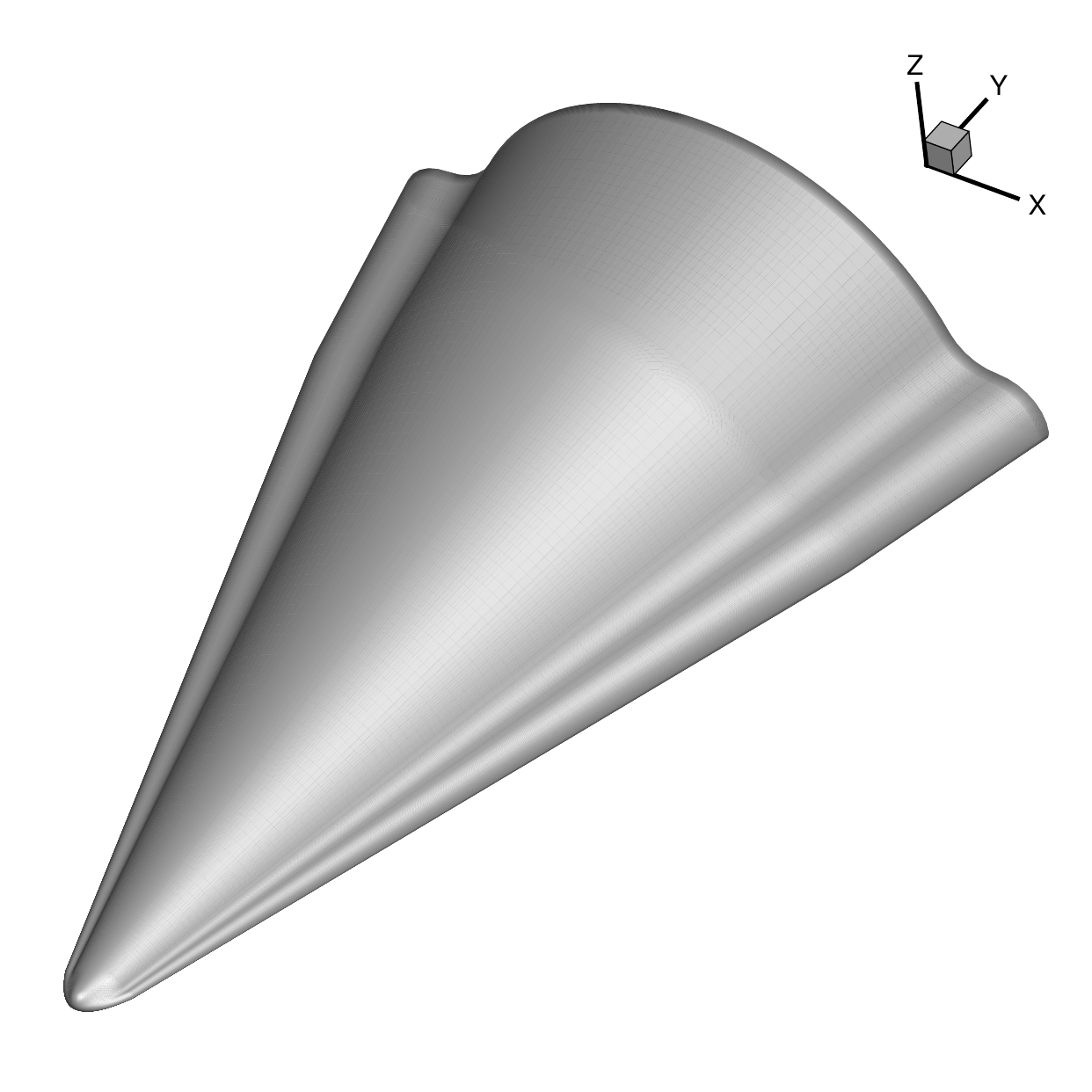}
		}
    \subfigure[Top view]
        {
    		\includegraphics[height=0.3333 \textwidth]{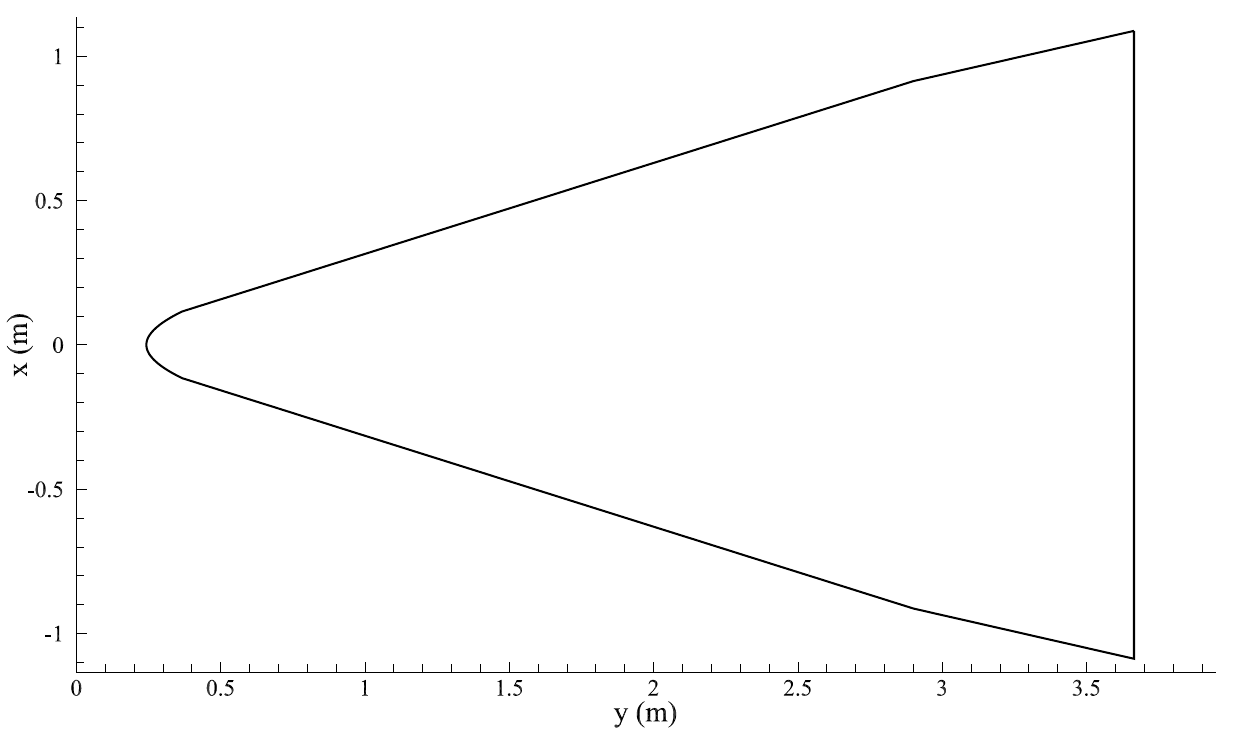}
    	}
	\caption{Geometric layout of the HTV-2 type baseline configuration.}
    \label{fig: baseline three views}
\end{figure}

The baseline configuration in this study is patterned after the HTV-2 type geometry described by Niu et al.~\citep{qinglin2019infrared}. The principal geometric dimensions, which maintain high fidelity to the HTV-2 platform featuring a sharp leading edge and a pronounced sweep-back angle, are detailed in Tab.~\ref{tab: geometric_parameters}, with the corresponding three-dimensional geometry and three-view projections illustrated in Fig.~\ref{fig: baseline three views}.

\begin{table}
\begin{center}
\begin{tabular*}{0.6\textwidth}{@{\extracolsep{\fill}}lr}
Main geometric parameters             & Values \\
Width of fuselage $l_1$               & $2.20\,\mathrm{m}$ \\
Length of fuselage $l_2$              & $3.67\,\mathrm{m}$ \\
Height of fuselage $l_3$              & $0.88\,\mathrm{m}$ \\
Base width $l_4$                      & $1.80\,\mathrm{m}$ \\
Windward surface radius $R_1$         & $5.70\,\mathrm{m}$ \\
Wing tip bluntness radius $R_2$       & $0.16\,\mathrm{m}$ \\
Leeward surface radius $R_3$          & $0.88\,\mathrm{m}$ \\
Forebody compression angle $\theta_1$ & $11.3^\circ$ \\
Leeward expansion angle $\theta_2$    & $7.6^\circ$ \\
Side slope angle $\theta_3$           & $80^\circ$ \\
\end{tabular*}
\end{center}
\caption{The main geometric parameters of the HTV-2 type baseline configuration.}
\label{tab: geometric_parameters}
\end{table}

In this section, four altitudes (70, 85, 100, and 120~km) are selected to evaluate the aerodynamic performance of the baseline configuration. Numerical simulations are performed using the multiscale UGKS method described in Sec.~\ref{sec: Numerical method} at a constant Mach number of 8. For the gas properties, the molecular weight is 28, and the Prandtl number is set to 0.72. Additionally, the variable hard sphere (VHS) model is employed with an exponent of 0.74, where the reference temperature is $273\,\mathrm{K}$, and the reference viscosity is $1.66 \times 10^{-5}\,\mathrm{kg/(m\cdot s)}$. The detailed freestream conditions~\citep{atmosphere1976us} are listed in Tab.~\ref{tab: freestream conditions}, with 85~km defined as the design point. As observed from the table, the freestream Knudsen number across these altitudes spans nearly four orders of magnitude.

\begin{table}
\begin{center}
\begin{tabular*}{0.8\textwidth}{@{\extracolsep{\fill}}c cccc}
Altitudes ($\rm{km}$) & $U_{\infty} \ (\rm{m/s})$ & $\rho_{\infty} \ (\rm{kg/m^3})$ & $T_{\infty} \ (\rm{K})$ & $\rm{Kn}_{\infty}$ \\
$70$  & 2401.48 & $7.42 \times 10^{-5}$ & $217$ & $8.74 \times 10^{-4}$ \\
$85$  & 2241.68 & $8.22 \times 10^{-6}$ & $189$ & $7.63 \times 10^{-3}$ \\
$100$ & 2274.61 & $5.60 \times 10^{-7}$ & $195$ & $0.11$ \\
$120$ & 3089.95 & $2.22 \times 10^{-8}$ & $360$ & $3.29$ \\
\end{tabular*}
\end{center}
\caption{Freestream conditions at different altitudes.}
\label{tab: freestream conditions}
\end{table}

\begin{figure}
	\centering
	\subfigure[$C_L$ and $C_D$]
        {\label{fig: l,d}
			\includegraphics[height=0.35 \textwidth]{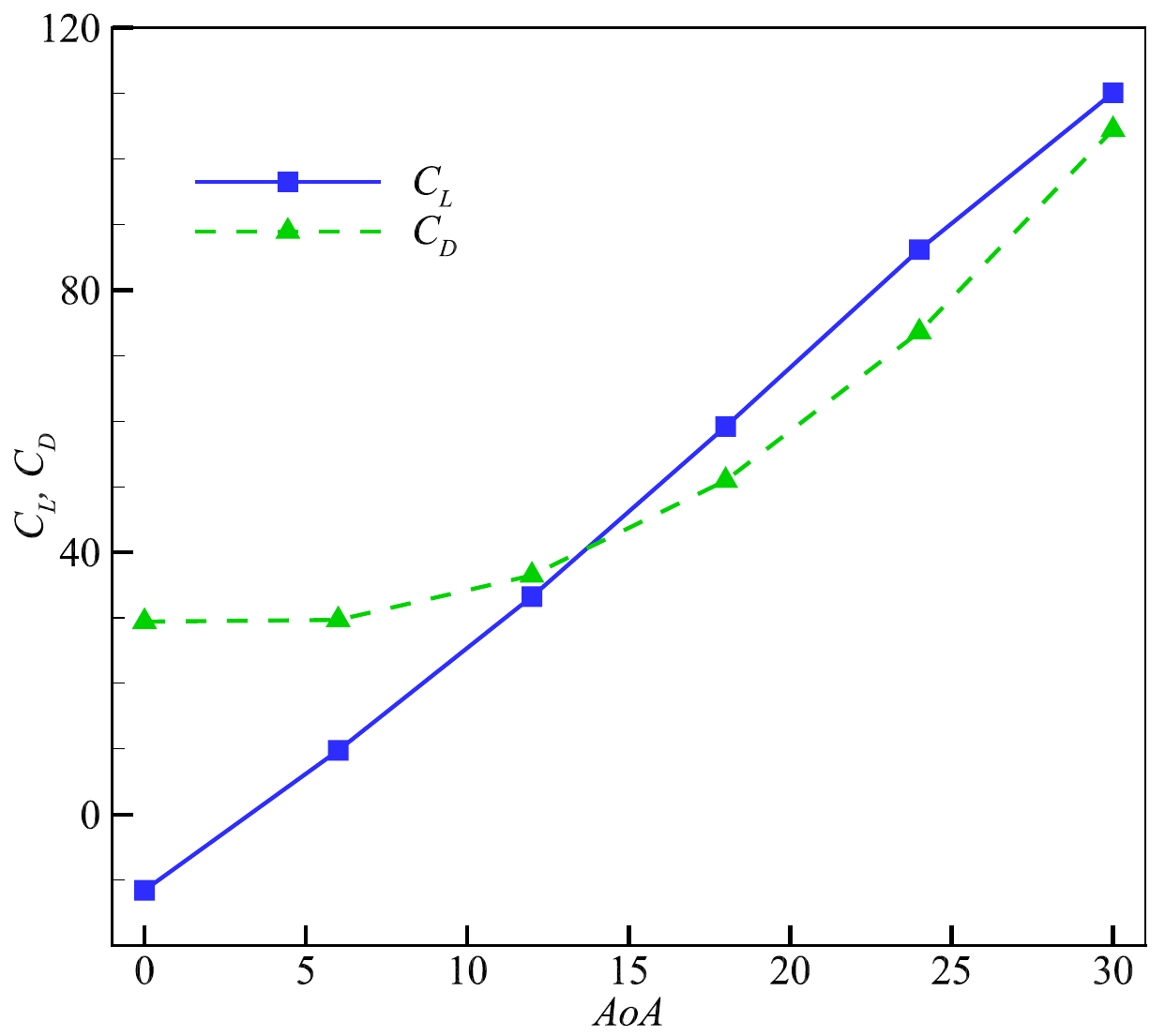}
		}
    \subfigure[$L/D$]
        {\label{fig: l/d}
    		\includegraphics[height=0.35 \textwidth]{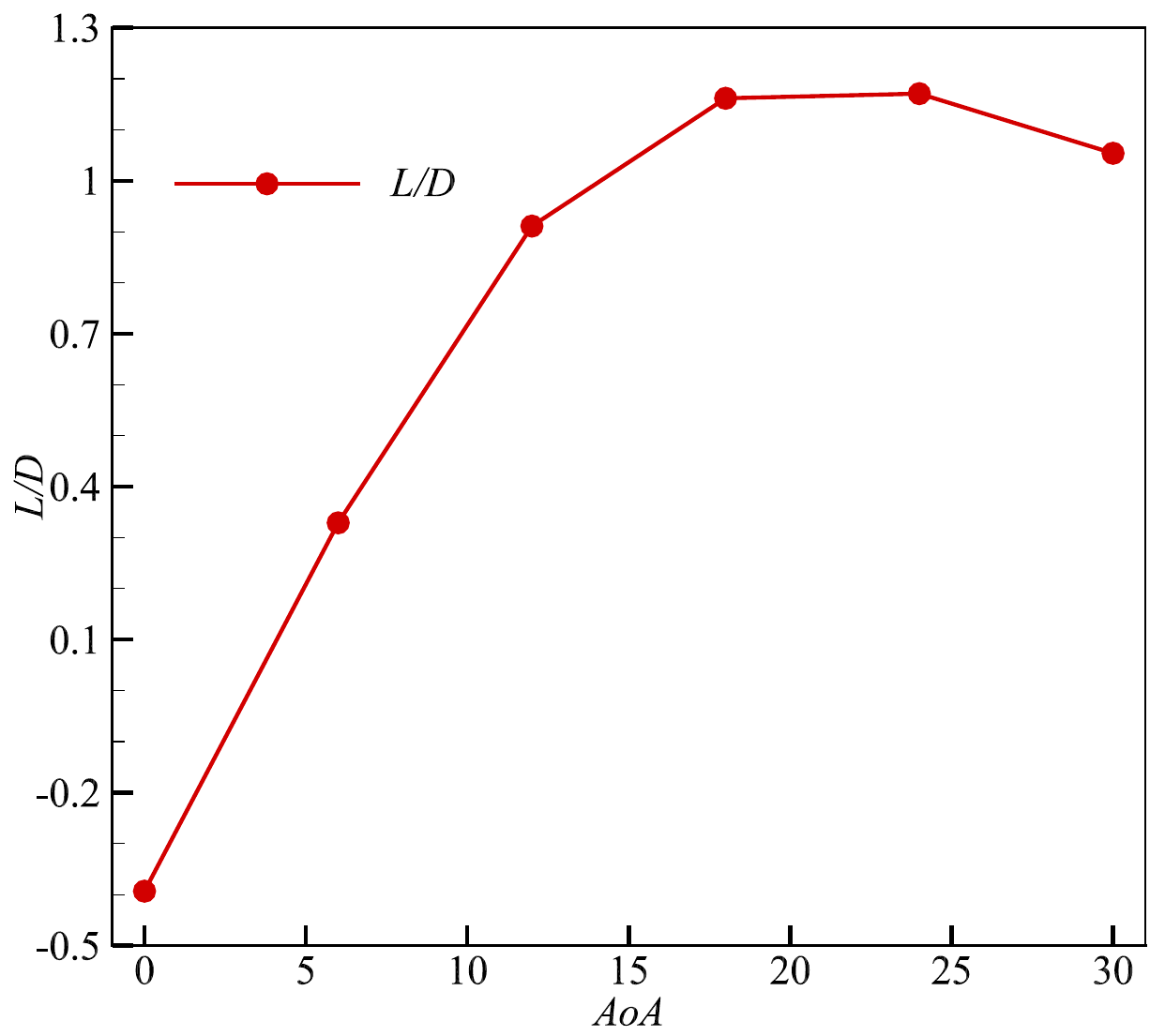}
    	}
	\caption{Effect of $AoA$ on the aerodynamic performance of the HTV-2 type baseline configuration at 85 km.}
    \label{fig: aerodynamic forces with aoa}
\end{figure}

To determine the specific flight attitude for the aerodynamic shape optimization, the influence of the angle of attack ($AoA$) on the aerodynamic forces is systematically investigated at the 85 km design point. Fig.~\ref{fig: aerodynamic forces with aoa} illustrates the variations of the lift coefficient ($C_L$), drag coefficient ($C_D$), and $L/D$ across a range of $AoA$. As depicted in the figure, to maintain an adequate safety margin, an $AoA$ of $18^\circ$ is conservatively selected as the target condition for the subsequent shape optimization.

\begin{figure}
	\centering
	\subfigure[$C_p$ at 70~km]
        {
			\includegraphics[height=0.2 \textwidth]{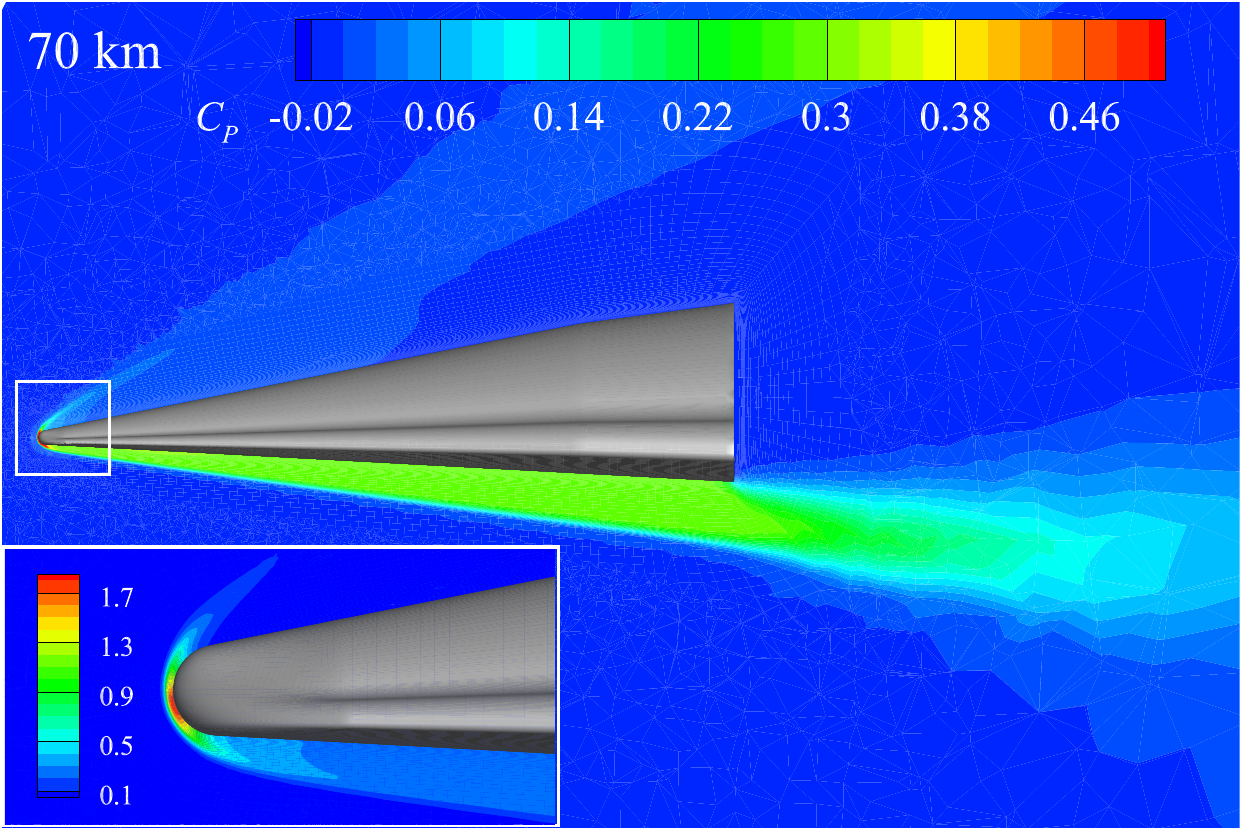}
            \label{fig: 3a}
		}
    \subfigure[$T$ at 70~km]
        {
    		\includegraphics[height=0.2 \textwidth]{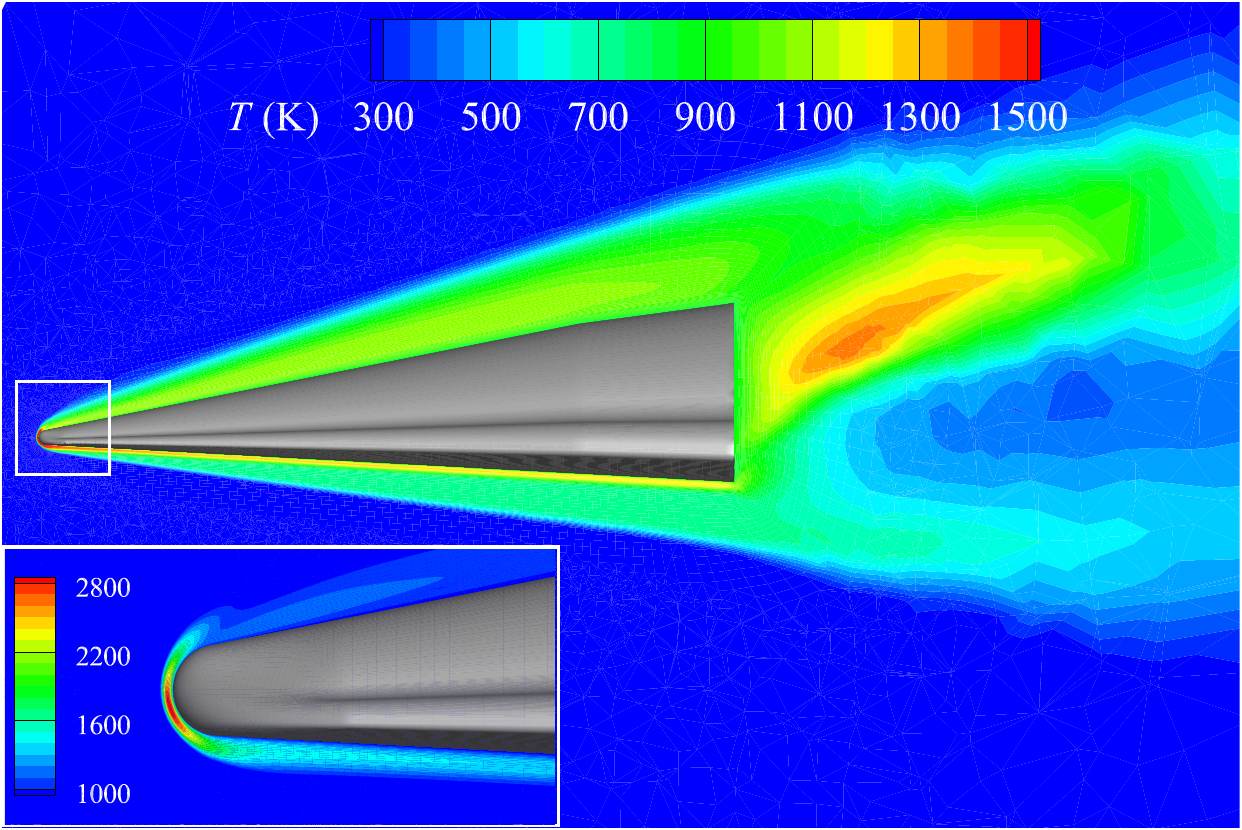}
            \label{fig: 3b}
    	}
    \subfigure[$\mathrm{Kn}_{\mathrm{GLL}}$ at 70~km]
        {
			\includegraphics[height=0.2 \textwidth]{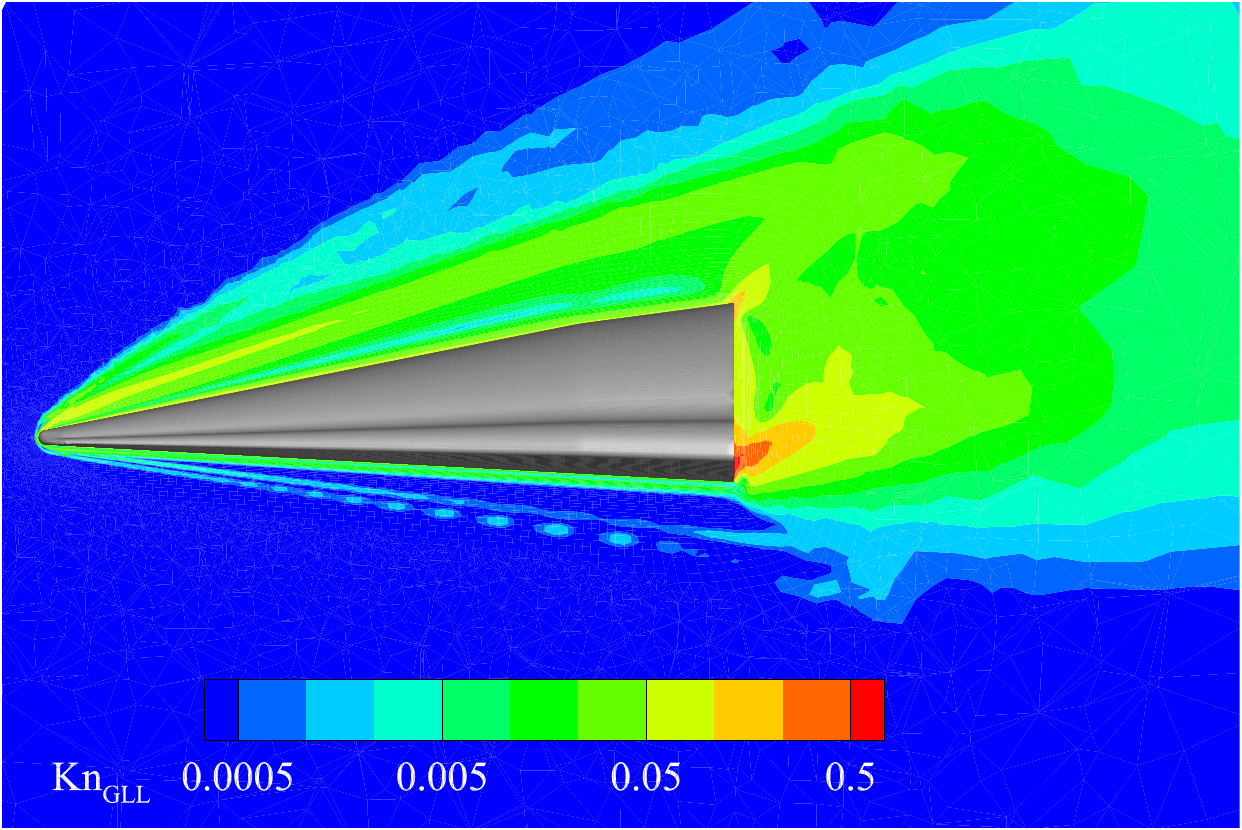}
            \label{fig: 3c}
		}
    \subfigure[$C_p$ at 85 km]
        {
			\includegraphics[height=0.2 \textwidth]{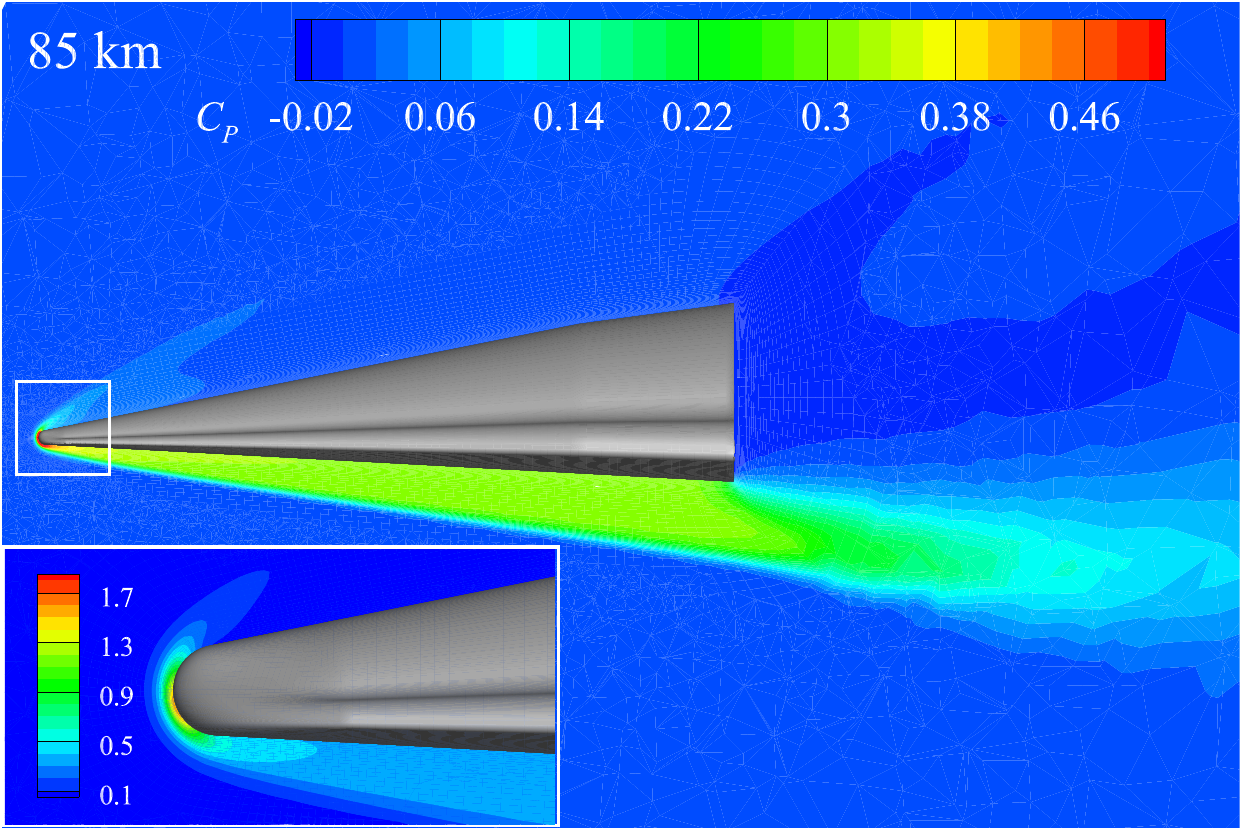}
            \label{fig: 3d}
		}
    \subfigure[$T$ at 85 km]
        {
    		\includegraphics[height=0.2 \textwidth]{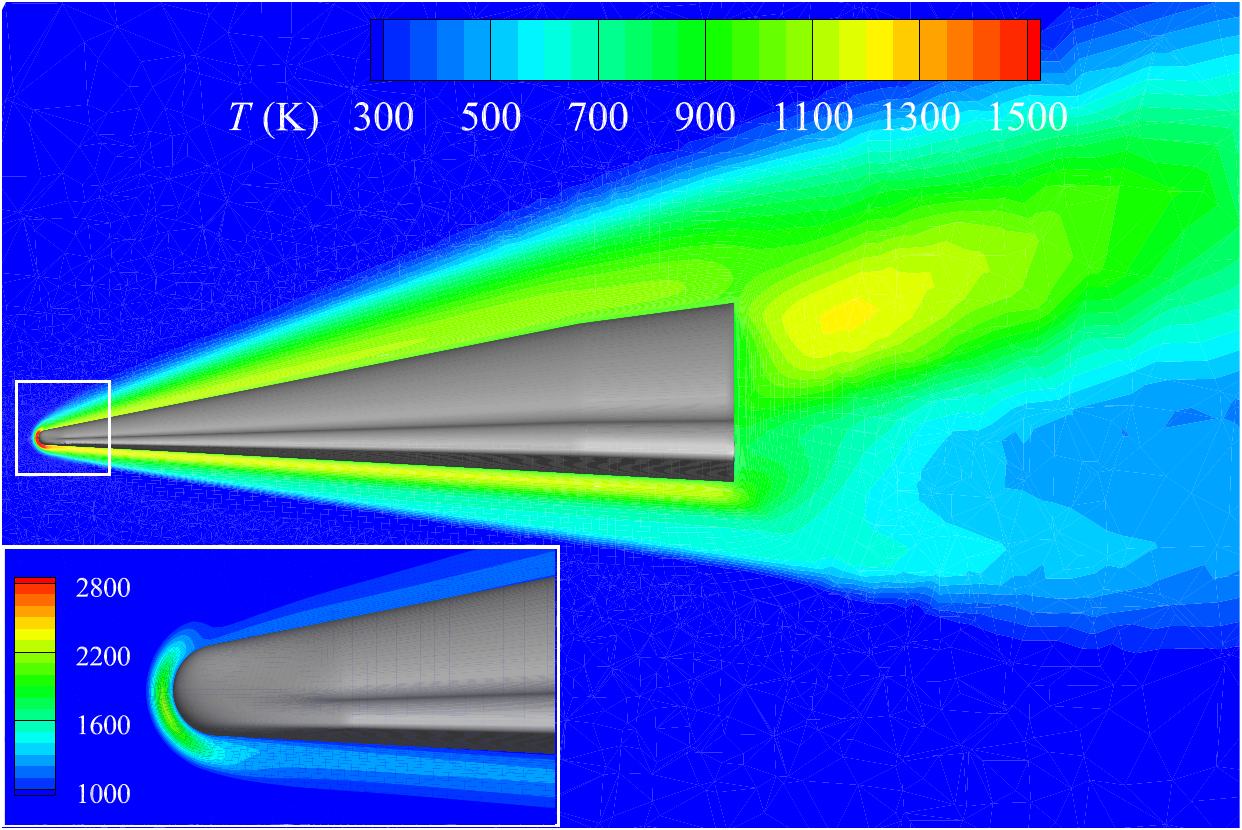}
            \label{fig: 3e}
    	}
    \subfigure[$\mathrm{Kn}_{\mathrm{GLL}}$ at 85 km]
        {
			\includegraphics[height=0.2 \textwidth]{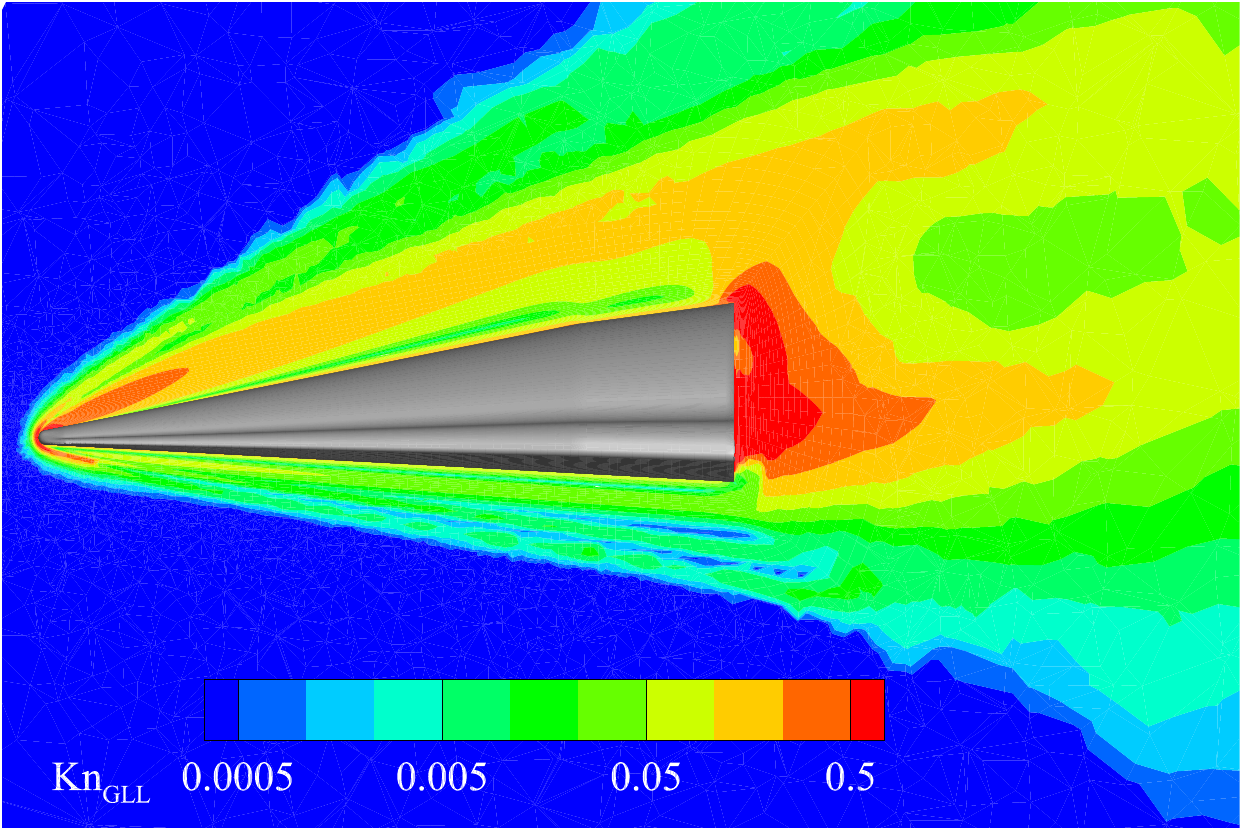}
            \label{fig: 3f}
		}
    \subfigure[$C_p$ at 100 km]
        {
			\includegraphics[height=0.2 \textwidth]{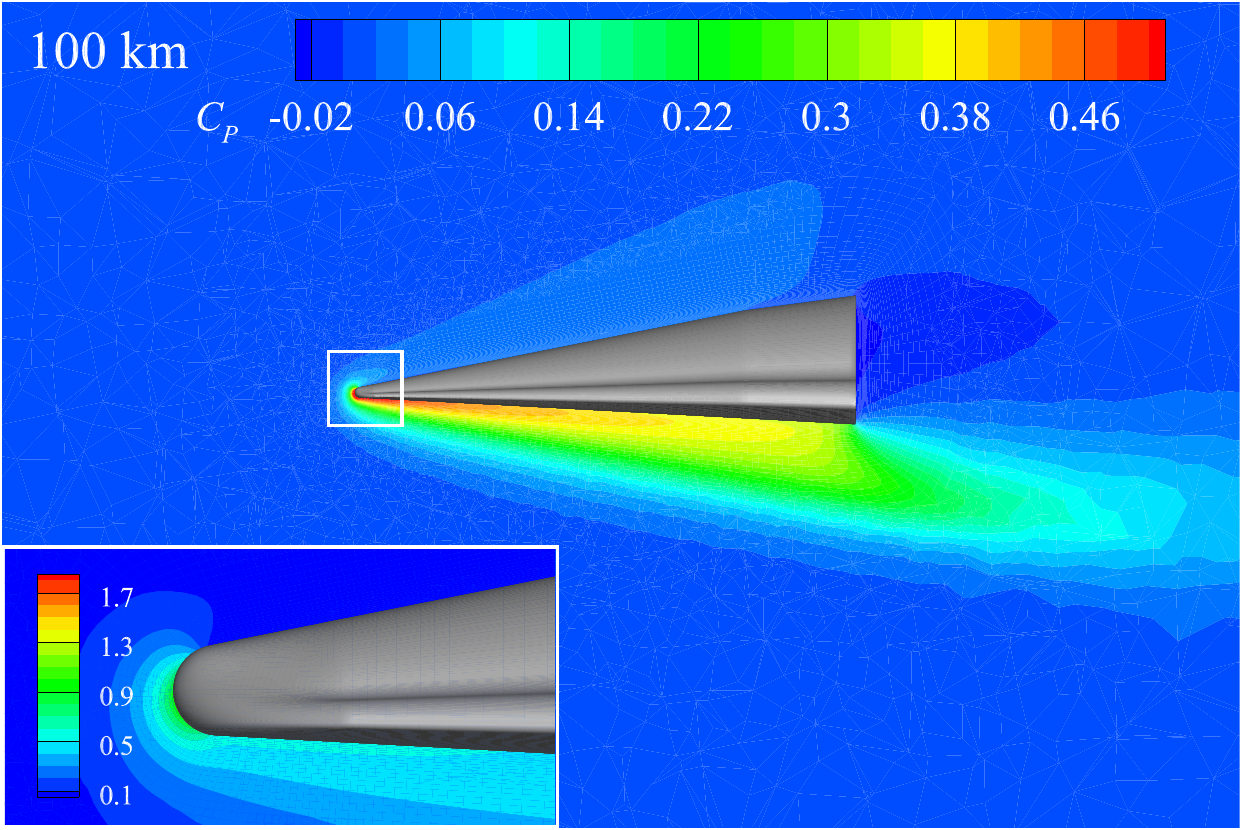}
            \label{fig: 3g}
		}
    \subfigure[$T$ at 100 km]
        {
    		\includegraphics[height=0.2 \textwidth]{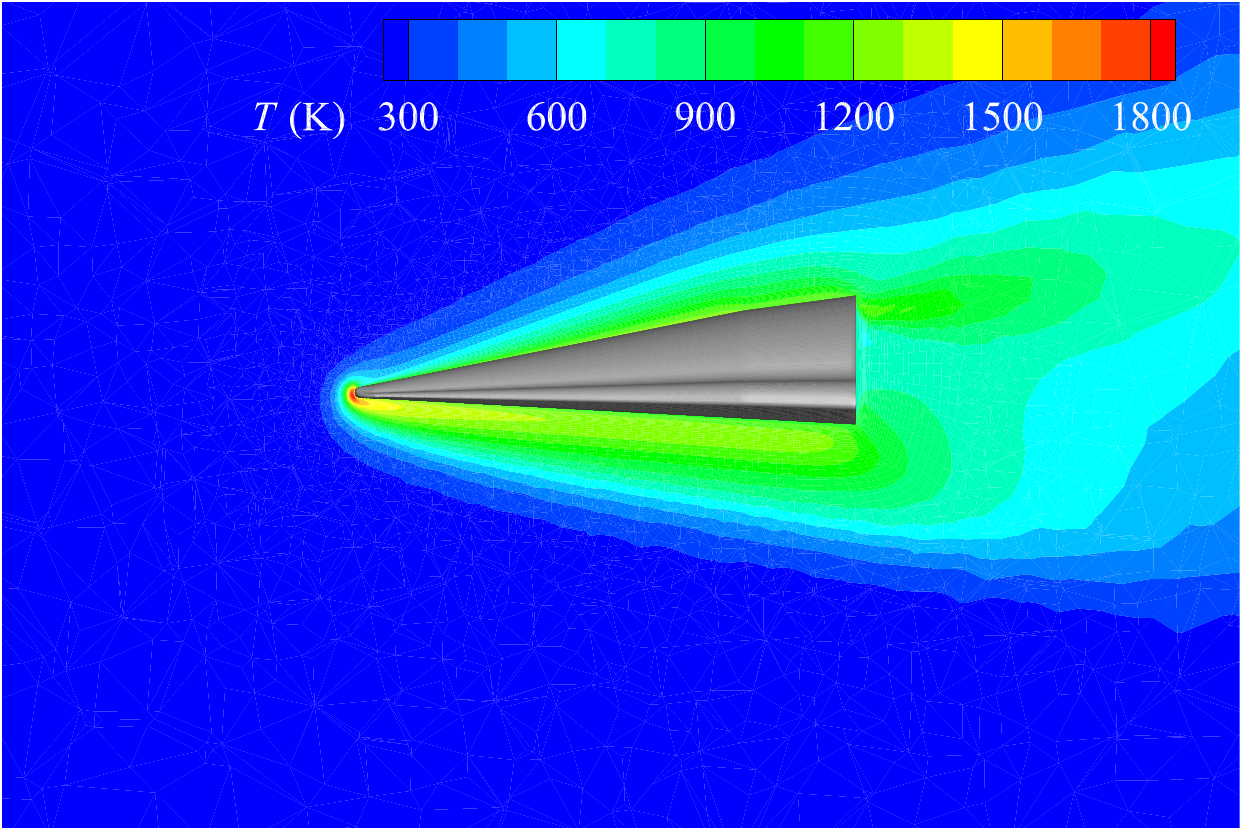}
            \label{fig: 3h}
    	}
    \subfigure[$\mathrm{Kn}_{\mathrm{GLL}}$ at 100 km]
        {
			\includegraphics[height=0.2 \textwidth]{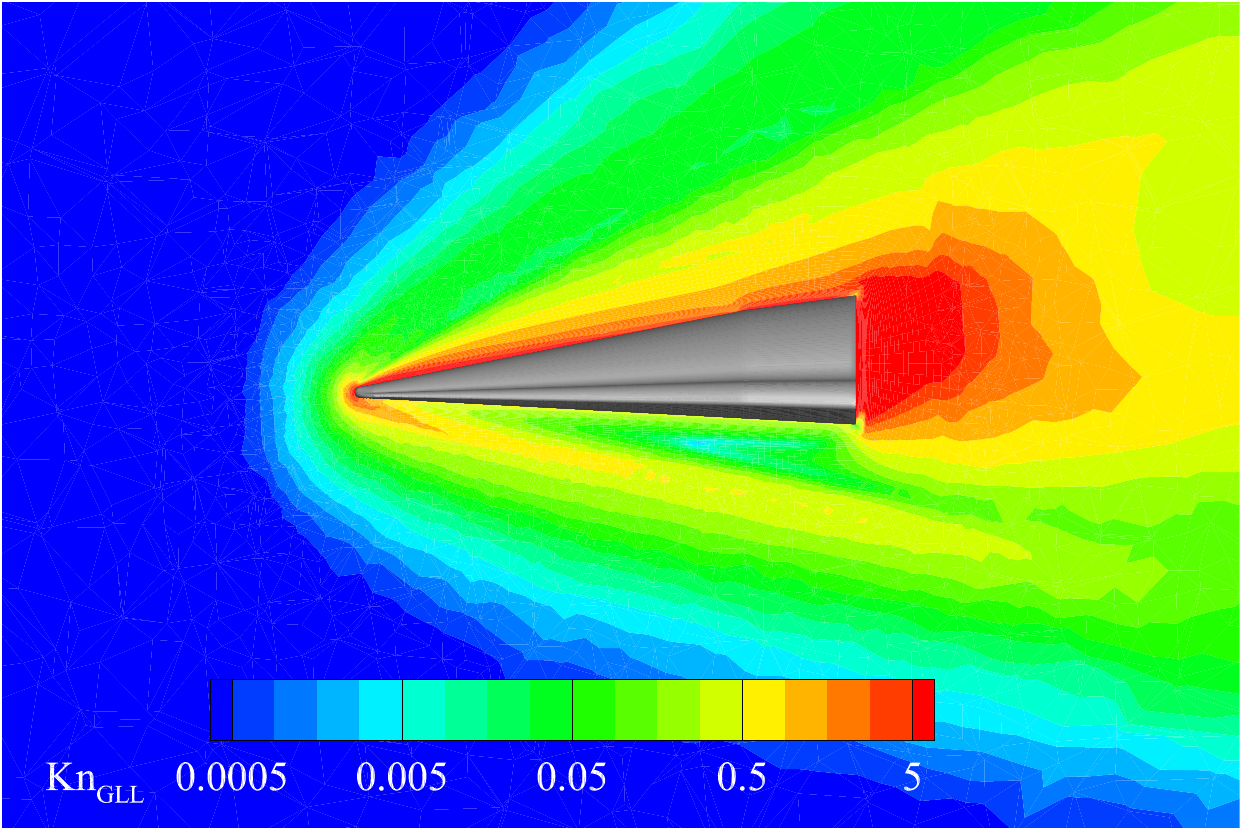}
            \label{fig: 3i}
		}
    \subfigure[$C_p$ at 120 km]
        {
			\includegraphics[height=0.2 \textwidth]{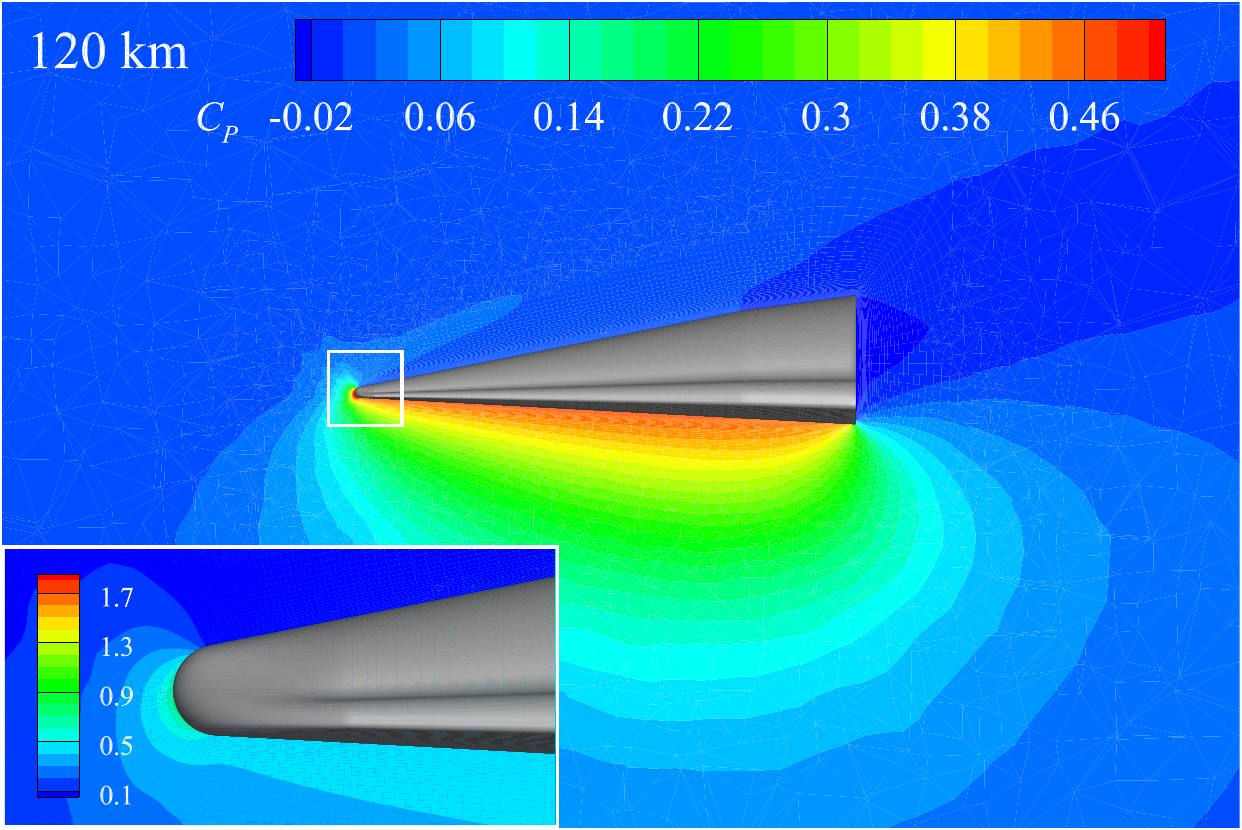}
            \label{fig: 3j}
		}
    \subfigure[$T$ at 120 km]
        {
    		\includegraphics[height=0.2 \textwidth]{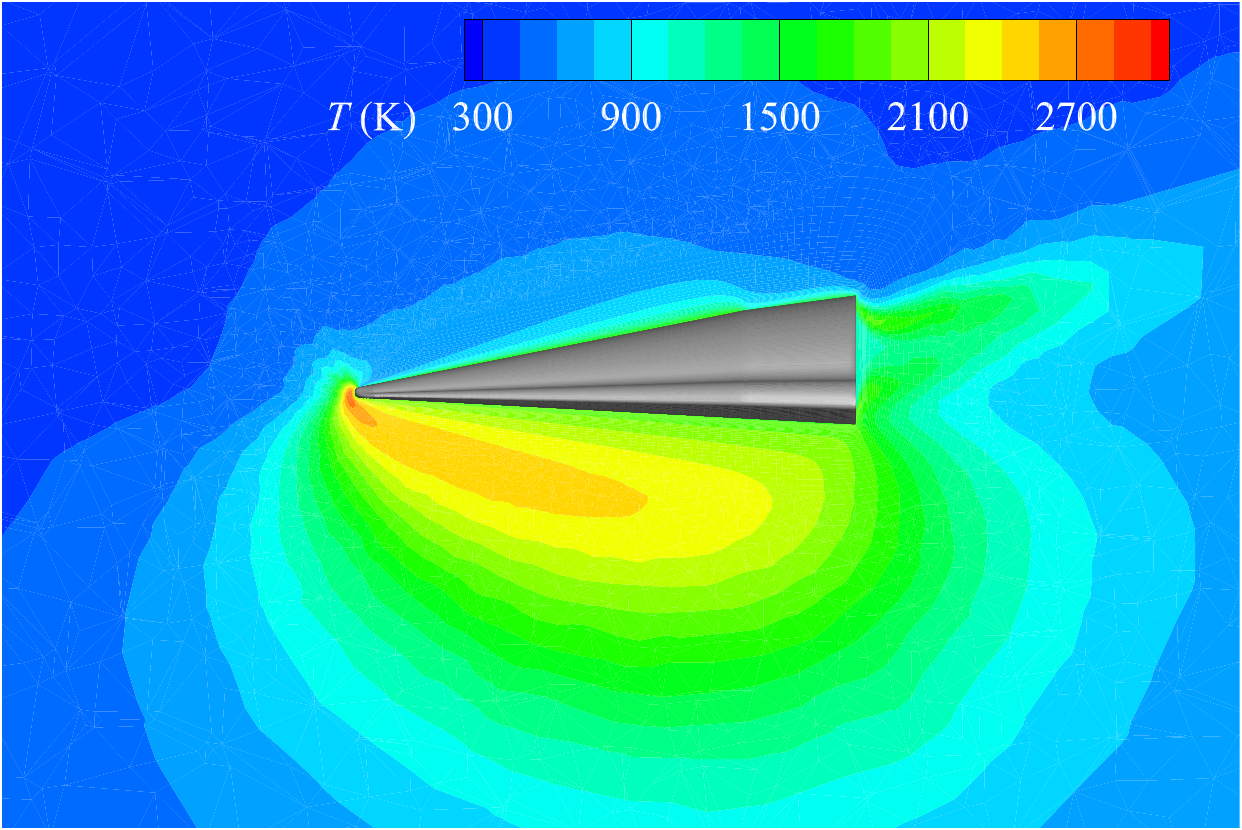}
            \label{fig: 3k}
    	}
    \subfigure[$\mathrm{Kn}_{\mathrm{GLL}}$ at 120 km]
        {
			\includegraphics[height=0.2 \textwidth]{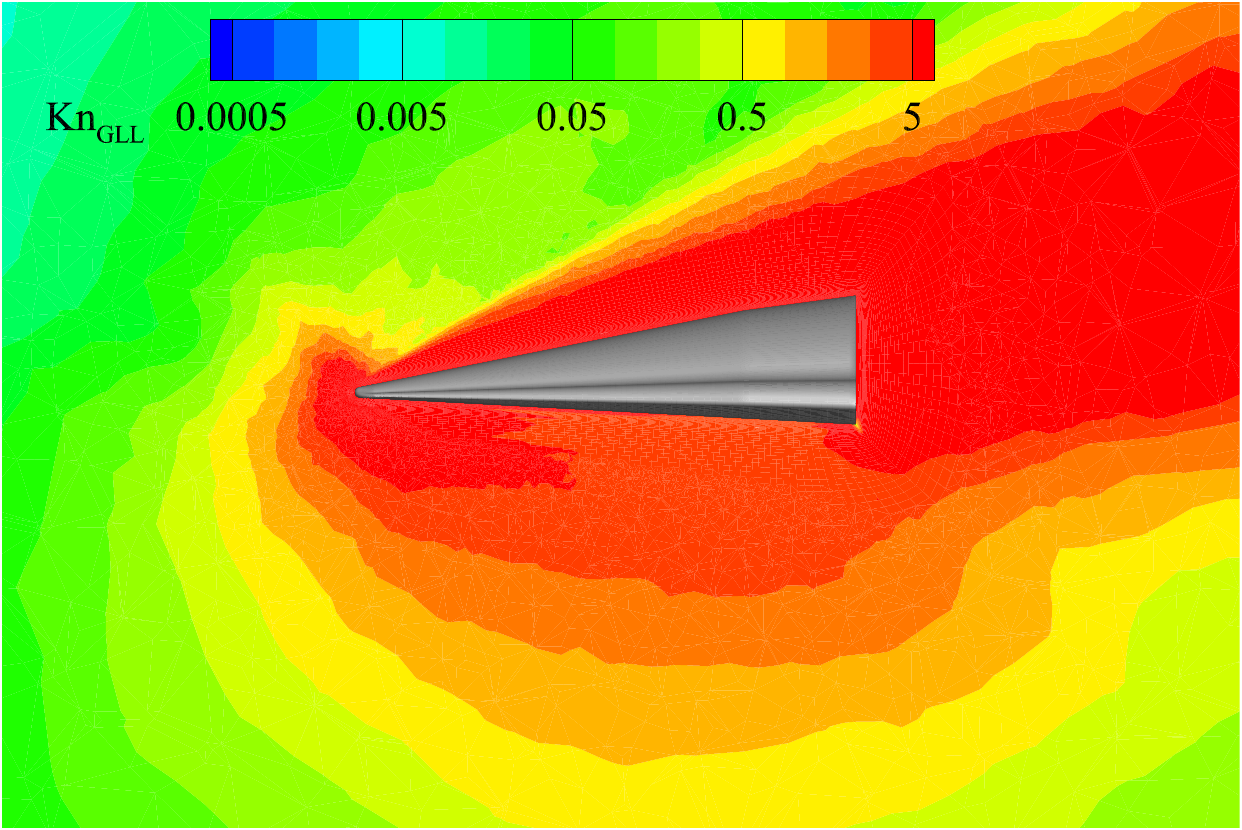}
		}
	\caption{Comparison of the aerodynamic performance on the symmetry plane of the HTV-2 type baseline configuration at varying altitudes.}
    \label{fig: compare}
\end{figure}

To further investigate the aerodynamic characteristics of HTV-2 type aircraft, the flow field around the configuration at a freestream Mach number of 8 and an AoA of $18^\circ$ is analyzed. Fig.~\ref{fig: compare} illustrates the pressure coefficient ($C_p$) distributions on the symmetry plane across different altitudes.
As shown in the figure, due to the relatively large incident angle, a strong shock wave forms on the windward side, while an expansion wave develops over the leeward surface. Furthermore, as the flight altitude increases and the freestream density drops, the shock wave layer becomes noticeably thicker and more diffuse, which is a typical feature of high-altitude rarefied aerodynamics. Consequently, in highly rarefied conditions, $C_p$ near the windward side is generally higher than that at lower altitudes.

\begin{figure}
    \centering
    \includegraphics[width=0.6 \textwidth]{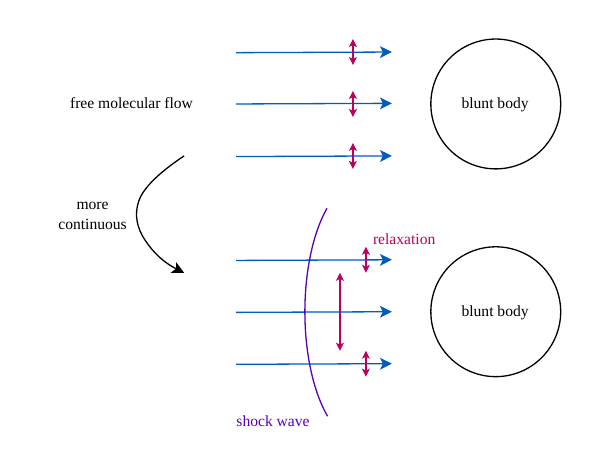}
    \caption{Schematic of the $C_p$ variation mechanism.}
    \label{fig: shock_relaxation_mechanism}
\end{figure}

This phenomenon is fundamentally driven by the gas relaxation process. As the altitude decreases from 120~km to 85~km, $C_p$ on the windward surface drops significantly. This mechanism can be understood by taking the free molecular flow as a baseline, as illustrated in the upper schematic of Fig.~\ref{fig: shock_relaxation_mechanism}. In the collisionless limit, $C_p$ at the wall is strictly proportional to the number flux of the freestream particles. Although particles exhibit random thermal motion, the uniformity of the incoming flow ensures that the net particle flux toward the wall remains unaffected.
However, as $\mathrm{Kn}$ decreases, intermolecular collisions drive the gas toward a local near-equilibrium state behind the bow shock. Because the shock is curved, the macroscopic temperature is highest at the central stagnation region and decreases radially outward. Consequently, particles near the center acquire a higher thermal velocity. Driven by this temperature gradient, a net flux of particles migrates radially outward from the high-temperature center. This outward migration reduces the local particle density impacting the central windward surface, thereby lowering $C_p$ compared to the large-$\mathrm{Kn}$ case (as shown in the lower schematic).
As the altitude drops further to 70~km, the difference in the $C_p$ distribution becomes marginal, because the flow approaches the continuum regime where inviscid mechanisms dominate the surface pressure. Ultimately, these altitude-dependent physical mechanisms dictate the distinct, cross-regime optimization strategies discussed in Sec.~\ref{sec: Results and analyses}.

In this study, the degree of local thermal non-equilibrium is characterized by the gradient length local Knudsen number, $\rm{Kn_{GLL}}$. Following the criteria established in Refs.~\citep{wei2023adaptive, hash1996assessment, sun2004hybrid, schwartzentruber2007modular}, $\rm{Kn_{GLL}}$ is defined as:
\begin{equation}
    \mathrm{Kn}_{\mathrm{GLL}} = \frac{\lambda_\mathrm{mfp}}{\rho / |\nabla \rho|},
    \label{eq: kngll}
\end{equation}
where $\lambda_\mathrm{mfp}$ represents the local mean free path. This parameter is employed to identify non-equilibrium regions by coupling the local macroscopic gradient with the microscopic characteristic length. Generally, flow regions where $\mathrm{Kn}_{\mathrm{GLL}} > 0.05$ are considered to be in a highly rarefied, non-equilibrium state.

Fig.~\ref{fig: compare} illustrates the $\mathrm{Kn}_{\mathrm{GLL}}$ distributions on the symmetry plane. The contours clearly demonstrate a significant magnification of multiscale effects as the altitude rises. At 70~km, the rarefied region is strictly confined to a small area in the wake of the fuselage. At 85 km, the non-equilibrium multiscale effects become more pronounced around the fuselage. By 100 km, almost the entire flow field is highly rarefied, except for a small compression region near the windward surface. Ultimately, at 120 km, the aircraft is completely immersed in a free-molecular-like rarefied environment. This phenomenon indicates the necessity and validity of employing a multiscale kinetic solver for the current study.

It is worth noting that at an altitude of 120 km, the flow is extremely rarefied, resulting in minimal aerodynamic loads. In this regime, the generated lift is entirely insufficient to counteract gravity, whereas the drag penalty remains prominent. Consequently, aerodynamic shape optimization at this altitude requires not only maximizing the $L/D$. As this falls outside the scope of the present study, it will be addressed in dedicated future research. Therefore, the subsequent shape optimization and performance analyses in this paper will focus exclusively on the altitudes of 70, 85, and 100~km.

All numerical simulations are executed on a high-performance computing cluster utilizing 96 cores across multiple Intel 8358 nodes via cross-node parallelization. For the various grid resolutions evaluated, the computational cost for a single converged case at the 85 km altitude is approximately 4 hours of wall-clock time, and the calculation time for the 100 km cases ranges from 4 to 5 hours.

\subsection{Grid independence verification}

Considering the varying Knudsen numbers across the flight trajectory, the flow physics span from the slip regime to the highly rarefied regime. At an altitude of 100 km, the flow is highly rarefied, and the numerical accuracy is predominately governed by the velocity space resolution rather than the physical mesh density. Therefore, the physical grid independence study is systematically conducted at the altitudes of 70~km and 85 km.

To ensure both the accuracy and efficiency of the numerical simulations, three different physical grid densities—categorized as coarse, medium, and refined—are generated for the HTV-2 type configuration. These grids share identical topological structures, with the thickness of the first near-wall cell ($\delta$) set to $1.0 \times 10^{-3}\,\mathrm{m}$, $5.0 \times 10^{-4}\,\mathrm{m}$, and $1.0 \times 10^{-4}\,\mathrm{m}$, respectively. The computational conditions are listed in Tab.~\ref{tab: freestream conditions}, specifically utilizing an angle of attack ($AoA$) of $18^\circ$ at 70~km and $5^\circ$ at 85~km. The dimensionless aerodynamic forces obtained using these different physical grid sizes are summarized in Tab.~\ref{tab: physical grid indep}.

\begin{table}
\begin{center}
\begin{tabular*}{0.8\textwidth}{@{\extracolsep{\fill}}c ccccccc}
Altitudes ($\rm{km}$) & $\delta$ ($\rm{m}$) & $F_y$ & $\upDelta F_y$ & $F_z$ & $\upDelta F_z$ & $M_x$ & $\upDelta M_x$ \\
$70 $   & $1.0 \times 10^{-3}$ & $56.4$ & $0.48\%$ & $274$  & $-0.01\%$ & $531$  & $-0.01\%$ \\
        & $5.0 \times 10^{-4}$ & $56.2$ & $0.18\%$ & $274$  & $0.01\%$  & $531$  & $0.01\%$  \\
        & $1.0 \times 10^{-4}$ & $56.1$ &          & $274$  &           & $531$  &           \\
$85$    & $1.0 \times 10^{-3}$ & $10.7$ & $0.03\%$ & $3.17$ & $-0.09\%$ & $5.53$ & $-0.10\%$ \\
        & $5.0 \times 10^{-4}$ & $10.7$ &          & $3.17$ &           & $5.53$ &           \\
\end{tabular*}
\end{center}
\caption{Aerodynamic forces of the HTV-2 type configuration with different physical grids.}
\label{tab: physical grid indep}
\end{table}

At 70~km, compared to the refined grid, the aerodynamic forces calculated by the medium grid display minor variances, with a maximum relative difference of only 0.18\%. Similarly, at 85 km, the coarse grid demonstrates negligible discrepancies compared to the medium grid, with the maximum difference being just 0.14\%. Consequently, to strike an optimal balance between computational fidelity and efficiency, the medium grid (with $\delta = 5.0 \times 10^{-4}\,\rm{m}$, containing approximately 1 million unstructured cells) is selected for the physical space discretization at 70~km, whereas the coarse grid (with $\delta = 1.0 \times 10^{-3}\,\rm{m}$) is adopted for the 85 km cases.

\begin{figure}
	\centering
	\subfigure[Physical mesh scale at symmetry plane at 70~km]
        {\label{fig: scale1}
			\includegraphics[height=0.4 \textwidth]{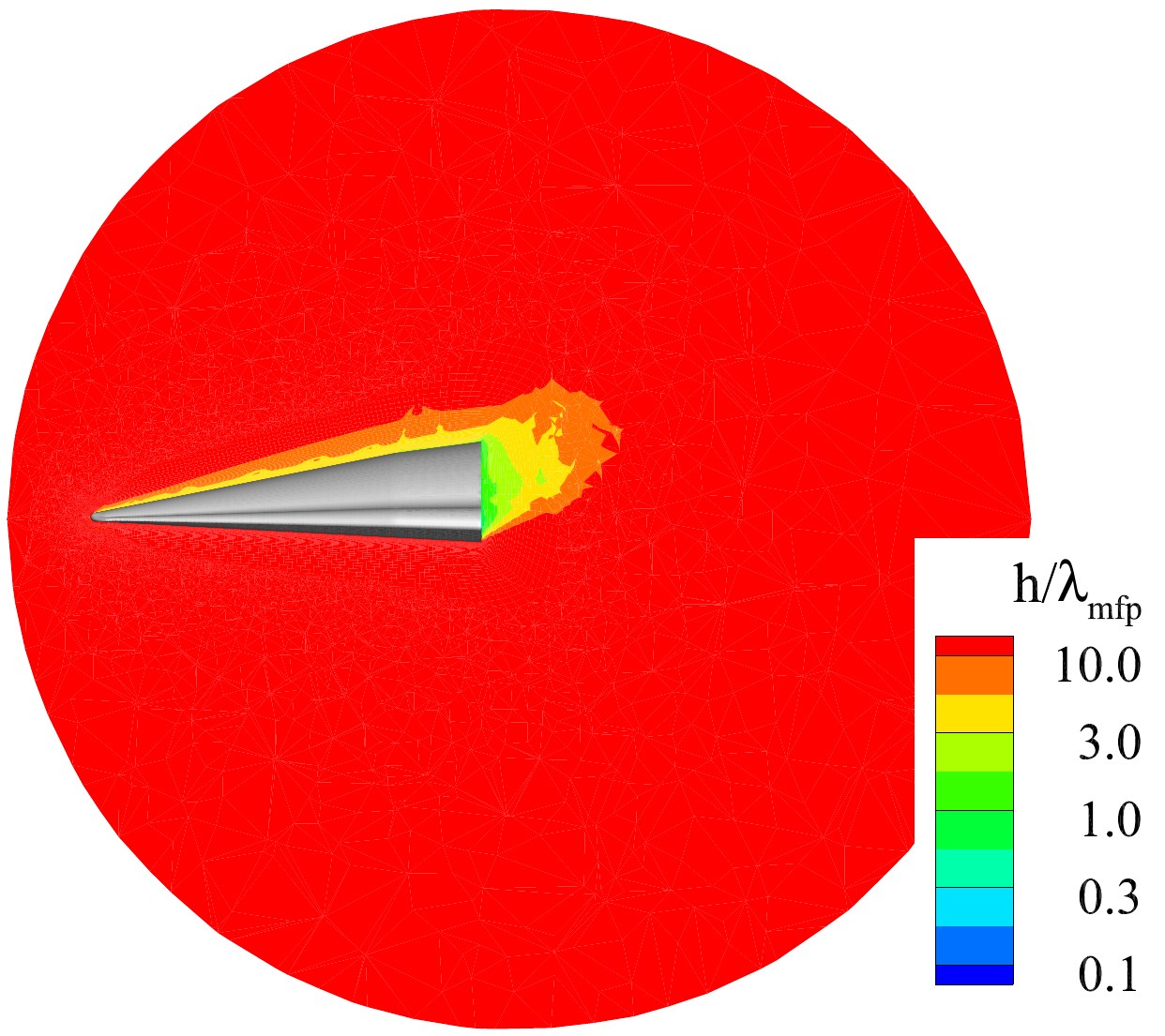}
		}
    \subfigure[The first near-wall cell scale at 70~km]
        {\label{fig: scale2}
			\includegraphics[height=0.4 \textwidth]{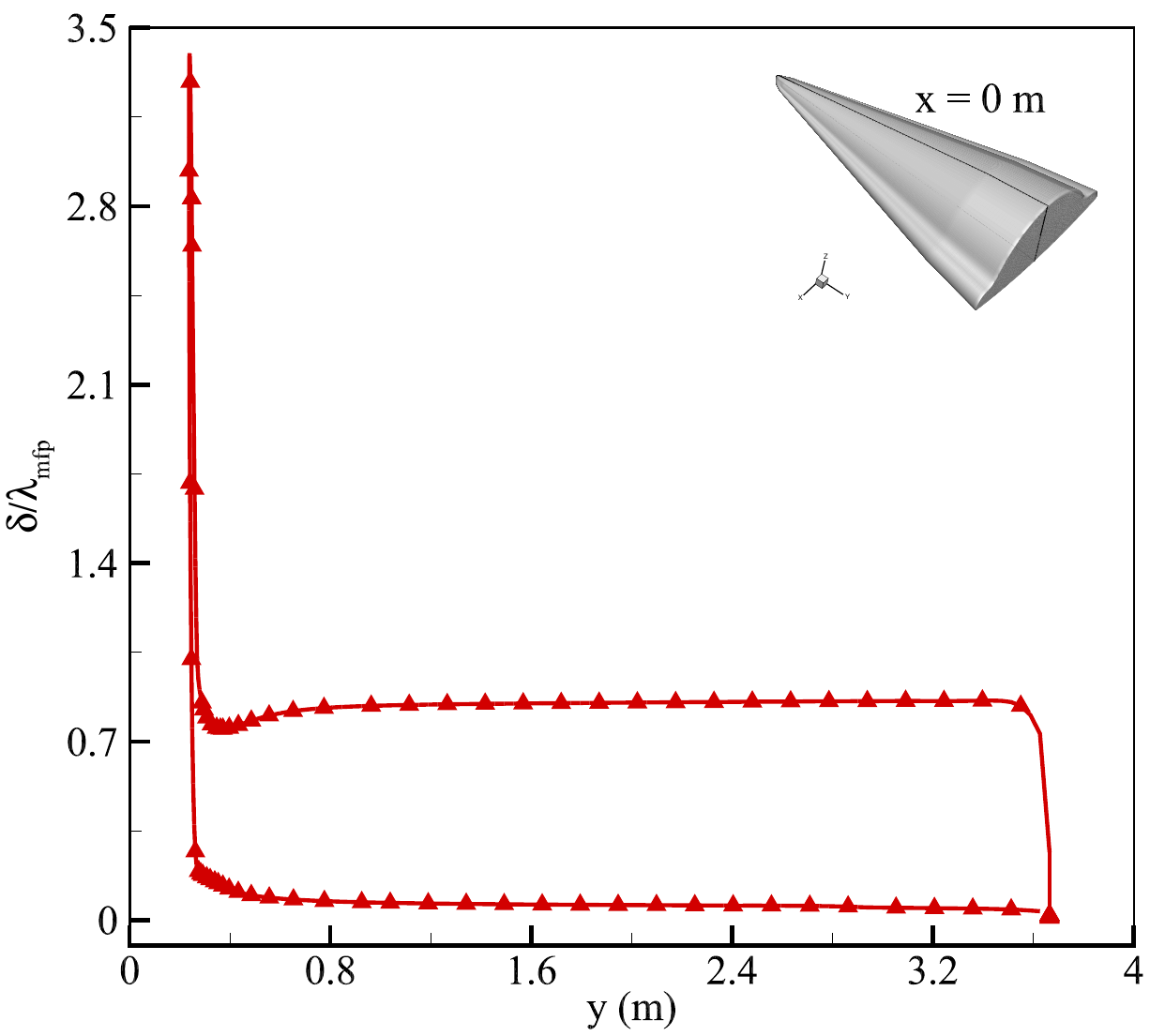}
		}
    \subfigure[Physical mesh consisting of 987796 cells at 85~km]
        {\label{fig: physical space}
			\includegraphics[height=0.4 \textwidth]{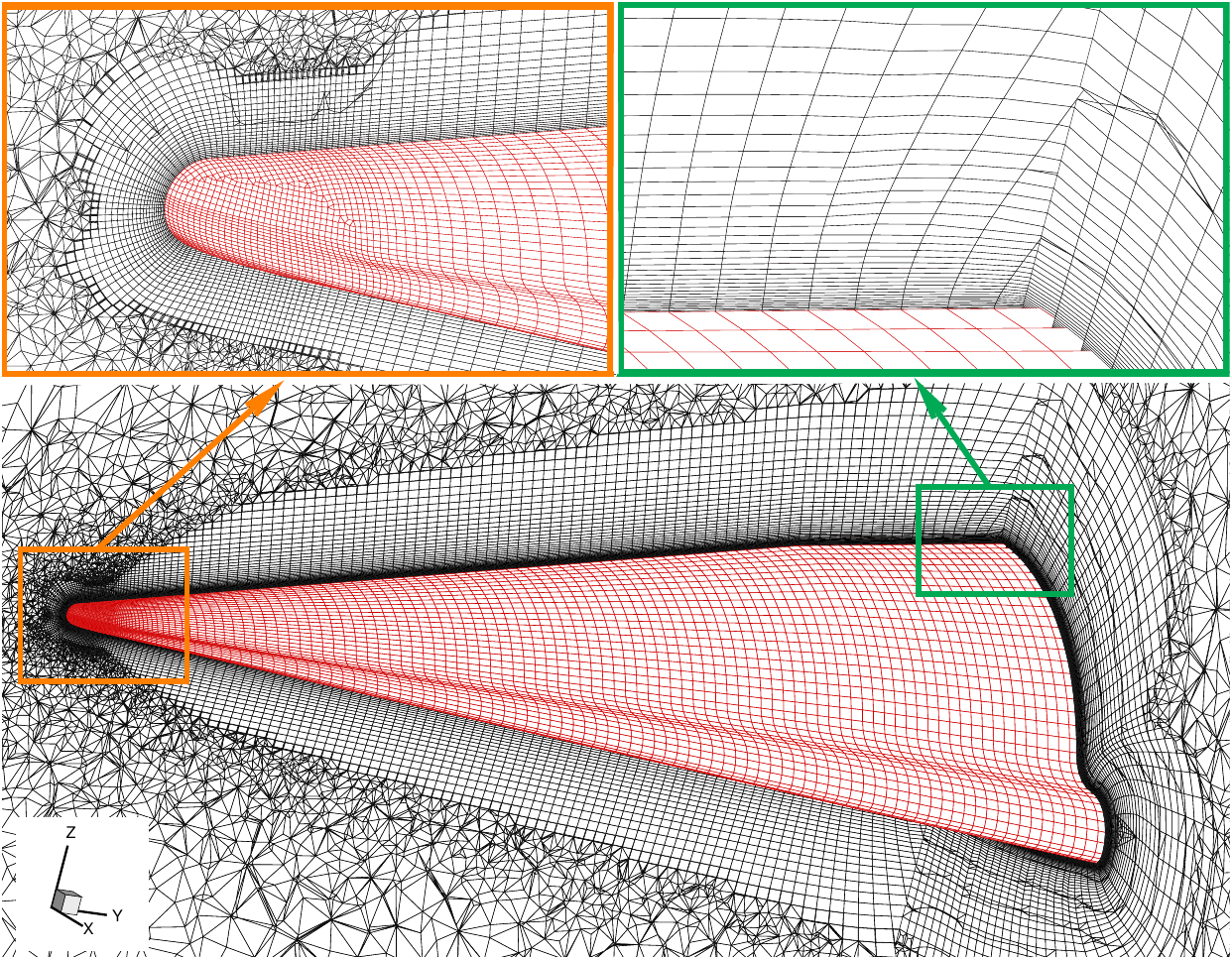}
		}
    \subfigure[Unstructured DVS mesh consisting of 13429 cells at 85~km]
        {\label{fig: velocity space}
    		\includegraphics[height=0.4 \textwidth]{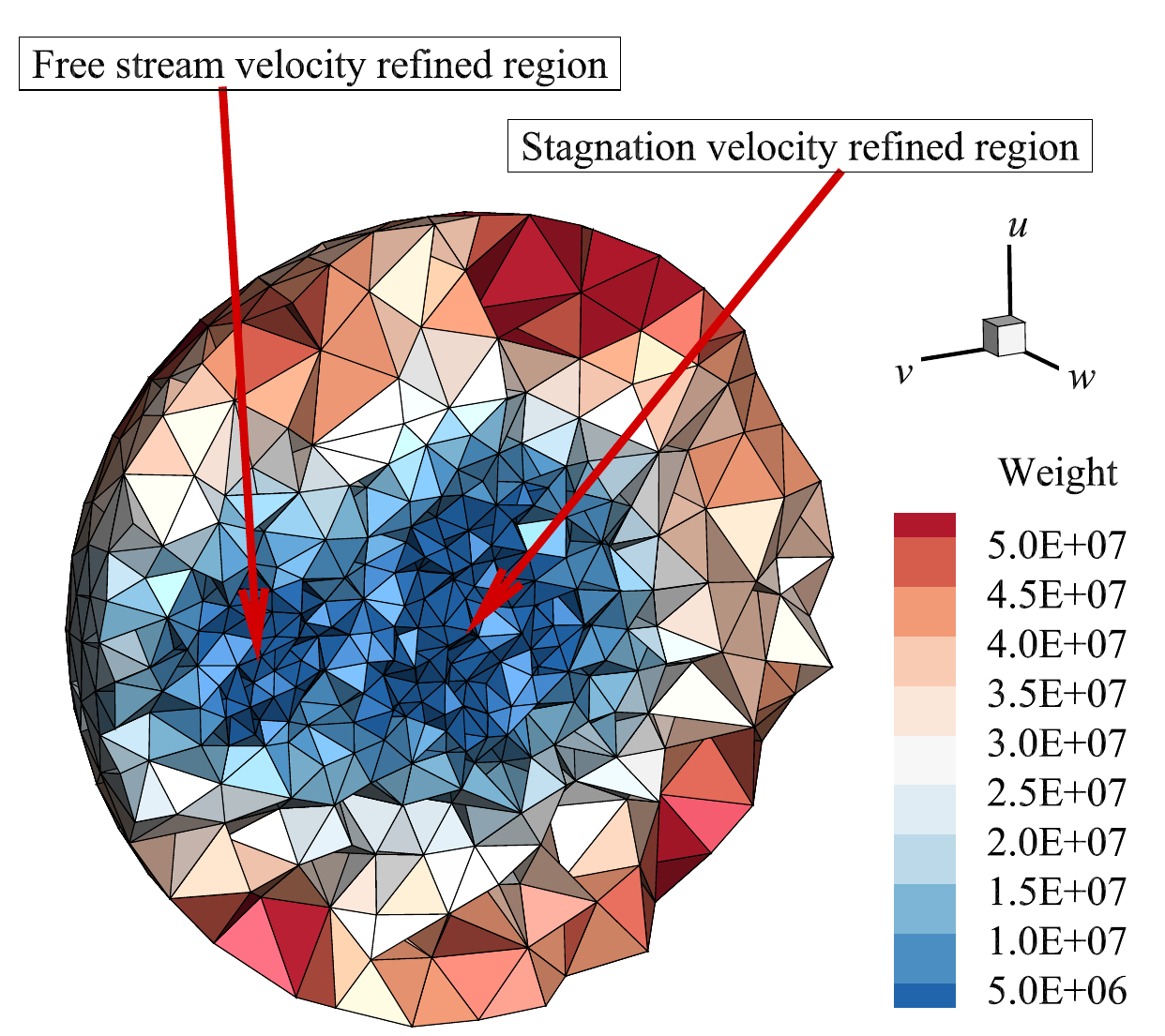}
    	}
	\caption{Schematic diagrams of the adopted discretization strategies.}
    \label{fig: grid independent test}
\end{figure}

Furthermore, Fig.~\ref{fig: scale1} and Fig.~\ref{fig: scale2} present the scale distribution of the medium grid on the symmetry plane at 70~km. For single-scale microscopic method, the cell size $h$ is limited by the $\lambda_{mfp}$. For example, in DSMC, $h$ is commonly required to be less than 1/3 $\lambda_{mfp}$. In Fig.~\ref{fig: scale1}, concentrating on the flow field, $h$ is calculated by root of cell volume. While in Fig.~\ref{fig: scale2}, for the detailed mesh at the wall, it is obvious that the mesh independence is achieved under the case that most mesh size is much larger than 1/3 $\lambda_{mfp}$, including the windward side and farfield, which is a significant efficiency advantage of the multiscale UGKS.

Due to the kinetic nature of the flow solver in the rarefied regime, an unstructured discrete velocity space (DVS) is adopted. An independence test for the DVS is equally crucial. Tab.~\ref{tab: velocity grid indep} presents the aerodynamic forces computed at 85 km using three different discrete velocity space sizes: 8213, 13429, and 17343 points. The data indicates that the intermediate DVS (13429 points) yields a maximum variation of 0.815\% compared to the finest velocity grid, whereas the coarsest DVS (8213 points) produces a larger deviation of up to 2.68\% in the $F_z$ component. Therefore, the DVS with 13429 discrete points is adopted.

\begin{table}
\begin{center}
\begin{tabular*}{0.8\textwidth}{@{\extracolsep{\fill}}c cccccc}
Velocity space & $F_y$ & $\upDelta F_y$ & $F_z$ & $\upDelta F_z$ & $M_x$ & $\upDelta M_x$ \\
8213    & 10.7 & -1.03\% & 3.17 & -2.21\%  & 5.54 & -2.37\% \\
13429   & 10.7 & -0.63\% & 3.26 & 0.4\%    & 5.70 & 0.48\%  \\
17343   & 10.8 &         & 3.24 &          & 5.67 &         \\
\end{tabular*}
\end{center}
\caption{Aerodynamic forces of the HTV-2 type configuration with different discrete velocity spaces at 85 km.}
\label{tab: velocity grid indep}
\end{table}

Fig.~\ref{fig: physical space} and Fig.~\ref{fig: velocity space} visually depict the selected coarse physical computational grid alongside the corresponding unstructured discrete velocity space.

\section{Optimization and sensitivity model}\label{sec: Optimization and sensitivity analysis}

\subsection{Surrogate-based optimization}\label{sec: SBO}

\begin{figure}
	\centering
    \includegraphics[width=1.0 \textwidth]{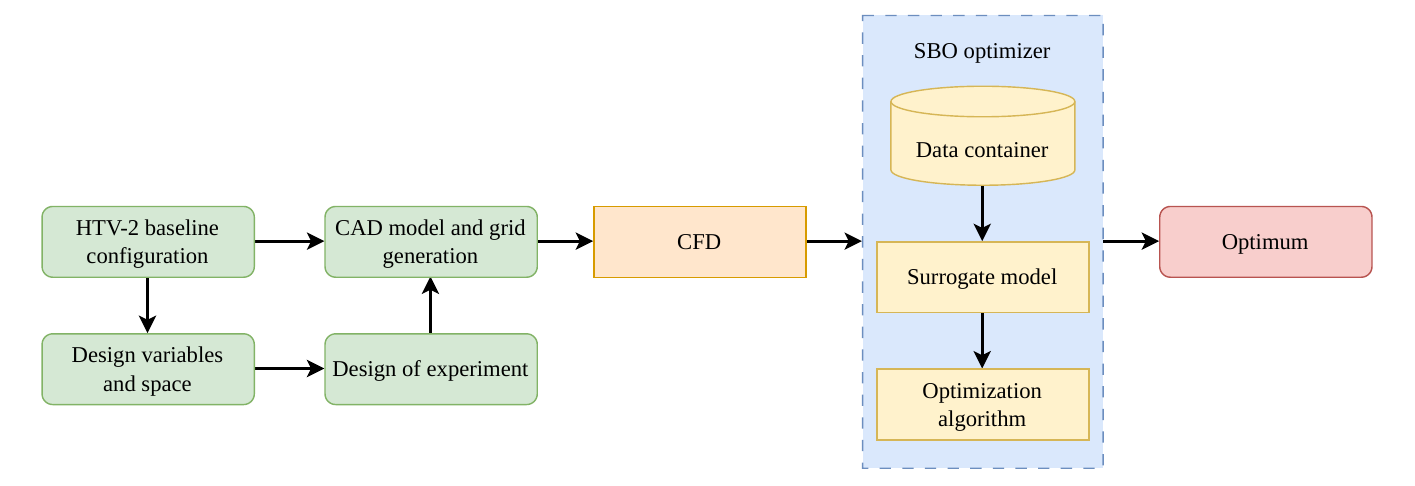}
	\caption{Framework of SBO process.}
    \label{fig: Framework of SBO process.}
\end{figure}

To overcome the prohibitive computational expense of high-fidelity aerodynamic analysis of hypersonic configurations, a SBO~\citep{forrester2009recent, forrester2008engineering} framework is established. As illustrated in Fig.~\ref{fig: Framework of SBO process.}, this automated procedure aims to decouple expensive kinetic simulations from iterative design loops. Beginning with the baseline configuration, the SBO algorithm proceeds by generating the dataset for optimization, constructing the surrogate model, and then utilizing the optimization algorithm to attain the final optimization outcome.

To efficiently explore the multi-dimensional design space, an optimal latin hypercube sampling (OLHS)~\citep{morris1995exploratory} strategy is implemented for the design of experiments (DoE), ensuring a space-filling distribution of training samples with minimal correlation.

To circumvent the curse of dimensionality, the parameterization is restricted to a set of variables. Guided by the HTV-2 design principles (Sec.~\ref{sec: Geometric model}), five relatively independent parameters are chosen: base width, windward surface radius, wing tip bluntness radius, leeward surface radius, and side slope angle. Their prescribed bounds are detailed in Tab.~\ref{tab: Design space of variables}. Using the OLHS algorithm, this design space is iteratively permuted to extract a strictly orthogonal, space-filling dataset, yielding 51 training samples for surrogate construction and 30 independent samples for accuracy validation.

\begin{table}
\begin{center}
\begin{tabular*}{0.8\textwidth}{@{\extracolsep{\fill}}c ccccc}
Design parameters & $l_4$ ($\rm{m}$) & $R_1$ ($\rm{m}$) & $R_2$ ($\rm{m}$) & $R_3$ ($\rm{m}$) & $\theta_3$ ($^\circ$) \\
Lower bound             & 1100 & 2700 & 80  & 400 & 45  \\
Baseline configuration  & 1800 & 5700 & 160 & 880 & 80  \\
Upper bound             & 1820 & 8700 & 200 & 900 & 125 \\
\end{tabular*}
\end{center}
\caption{Design space of variables.}
\label{tab: Design space of variables}
\end{table}

To establish a high-fidelity mapping from the design space to the aerodynamic responses, the Kriging method~\citep{hoerl1970ridge, tikhonov1977solutions} is employed. This model assumes that the observed aerodynamic responses $\boldsymbol{y} = \{y^{(1)}, y^{(2)}, \dots, y^{(n)}\}^T$ from the sampling matrix $\boldsymbol{X} = \{\boldsymbol{x}^{(1)}, \boldsymbol{x}^{(2)}, \dots, \boldsymbol{x}^{(n)}\}^T$ are realizations of a stochastic process with a global mean $\boldsymbol{1}\mu$ and covariance $\sigma^2 \boldsymbol{\Psi}$. The spatial correlation between any two sample points $\boldsymbol{x}^{(i)}$ and $\boldsymbol{x}^{(l)}$ is defined as
\begin{equation}
    \rm{cor}[Y(\boldsymbol{x}^{(i)}), Y(\boldsymbol{x}^{(l)})] = \exp \left( -\sum_{j=1}^d \theta_j |x_j^{(i)} - x_j^{(l)}|^{p_j} \right),
\end{equation}
where $\theta_j$ and $p_j$ denote the correlation decay rate and smoothness parameter, respectively, populating the $n \times n$ symmetric correlation matrix $\boldsymbol{\Psi}$.
To filter potential numerical noise from high-fidelity CFD simulations and prevent overfitting, a regression constant $\lambda$ is added to the main diagonal of $\boldsymbol{\Psi}$. Consequently, the regression-type Kriging predictor at an untried point $\boldsymbol{x}$ is formulated as
\begin{equation}
    \hat{y}_r = \hat{\mu}_r + \boldsymbol{\psi}^T (\boldsymbol{\Psi} + \lambda \boldsymbol{I})^{-1} (\boldsymbol{y} - \boldsymbol{1}\hat{\mu}_r),
\end{equation}
where $\boldsymbol{\psi}$ is the correlation vector between $\boldsymbol{x}$ and the training data, $\boldsymbol{I}$ is the identity matrix. And the optimal estimate of the global mean $\hat{\mu}_r$ is
\begin{equation}
    \hat{\mu}_r = \frac{\boldsymbol{1}^T (\boldsymbol{\Psi} + \lambda \boldsymbol{I})^{-1} \boldsymbol{y}}{\boldsymbol{1}^T (\boldsymbol{\Psi} + \lambda \boldsymbol{I})^{-1} \boldsymbol{1}}.
\end{equation}
Before executing the optimization search, the generalization capability of the established surrogate is quantified using independent testing samples via the coefficient of determination ($R^2$)
\begin{equation}
    R^2 = 1 - \frac{\sum_{i=1}^{n_{\text{test}}} (y_i - \hat{y}_r^{(i)})^2}{\sum_{i=1}^{n_{\text{test}}} (y_i - \bar{y})^2},
\end{equation}
where $\hat{y}_r^{(i)}$ and $y_i$ denote the predicted and actual responses for the testing set.

In this research, since the aerodynamic optimization of the HTV-2 type configuration involves multiple coupled objectives, including $L/D$, center of pressure ($X_{cp}$), and volumetric efficiency ($\eta$) etc., it is necessary to establish individual regression Kriging models for each response.
The $\eta$ serves as a critical geometric constraint for vehicle capacity, which is defined as
\begin{equation}
    \eta = \frac{V^{2/3}}{S_{\text{wet}}},
\end{equation}
where $S_{\text{wet}}$ represents the wet area of the configuration.
To fully utilize the extracted sampling plan and assess the generalization capability, the accuracy of these established models is rigorously examined at the design condition of 85 km altitude. As detailed in Tab.~\ref{tab: Test error of surrogate model}, validation results for the dimensionless parameters demonstrate high fitting accuracy, with a maximum relative error strictly below 0.31\% and an $R^2$ approaching 0.99. These stringent metrics confirm the reliability of the established Kriging models, fully justifying their use as efficient replacements for the expensive kinetic solver in the subsequent optimization search.

\begin{table}
\begin{center}
\begin{tabular*}{0.8\textwidth}{@{\extracolsep{\fill}}c ccc}
Target parameters & Mean relative error (\%) & Maximum relative error (\%) & $R^2$ \\
$C_L$          & 0.04160 & 0.1433  & 0.9976 \\
$C_D$          & 0.04677 & 0.1486  & 0.9981 \\
$L/D$          & 0.05757 & 0.1491  & 0.9991 \\
$X_{cp}$       & 0.02194 & 0.06094 & 0.9992 \\
$\eta$         & 0.07817 & 0.3072  & 0.9991 \\
\end{tabular*}
\end{center}
\caption{Test error of surrogate model at 85 km.}
\label{tab: Test error of surrogate model}
\end{table}

To efficiently and accurately locate the global extrema over the highly nonlinear aerodynamic response surface, a gradient-free hybrid optimization strategy is proposed. Relying solely on local search methods may lead to premature convergence, while pure heuristic algorithms often suffer from limited solution precision. Therefore, a cascaded framework combining a Genetic Algorithm (GA)~\citep{holland1992adaptation} and the Nelder-Mead (NM) simplex method~\citep{ja1965simplex} is adopted. 

In this approach, the GA performs a broad global exploration guided by the Kriging surrogates, handling volumetric constraints via a penalty function to locate the optimal basin. Subsequently, the NM algorithm utilizes the best GA solution as its initial vertex to conduct local refinement, precisely pinpointing the global optimum. The detailed execution of this hybrid procedure is summarized in Algorithm~\ref{alg:hybrid_opt}.

\begin{algorithm}
\caption{Gradient-Free Hybrid Optimization Strategy (GA-NM)}
\label{alg:hybrid_opt}

\vspace{3mm}

\noindent \textbf{Input:} Population $N$, max gen $G$, crossover prob. $p_c$, mutation prob. $p_m$, bounds $[\mathbf{x}_{\min}, \mathbf{x}_{\max}]$. \\
\noindent \textbf{Output:} Global optimum $\mathbf{x}^*_{\text{opt}}$.

\vspace{3mm}

\begin{tabbing}
\hspace*{1.5em} \= \hspace*{1.5em} \= \hspace*{1.5em} \= \hspace*{1.5em} \= \kill

\textbf{\textit{Phase 1: Global Exploration via Genetic Algorithm}} \\
Initialize population $P_0$ within bounds \\
\textbf{for} $g = 1 \to G$ \textbf{do} \\
\> \textbf{for} $i = 1 \to N$ \textbf{do} \\
\> \> $(L/D)_i, \eta_i \leftarrow \text{Kriging}(\mathbf{x}_i)$ \\
\> \> $F(\mathbf{x}_i) \leftarrow -(L/D)_i + \max(0, \eta_{0} - \eta_i) \times \lambda$ \\
\> \textbf{end for} \\
\> Best individual $\mathbf{x}_{\text{elite}} \leftarrow \arg\min_{\mathbf{x}} F(\mathbf{x})$  \\
\> $P_{\text{parent}} \leftarrow \text{Tournament}(P_{g-1})$  \\
\> $P_{\text{child}} \leftarrow \text{Crossover}(P_{\text{parent}}, p_c)$  \\
\> $P_{\text{child}} \leftarrow \text{Mutate}(P_{\text{child}}, p_m)$  \\
\> $P_g \leftarrow P_{\text{child}} \cup \{\mathbf{x}_{\text{elite}}\}$  \\
\textbf{end for} \\
$\mathbf{x}^*_{\text{GA}} \leftarrow \mathbf{x}_{\text{elite}}$  \\[3mm]

\textbf{\textit{Phase 2: Local Refinement via Nelder-Mead Simplex}} \\
Initialize simplex $\mathbf{S} = \{\mathbf{v}_1, \dots, \mathbf{v}_{d+1}\}$ around $\mathbf{x}^*_{\text{GA}}$ \\
\textbf{while} NM convergence criteria not met \textbf{do} \\
\> Sort $\mathbf{S}$ such that $F(\mathbf{v}_1) \le F(\mathbf{v}_2) \le \dots \le F(\mathbf{v}_{d+1})$ \\
\> $\bar{\mathbf{v}} \leftarrow \frac{1}{d} \sum_{i=1}^{d} \mathbf{v}_i$  \\
\> Reflection point $\mathbf{v}_r \leftarrow \bar{\mathbf{v}} + \alpha (\bar{\mathbf{v}} - \mathbf{v}_{d+1})$  \\
\> \textbf{if} $F(\mathbf{v}_r) < F(\mathbf{v}_1)$ \textbf{then} \\
\> \> Expansion point $\mathbf{v}_e \leftarrow \bar{\mathbf{v}} + \gamma (\mathbf{v}_r - \bar{\mathbf{v}})$ \\
\> \> $\mathbf{v}_{d+1} \leftarrow \mathbf{v}_e$ \text{if} $F(\mathbf{v}_e) < F(\mathbf{v}_r)$ \textbf{else} $\mathbf{v}_r$ \\
\> \textbf{else if} $F(\mathbf{v}_r) \ge F(\mathbf{v}_d)$ \textbf{then} \\
\> \> Contraction point $\mathbf{v}_c \leftarrow \bar{\mathbf{v}} + \beta (\mathbf{v}_{\text{worst}} - \bar{\mathbf{v}})$  \\
\> \> \textbf{if} $F(\mathbf{v}_c) < F(\mathbf{v}_{\text{worst}})$ \textbf{then} $\mathbf{v}_{d+1} \leftarrow \mathbf{v}_c$ \\
\> \>  \textbf{else}  Shrink simplex $\mathbf{v}_i \leftarrow \mathbf{v}_1 + \delta (\mathbf{v}_i - \mathbf{v}_1)$ for $i > 1$  \\
\> \textbf{else} \\
\> \> $\mathbf{v}_{d+1} \leftarrow \mathbf{v}_r$ \\
\> \textbf{end if} \\
\textbf{end while} \\
\textbf{Return} $\mathbf{x}^*_{\text{opt}} \leftarrow \mathbf{v}_1$
\end{tabbing}

\vspace{1mm}
\end{algorithm}

\subsection{Sensitivity model}

Sensitivity analysis investigates the impact of variations in model inputs on outputs. To overcome the limitations of local derivatives, the Sobol global sensitivity method~\citep{sobol2001global, xiaozhe2024surrogate} decomposes the total variance of outputs into the sum of variances of the individual parameters and groups of parameters, evaluating by each parameter's contribution ratio to the output variance.

Consider an aerodynamic response function $y = f(\boldsymbol{x})$ defined within an $n$-dimensional unit hypercube, where the input vector is $\boldsymbol{x} = \{x_1, \cdots, x_n\}$. This function can be decomposed into summands of increasing dimensionality
\begin{equation}
    f(\boldsymbol{x}) = f_0 + \sum_{s=1}^{n} \sum_{1 \le i_1 < \cdots < i_s \le n} f_{i_1 \cdots i_s}(x_{i_1}, \cdots, x_{i_s}),
    \label{eq: sobol_decomposition}
\end{equation}
where $f_0$ is a constant representing the expected value of the response, and each subsequent term depends only on its explicitly specified subset of variables. This decomposition is strictly unique if the following orthogonality condition is established:
\begin{equation}
    \int_0^1 f_{i_1 \cdots i_s}(x_{i_1}, \cdots, x_{i_s})\mathrm{d} x_k = 0 \quad \text{for } k = i_1, \cdots, i_s .
\end{equation}
Members in Eq.~\ref{eq: sobol_decomposition} can be expressed as integrals of $f(\boldsymbol{x})$,
\begin{equation}
    \begin{aligned}
        &\int f(\boldsymbol{x}) \mathrm{d} \boldsymbol{x} = f_0 \quad \text{for } \mathrm{d} \boldsymbol{x} = \mathrm{d} x_1 \cdots \mathrm{d} x_n , \\
        &\int f(\boldsymbol{x})\prod_{k \neq i}\mathrm{d} x_k = f_0 + f_i(x_i), \\
        &\int f(\boldsymbol{x})\prod_{k \neq i,j}\mathrm{d} x_k = f_0 + f_i(x_i) + f_j(x_j) + f_{ij}(x_i, x_j).
    \end{aligned}
\end{equation}
The variance of $y$ can be defined as
\begin{equation}
    \begin{aligned}
        \text{Var(Y)} &= \int f^2(\boldsymbol{x})d\boldsymbol{x} - f_0^2 \\
        &= \sum_{s=1}^{n} \sum_{i_1 < \cdots < i_s}^{n} \int f_{i_1 \cdots i_s}^2 \mathrm{d} x_{i_1} \cdots \mathrm{d} x_{i_s}.
    \end{aligned}
\end{equation}
And the variances based on effects of input $\boldsymbol{x}$ can be written as
\begin{equation}
    \text{Var}(\text{Y}_{i_1 \cdots i_s})= \int f_{i_1 \cdots i_s}^2 \mathrm{d} x_{i_1} \cdots \mathrm{d} x_{i_s} ,
\end{equation}
where $\text{Var(Y)} = \sum_{s=1}^{n} \sum_{i_1 < \cdots < i_s}^{n} \text{Var}(\text{Y}_{i_1 \cdots i_s})$. The corresponding Sobol global sensitivity indices can be expressed as
\begin{equation}
    S_{i_1 \cdots i_s} = \frac{\text{Var}(\text{Y}_{i_1 \cdots i_s})}{\text{Var(Y)}} .
\end{equation}
In this case, the first-order sensitivity indices which represent the contribution to the output variance of an individual input $x_i$, which can be written as:
\begin{equation}
    S_i = \frac{\text{Var}(\text{Y}_{i})}{\text{Var(Y)}} = \frac{\int f_i^2 \mathrm{d} x_i}{\int f^2(\boldsymbol{x})d\boldsymbol{x} - f_0^2}.
\end{equation}

\section{Results and analyses}\label{sec: Results and analyses}

In this section, based on the established Kriging surrogate models, the single-objective aerodynamic optimization of the HTV-2 type configuration is independently executed across three representative altitudes: 70~km, 85 km (the primary design condition), and 100 km. To ensure comparative consistency, the freestream conditions for the 70~km and 100 km cases are prescribed in reference to the 85 km baseline, as detailed in Sec.~\ref{sec: Geometric model}.

\subsection{Aerodynamic optimization at 85 km.}

The single-objective aerodynamic optimization under the primary design condition at the altitude of 85 km is mathematically expressed as follows:
\begin{equation}
    \begin{aligned}
        & \min \quad F(\boldsymbol{X}) = -L/D ,\\
        & \text{where} \quad \boldsymbol{X} = (l_4, R_1, R_2, R_3, \theta_3) ,\\
        & \text{s.t.} \quad
        \begin{cases}
            X_{cp} \in [0.9 X_{cg}, 1.1 X_{cg}] \\
            \eta \ge 0.92 \eta_0
        \end{cases},
    \end{aligned}
\end{equation}
where $\boldsymbol{X}$ is the design space shown in Tab.~\ref{tab: Design space of variables}, $X_{cg}$ denotes the center of gravity, empirically located at $0.63l_2$ based on engineering practice, and $\eta_0$ represents the volumetric efficiency of the baseline configuration. During the hypersonic cruising phase, the achievable range is positively correlated with the aerodynamic efficiency. Therefore, maximizing the $L/D$ is chosen as the optimization goal. To ensure the adequate ability to trim and control the aircraft, a constraint is imposed on the longitudinal shift of the center of pressure. Its allowable variation is restricted within a $\pm 10\%$ margin relative to the center of gravity. Furthermore, to guarantee sufficient internal space for payload packaging and subsystem integration, a volumetric constraint is incorporated, ensuring that the configuration's volume parameter reduces strictly less than $8 \%  \eta_0$. To enforce these constraints during the evolutionary search, an exterior penalty function is adopted. By heavily penalizing violations of both the minimum volume limit and the allowable $X_{cp}$ shift, the modified unconstrained objective function $F'(\boldsymbol{X})$ is formulated as:
\begin{equation}
    \min F'(\boldsymbol{X}) = F(\boldsymbol{X}) + \lambda \Big[ \max(0, 0.92\eta_0 - \eta) + \max(0, |X_{cp} - X_{cg}| - 0.1X_{cg}) \Big],
\end{equation}
where the penalty factor $\lambda \gg 0$ ensures strict constraint satisfaction.

\subsubsection{Optimization results}

Enabled by the Kriging surrogate models, the proposed hybrid strategy efficiently evaluates over 80,000 geometric configurations—a process that would demand months of computational time using multiscale UGKS solvers—in merely a few minutes. During the execution, the initial GA ($N=200$) successfully establishes a stable global basin around generation 100, followed by the NM refinement to swiftly pinpoint the optimum. This rapid and smooth convergence strongly demonstrates the exceptional computational efficiency and robustness of the framework in addressing strictly constrained aerodynamic design problems.

\begin{figure}
	\centering
    \includegraphics[width=0.5 \textwidth]{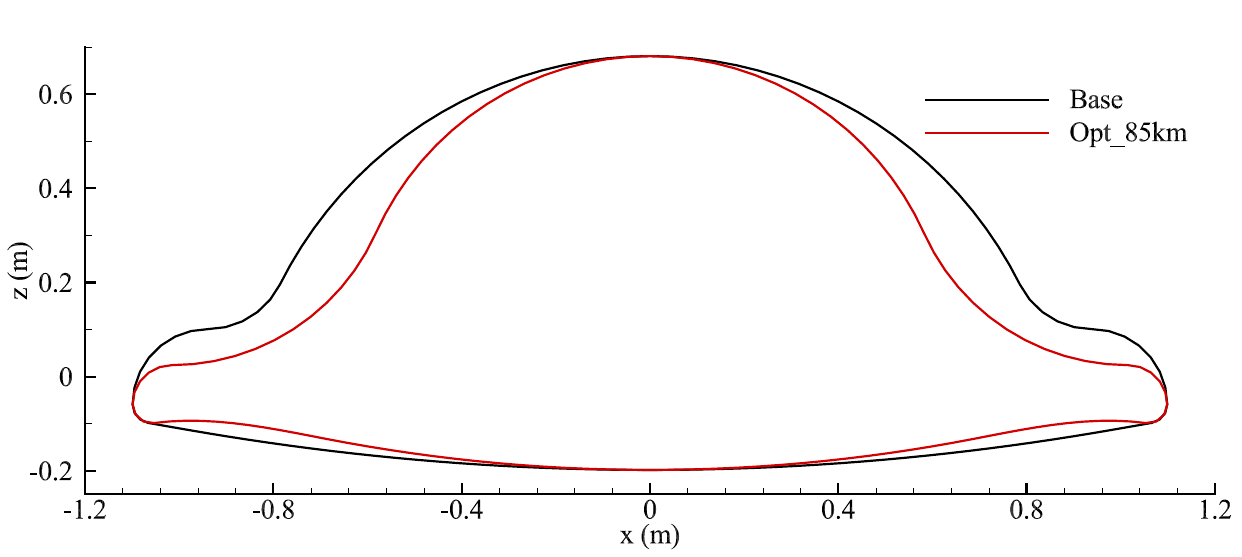}
	\caption{Geometric profile comparison between the baseline and optimized configurations at 85 km.}
    \label{fig: geometric_comparison_85}
\end{figure}

The design variables and aerodynamic forces before and after optimization are presented in Tab.~\ref{tab: Comparison between base & optimum 85}. As indicated in the table, compared with the baseline configuration, the base width and all radius dimensions of the optimum decrease, whereas the side slope angle increases. The relative variations of these geometric parameters generally range from 20\% to 50\%. Notably, the wing tip bluntness radius experiences the most significant reduction, reaching -47.84\%. As can be seen in the Fig.~\ref{fig: geometric_comparison_85}, the optimized geometry trends toward a flatter and more slender profile to maximize the $L/D$ in the rarefied regime. Meanwhile, all engineering constraints, including the trim control and volume, are strictly satisfied.

\begin{table}
\begin{center}
\begin{tabular*}{0.8\textwidth}{@{\extracolsep{\fill}}c ccc}
Parameters            & Baseline & Optimum & Variation \\
$l_4$ (m)             & 1800.00  & 1379.38 & -23.37\% \\
$R_1$ (m)             & 5700.00  & 3619.58 & -36.50\% \\
$R_2$ (m)             & 160.00   & 83.45   & -47.84\% \\
$R_3$ (m)             & 880.00   & 639.42  & -27.34\% \\
$\theta_3$ ($^\circ$) & 80.00    & 97.57   & 21.96\%  \\
$C_L$                 & 58.977   & 60.033  & 1.79\%   \\
$C_D$                 & 50.594   & 48.351  & -4.43\%  \\
$L/D$                 & 1.166    & 1.242   & 6.55\%   \\
$X_{cp}$              & 2.261    & 2.321   & 2.64\%   \\
$\eta$                & 0.1278   & 0.1176  & -7.99\%  \\
\end{tabular*}
\end{center}
\caption{Comparison of geometric parameters and aerodynamic performance between baseline and optimized configuration at 85 km.}
\label{tab: Comparison between base & optimum 85}
\end{table}

In terms of aerodynamic forces, the $C_L$ of the optimized configuration rises by 1.79\%, and the $C_D$ decreases by 4.43\%. Consequently, the $L/D$ is improved by 6.55\%. These results demonstrate that the optimization effectively enhances the cruising performance of the HTV-2 type configuration when strict volume constraints are applied. To assess the reliability of the optimum results, the optimized configuration is calculated by multiscale UGKS method. As demonstrated in Tab.~\ref{tab: Comparison between optimum & cfd 85}, the maximum relative deviation between the optimized and UGKS results is less than 0.22\%, indicating a high feasibility of the optimization outcomes.

\begin{table}
\begin{center}
\begin{tabular*}{0.8\textwidth}{@{\extracolsep{\fill}}c ccc}
Parameters & Kriging Predictions & Multiscale UGKS results & Relative error \\
$C_L$               & 60.033   & 59.921  & 0.19\%    \\
$C_D$               & 48.351   & 48.244  & 0.22\%    \\
$L/D$               & 1.242    & 1.242   & -0.0037\% \\
$X_{cp}$            & 2.321    & 2.320   & 0.022\%   \\
$\eta$              & 0.1176   & 0.1176  & -0.0057\% \\
\end{tabular*}
\end{center}
\caption{Validation of the optimized aerodynamic performance at 85 km.}
\label{tab: Comparison between optimum & cfd 85}
\end{table}

\subsubsection{Aerodynamic characteristic analysis}

\begin{figure}
	\centering
	\subfigure[$x = 0\, \rm{m}$]
        {
			\includegraphics[width=0.42 \textwidth]{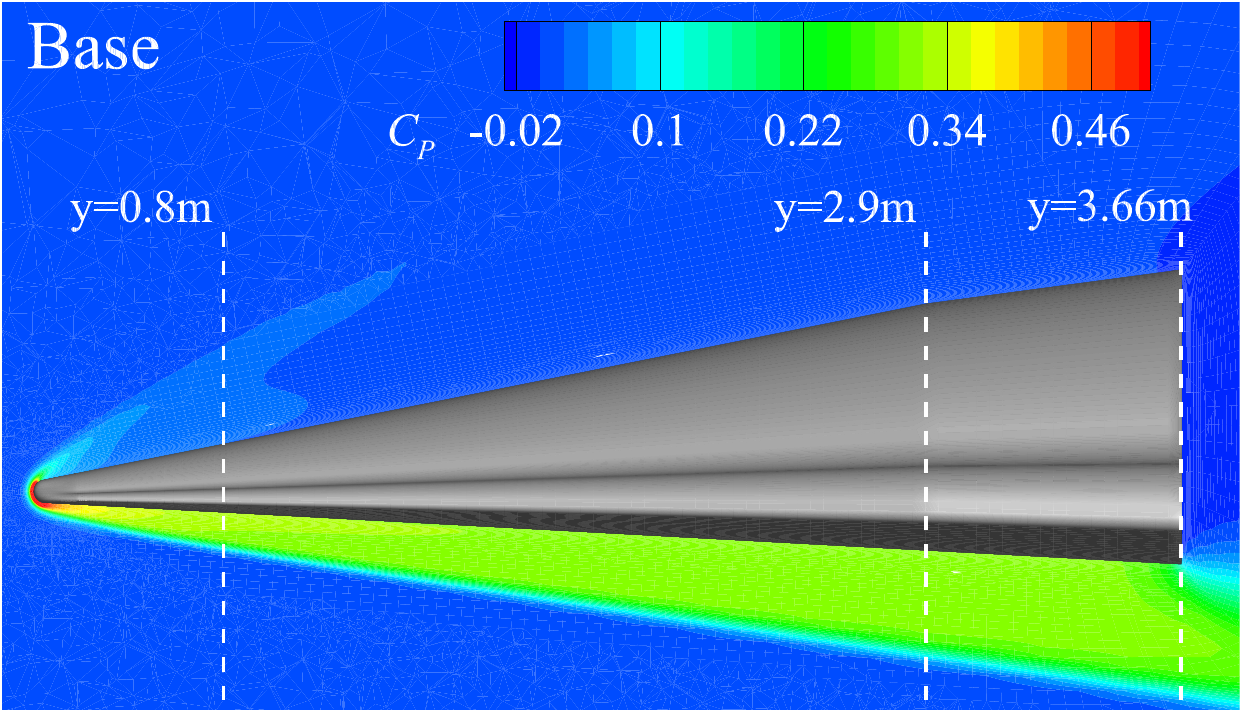}
		}
    \subfigure[$x = 0\, \rm{m}$]
        {
    		\includegraphics[width=0.42 \textwidth]{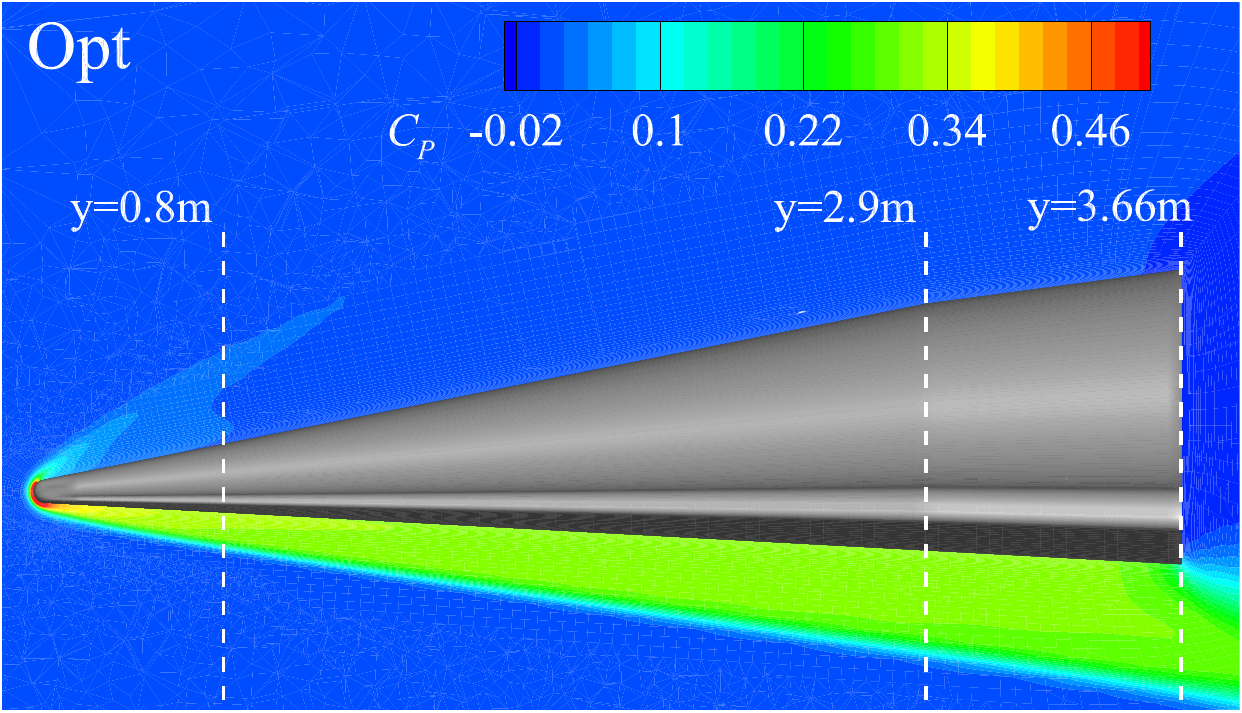}
    	}
    \subfigure[$y = 0.8\, \rm{m}$]
        {
    		\includegraphics[height=0.15 \textwidth]{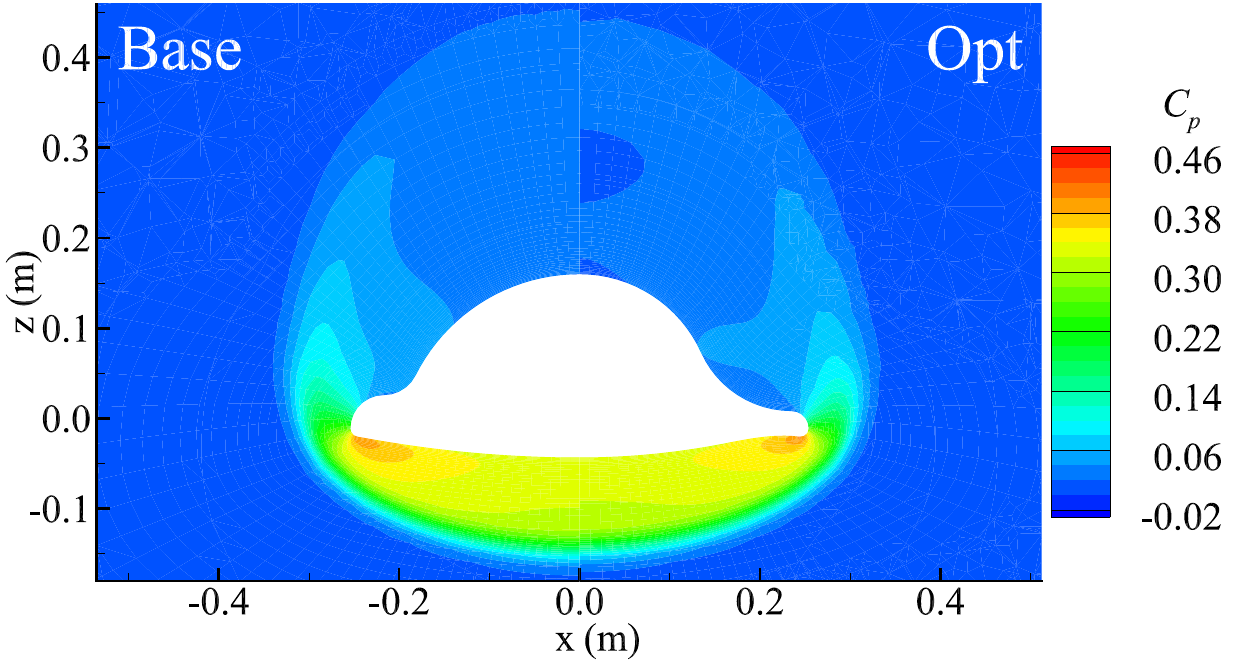}
    	}
    \subfigure[$y = 2.9\, \rm{m}$]
        {
    		\includegraphics[height=0.15 \textwidth]{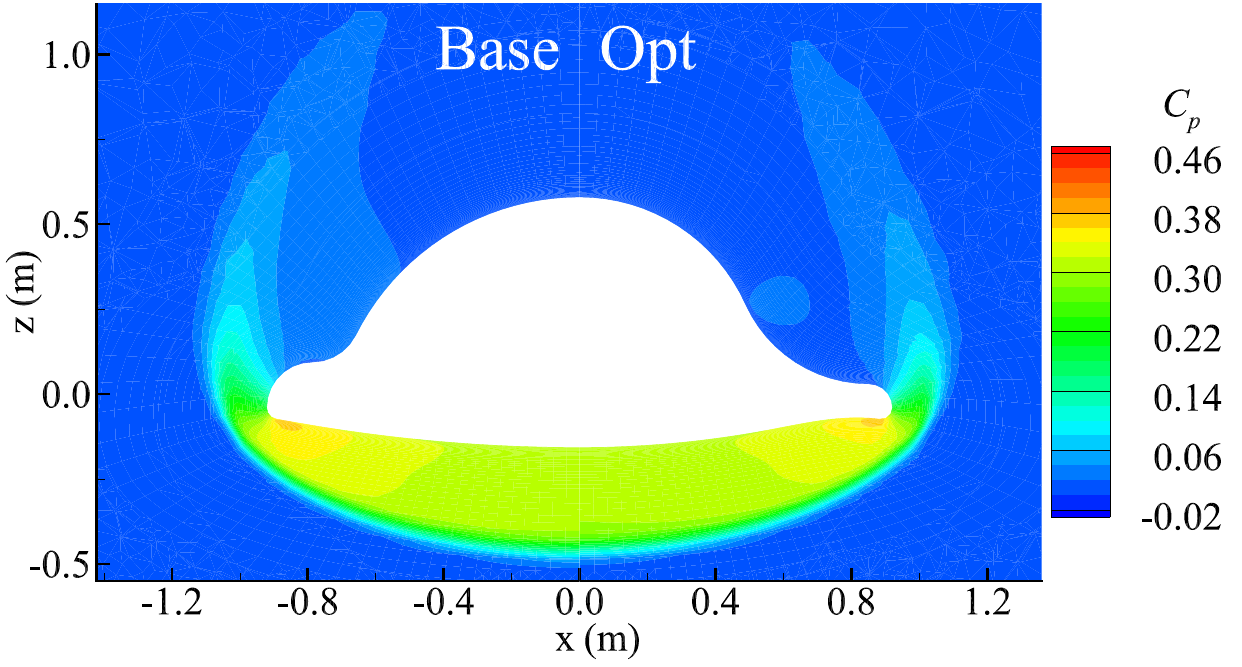}
    	}
    \subfigure[$y = 3.66\, \rm{m}$]
        {
			\includegraphics[height=0.15 \textwidth]{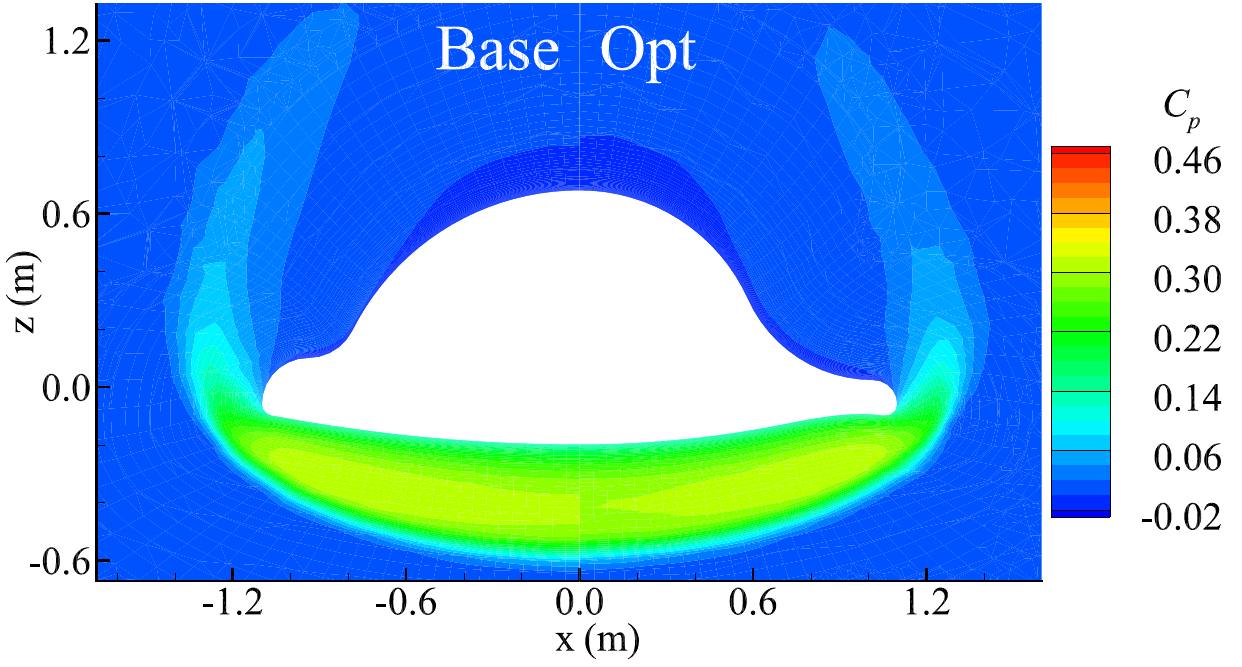}
		}
	\caption{Comparison of the pressure distribution of the HTV-2 type configurations along the longitudinal and axial cross-sections at 85 km.}
    \label{fig: cp counter at 85km}
\end{figure}

To further investigate the aerodynamic mechanisms, the flow field structures before and after the optimization are analyzed by extracting cross-sectional contours at various longitudinal and axial stations. As depicted in the lateral view of Fig.~\ref{fig: cp counter at 85km}, due to the relatively large freestream $AoA$, a strong compression shock wave is generated on the windward side of the vehicle, while a distinct expansion wave forms over the leeward surface.

\begin{figure}
	\centering
	\subfigure[$x = 0\, \rm{m}$]
        {
			\includegraphics[width=0.42 \textwidth]{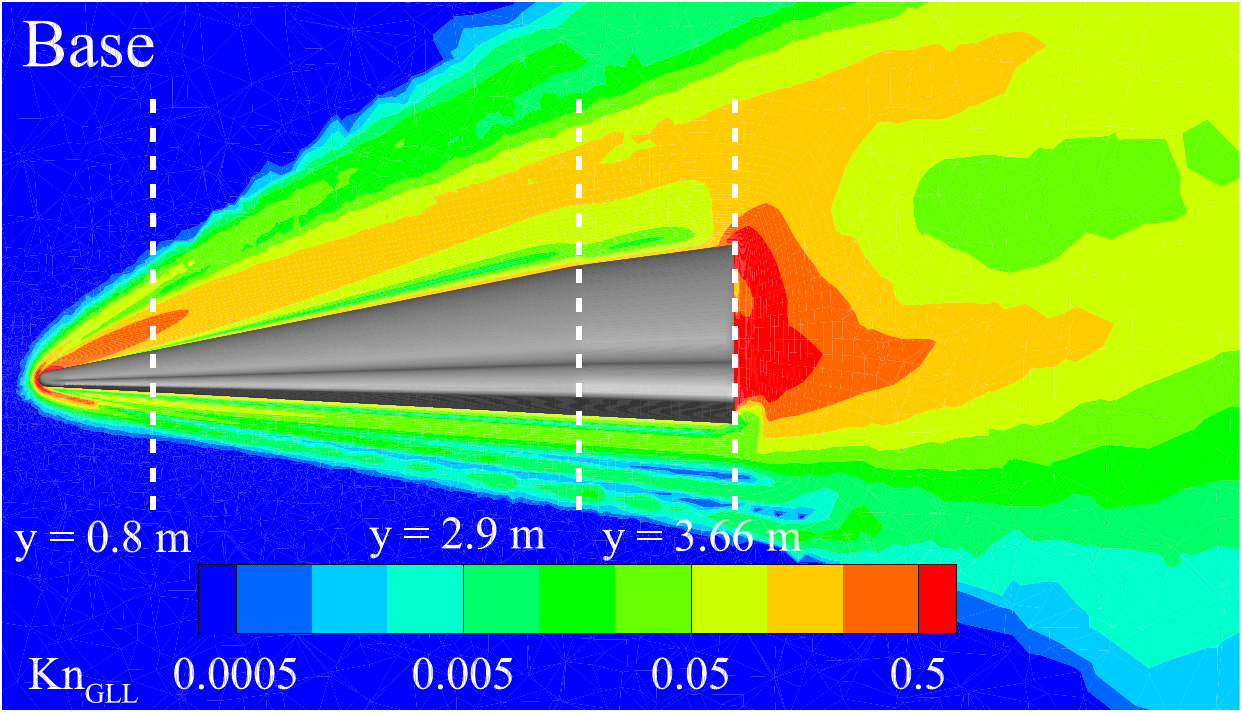}
		}
    \subfigure[$x = 0\, \rm{m}$]
        {
    		\includegraphics[width=0.42 \textwidth]{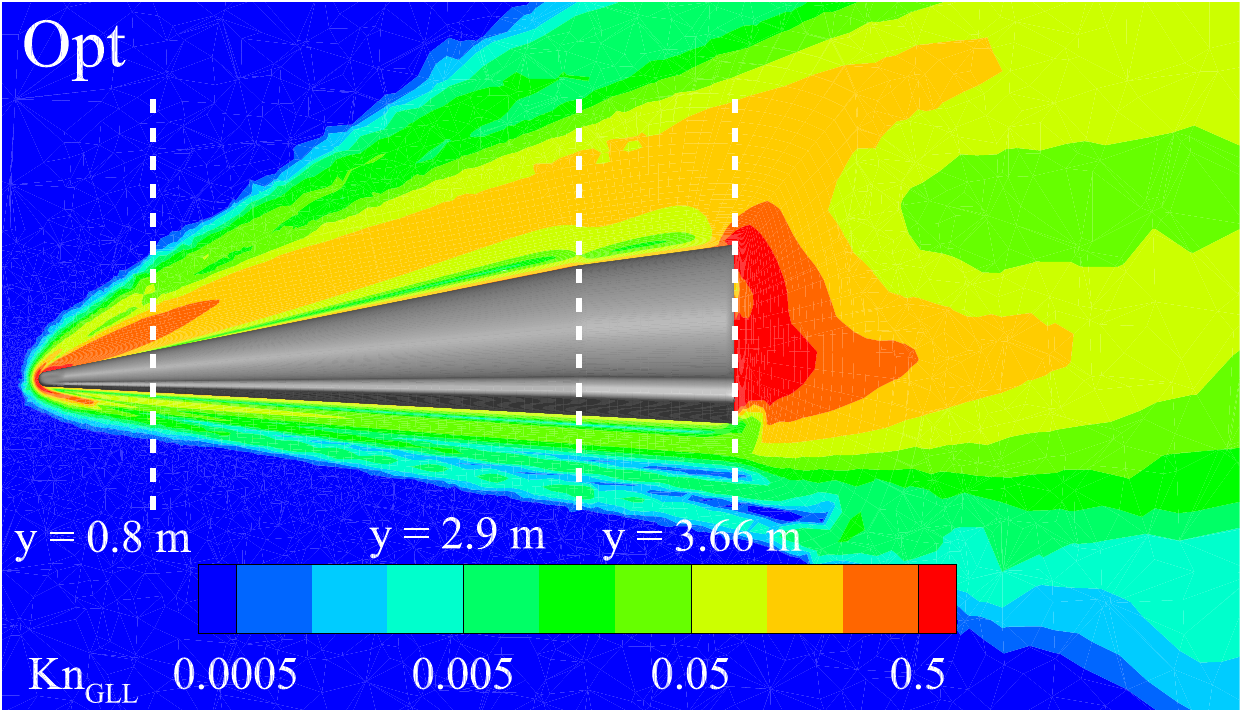}
    	}
    \subfigure[$y = 0.8\, \rm{m}$]
        {
    		\includegraphics[height=0.17 \textwidth]{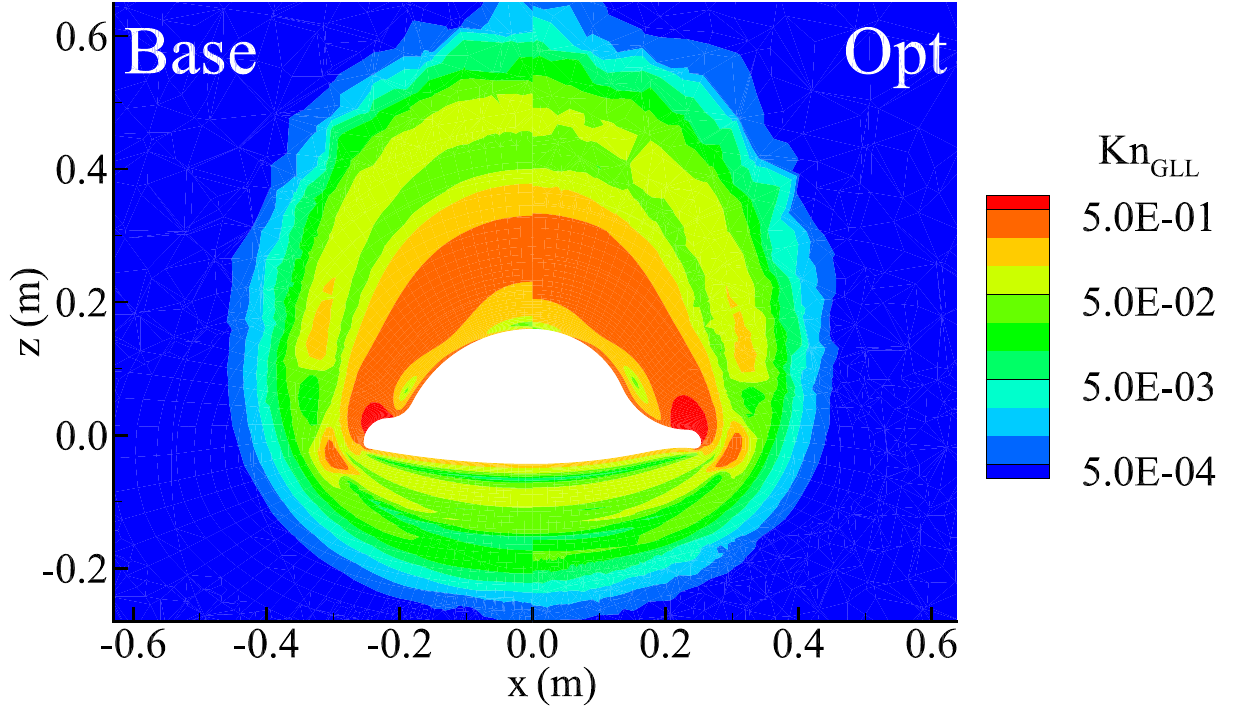}
    	}
    \subfigure[$y = 2.9\, \rm{m}$]
        {
    		\includegraphics[height=0.17 \textwidth]{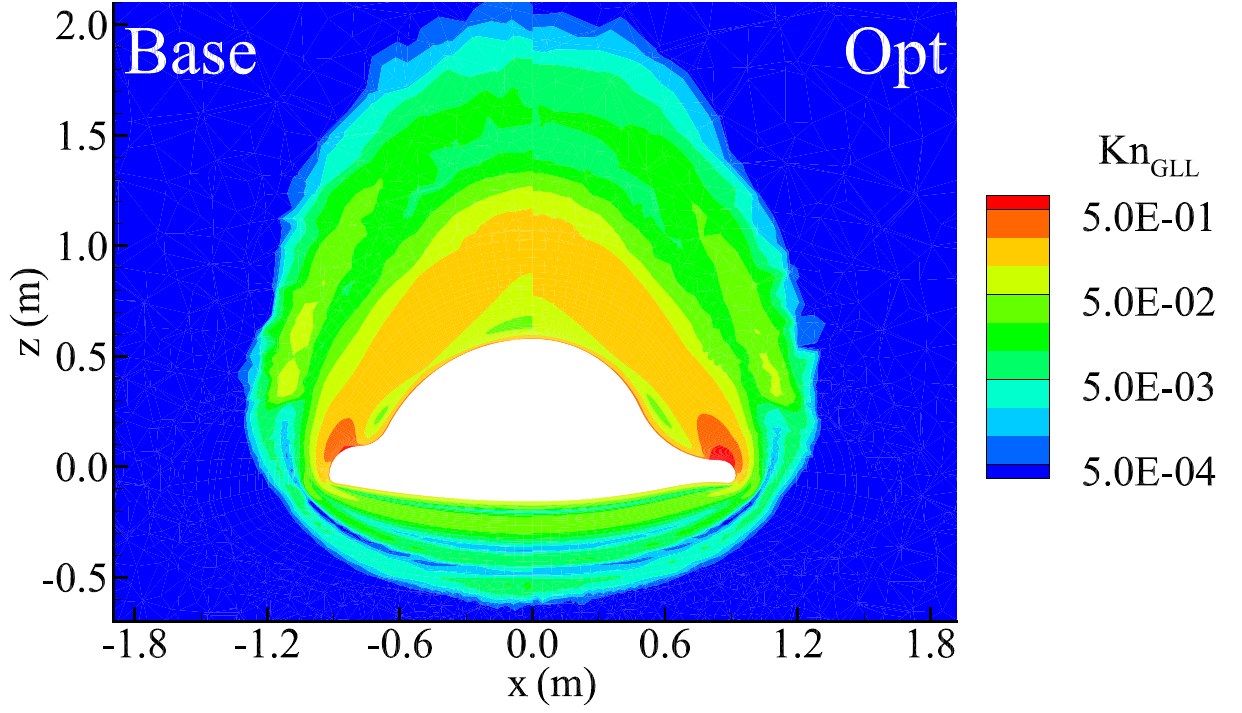}
    	}
    \subfigure[$y = 3.66\, \rm{m}$]
        {
			\includegraphics[height=0.17 \textwidth]{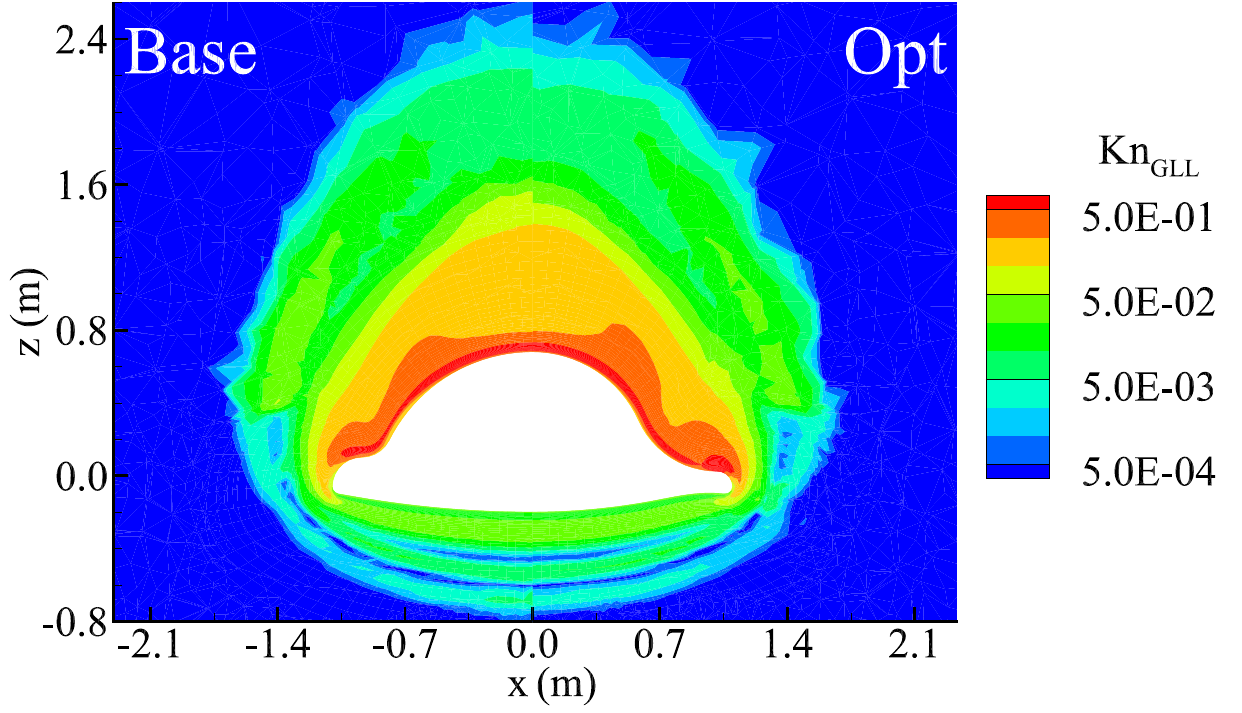}
		}
	\caption{Distributions of $\rm{Kn}_{GLL}$ for the baseline and optimized configurations at 85 km.}
    \label{fig: Kn_GLL at 85km}
\end{figure}

Moreover, the highly rarefied flow at this altitude exhibits strong multiscale effects. Fig.~\ref{fig: Kn_GLL at 85km} illustrates the $\rm{Kn}_{GLL}$ distributions on the symmetry plane and various cross-sectional views. Compared to the baseline, the optimized configuration demonstrates enhanced rarefaction around the fuselage, particularly in the vicinity of the wing surfaces. The contours reveal a more pronounced flow expansion on the leeward side of the vehicle.

\begin{figure}
	\centering
	\subfigure[$C_p$]
        {\label{fig: cp slice at 85km}
			\includegraphics[height=0.4 \textwidth]{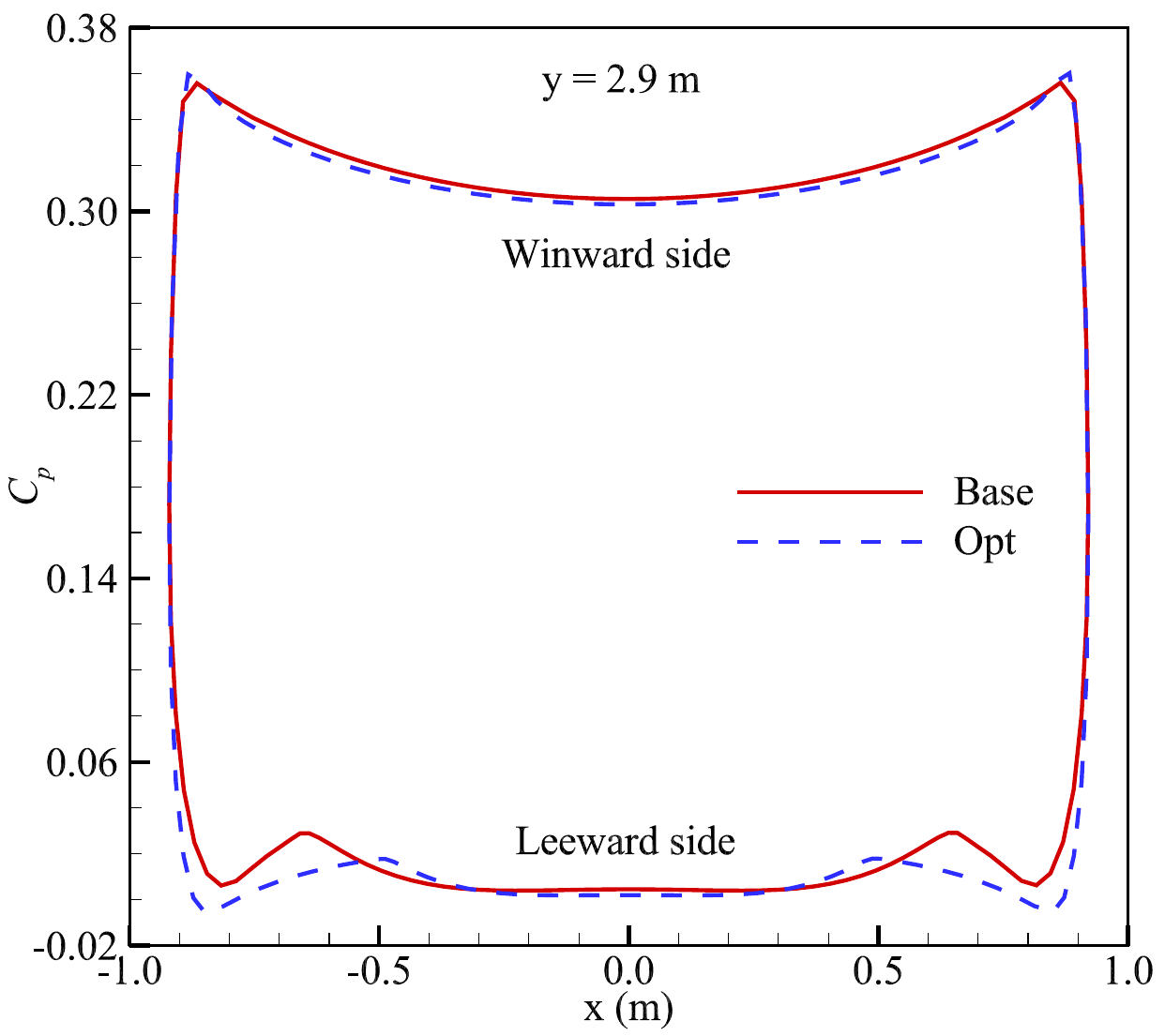}
		}
    \subfigure[$C_f$]
        {\label{fig: cf slice at 85km}
    		\includegraphics[height=0.4 \textwidth]{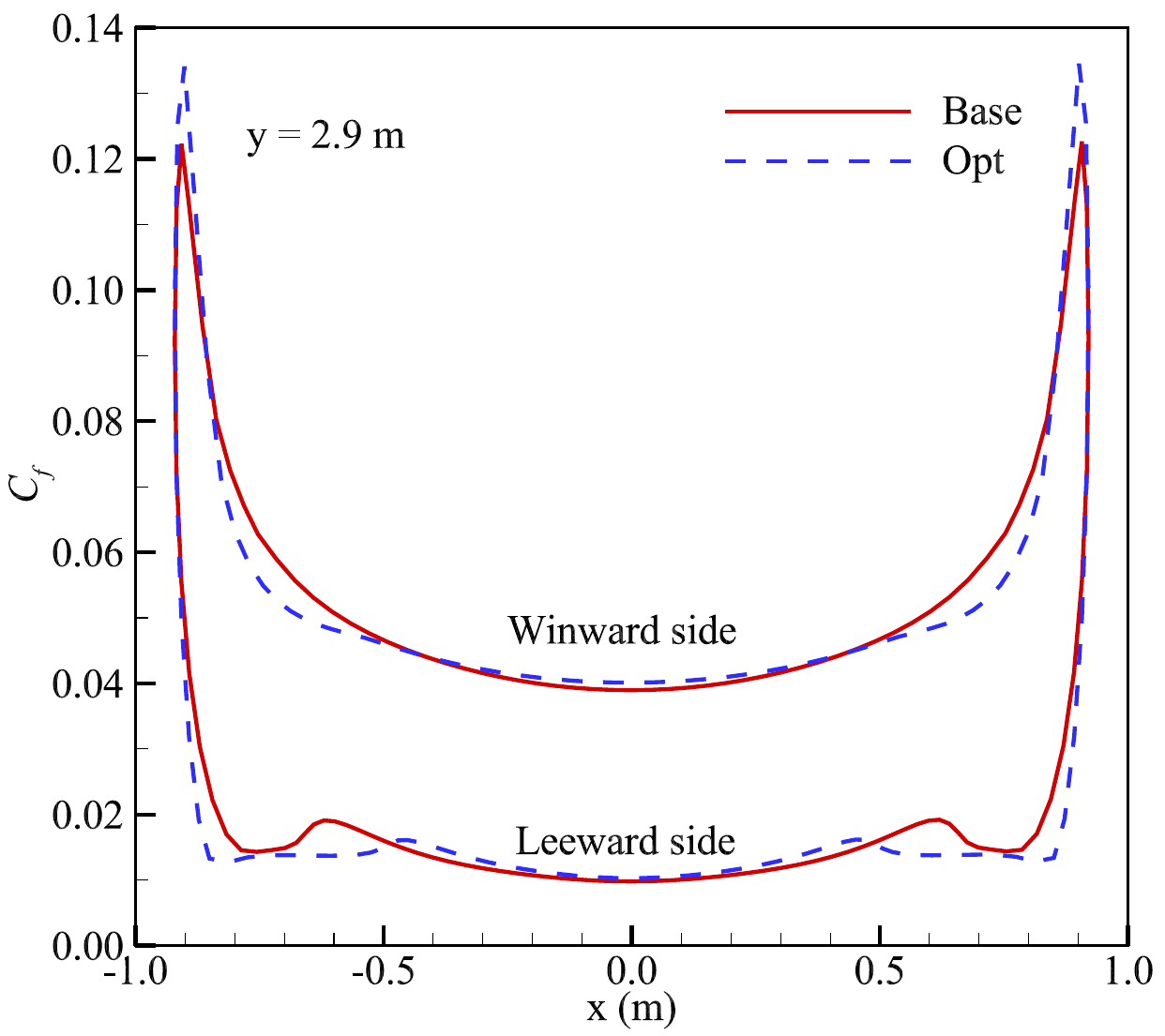}
    	}
	\caption{Aerodynamic coefficient distributions along x axis at $y = 2.9 \rm{m}$ at 85 km.}
    \label{}
\end{figure}

To understand the underlying physical mechanisms of the aerodynamic gains, the pressure distribution on the body surface is further analyzed. Fig.~\ref{fig: cp slice at 85km} illustrates the surface pressure coefficient ($C_p$) distribution at the spanwise slice of $y = 2.9 \rm{m}$. As shown in Fig.~\ref{fig: cp counter at 85km} and Fig.~\ref{fig: cp slice at 85km}, on the leeward surface, the reduced wing tip bluntness radius and the climb of the slide slope angle shift the wing shock downward and attenuate its intensity.
Furthermore, the decreased leeward surface radius creates a larger overall curvature on the leeward body, which significantly enhances the expansion waves. As a result, a more pronounced pressure drop is observed across the leeward region, directly contributing to the increase in $C_L$.

On the windward surface, the reductions in both the windward surface radius and base width increase the local surface curvature. This geometric modification weakens the windward compression shock, leading to a pressure reduction, particularly near the symmetry axis. Together, these localized pressure variations modulate the aerodynamic forces, ultimately leading to the overall improvement of the $C_L$.

Fig.~\ref{fig: cf slice at 85km} compares the skin friction coefficient ($C_f$) distributions at the spanwise slice of $y = 2.9 \rm{m}$. As previously noted, the heightened rarefaction in the leeward region leads to a localized reduction in skin friction, which is physically consistent with the diminished molecular collision frequency in these low-density zones. Notably, the optimum features characteristic drooped wing tips. This geometric adaptation successfully mitigates the local shear stress near the wing edges, further reducing the friction drag. Ultimately, the effective reduction in friction drag components, acting simultaneously with the enhancement of $C_L$, comprehensively explains the substantial rise in the overall $L/D$ of the optimized HTV-2 type configuration.

\subsubsection{Global sensitivity analysis}

\begin{figure}
	\centering
	\subfigure[]
        {\label{}
			\includegraphics[width=0.3 \textwidth]{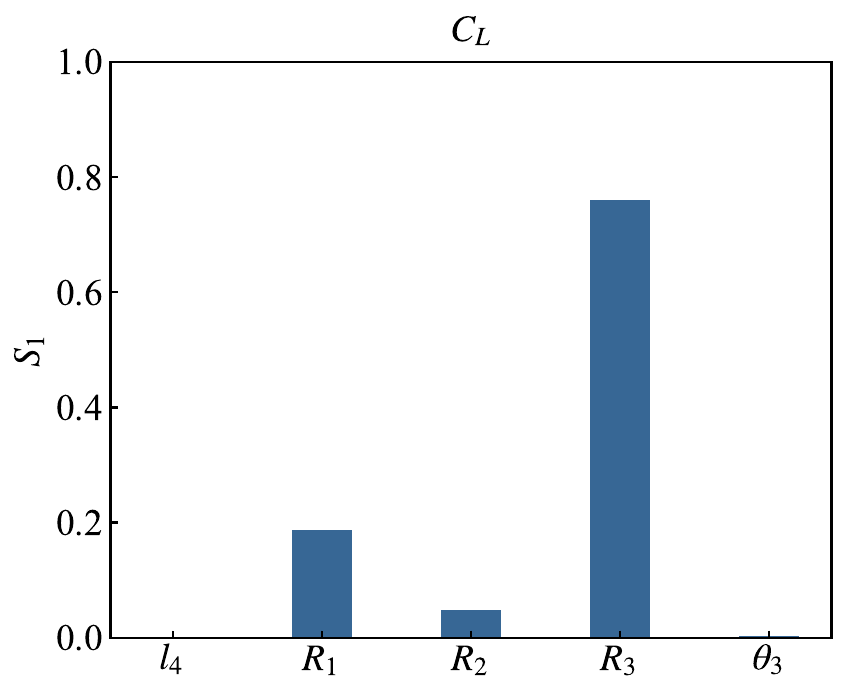}
		}
    \subfigure[]
        {\label{}
    		\includegraphics[width=0.3 \textwidth]{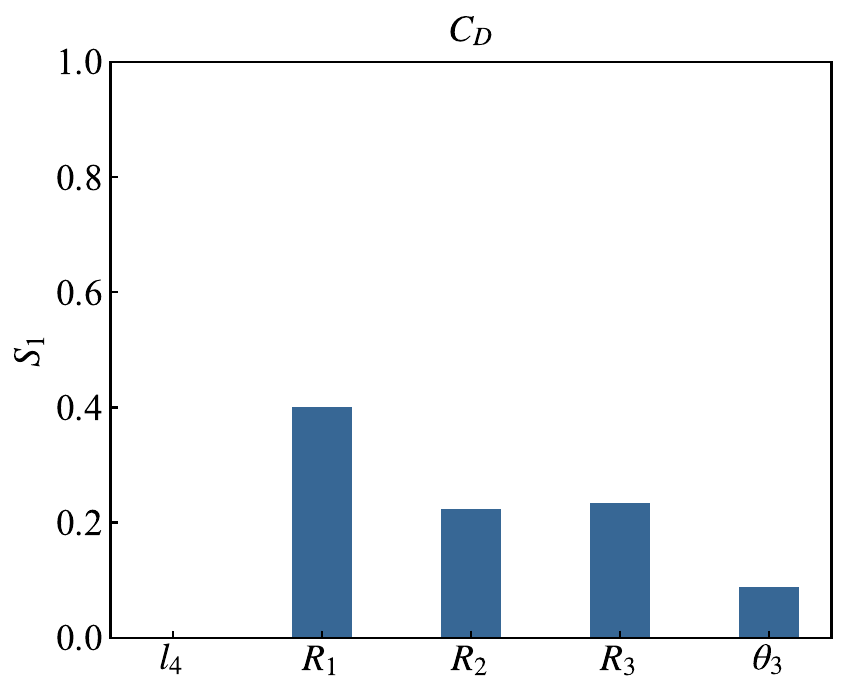}
    	}
    \subfigure[]
        {\label{}
    		\includegraphics[width=0.3 \textwidth]{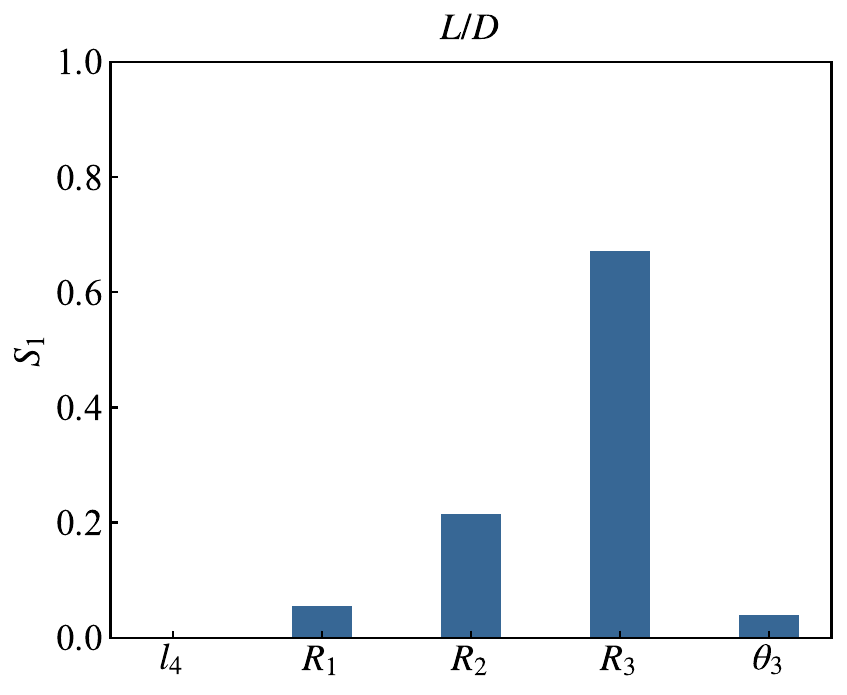}
    	}
	\caption{Sobol first-order sensitivity indices at 85 km.}
    \label{fig: s1 85}
\end{figure}

The first-order global sensitivity indices ($S_1$) of the geometric parameters within the design space are presented in Fig.~\ref{fig: s1 85}. As indicated in the figure, $R_3$ exhibits the most dominant influence on the $C_L$, with an $S_1$ value reaching approximately 0.76, far exceeding those of the other parameters. Physically, the reduction of $R_3$ enhances the flow expansion effect on the leeward surface, which lowers the pressure coefficient and predominantly drives the $C_L$ increment under the current freestream condition. Additionally, $R_1$ plays a secondary but crucial role in $C_L$ generation, primarily because its reduction mitigates the windward shock intensity.

In contrast to the $C_L$, the $C_D$ is more evenly influenced by multiple geometric parameters, with $R_1$, $R_3$, $R_2$, and $\theta_3$ all showing comparable sensitivities. Notably, $R_2$ contributes moderately to both $C_L$ and $C_D$, but emerges as the second most influential parameter governing the overall $L/D$. Furthermore, the sensitivity indices for the base width ($l_4$) are negligible across all aerodynamic responses. This suggests that, under the premise of a fixed overall fuselage width, variations in the base width have a minimal impact on the aerodynamic performance of the HTV-2 type configuration.

To conclude, the Sobol sensitivity analysis identifies the contour radii ($R_3$, $R_1$, and $R_2$) as the primary geometric drivers of aerodynamic performance. These sensitivity distributions physically confirm that a more aggressive configuration—characterized by optimized radii and flatter profiles—effectively yields superior aerodynamic efficiency within the design envelope at this altitude.

\subsection{Optimization Extensions to 70~km and 100 km}

Building upon the aforementioned results, the framework is further applied to the extended flight conditions at the altitudes of 70~km and 100 km. Due to the significant variations in freestream conditions, the$X_{cp}$ of the HTV-2 type baseline configuration at an $18^\circ$ $AoA$ deviates substantially from the center of gravity. To prevent over-constraining the problem and to maintain a sufficiently broad feasible design space, the strict trim constraint on the longitudinal pressure center shift is relaxed for these subsequent cases. Therefore, the optimization problem is reformulated to maximize the $L/D$ while exclusively enforcing the minimum volume requirement. Utilizing the exterior penalty method, the modified optimization problem is mathematically expressed as follows:
\begin{equation}
    \begin{aligned}
        & \min \quad F'(\boldsymbol{X}) = F(\boldsymbol{X}) + \lambda \max(0, 0.92\eta_0 - \eta) \\
        & \text{where} \quad F(\boldsymbol{X}) = -L/D \\
        & \quad \quad \quad \ \boldsymbol{X} = (l_4, R_1, R_2, R_3, \theta_3) \\
        & \text{s.t.} \quad \eta \ge 0.92 \eta_0
    \end{aligned}
\end{equation}

\subsubsection{Optimization results}

Tab.~\ref{tab: Optimization outcomes 70 and 100} provides a detailed quantitative comparison of the geometric parameters and aerodynamic performance before and after the optimization at 70~km and 100 km. Overall, the geometric variations at both altitudes exhibit a consistent trend, although the specific magnitudes of modification differ. For instance, the $l_4$ experiences a similar reduction, and the $R_2$ is extremely reduced by 50\% in both cases, hitting the lower boundary constraint. This indicates that minimizing the wing tip bluntness is universally beneficial for aerodynamic enhancement across these extended rarefied regimes. Moreover, the $\theta_3$ increases in both scenarios, with a more pronounced increment observed at 70~km (52.60\%) than at 100 km (33.81\%).

\begin{table}
\begin{center}
\begin{tabular*}{0.9\textwidth}{@{\extracolsep{\fill}}c cccccc}
\multirow{1}{*}{Parameters} & \multicolumn{3}{c}{70~km} & \multicolumn{3}{c}{100 km} \\
                    & Baseline & Optimum & Variation & Baseline & Optimum & Variation \\
$l_4$ (m)             & 1800.00  & 1174.40 & -34.76\% & 1800.00  & 1137.56 & -36.80\% \\
$R_1$ (m)             & 5700.00  & 2700.74 & -52.62\% & 5700.00  & 4572.37 & -19.78\% \\
$R_2$ (m)             & 160.00   & 80.00   & -50.00\% & 160.00   & 80.00   & -50.00\% \\
$R_3$ (m)             & 880.00   & 729.01  & -17.16\% & 880.00   & 640.29  & -27.24\% \\
$\theta_3$ ($^\circ$) & 80.00    & 122.08  & 52.60\%  & 80.00    & 107.05  & 33.81\%  \\
$C_L$                 & 62.390   & 62.103  & -0.46\%  & 54.809   & 56.446  & 2.99\%   \\
$C_D$                 & 35.401   & 33.554  & -5.22\%  & 98.909   & 93.422  & -5.55\%  \\
$L/D$                 & 1.762    & 1.857   & 5.37\%   & 0.554    & 0.604   & 9.00\%   \\
$\eta$                & 0.1278   & 0.1176  & -7.99\%  & 0.1278   & 0.1176  & -7.99\%  \\
\end{tabular*}
\end{center}
\caption{Optimization outcomes at 70~km and 100 km: Comparison of geometric and aerodynamic parameters.}
\label{tab: Optimization outcomes 70 and 100}
\end{table}

\begin{figure}
	\centering
	\subfigure[70~km]
        {\label{fig: 70km_base}
			\includegraphics[width=0.45 \textwidth]{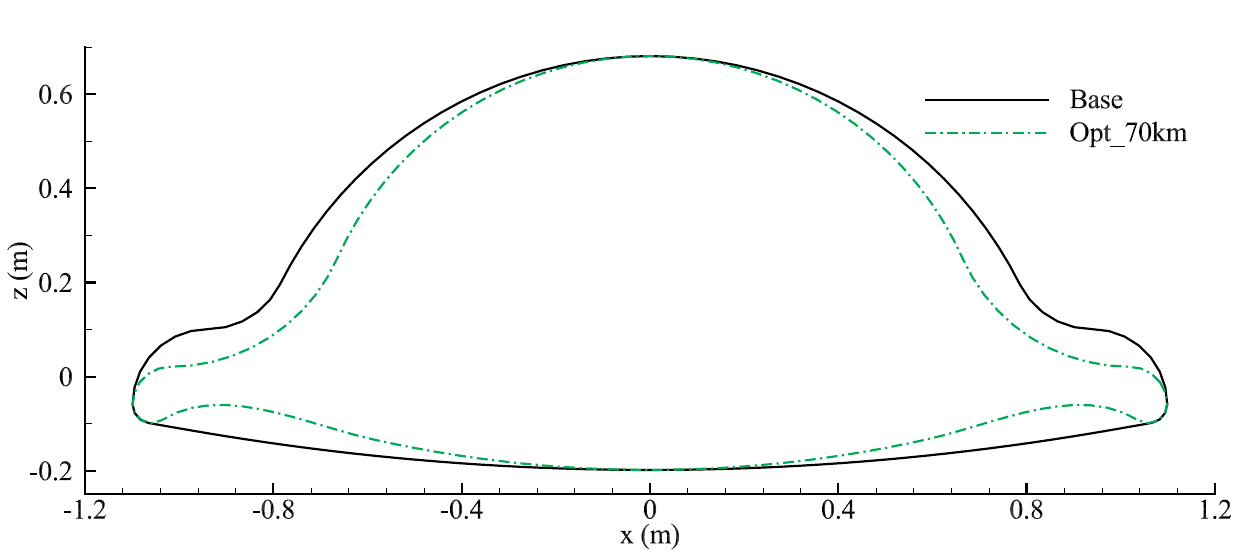}
		}
    \subfigure[100 km]
        {\label{fig: 100km_base}
    		\includegraphics[width=0.45 \textwidth]{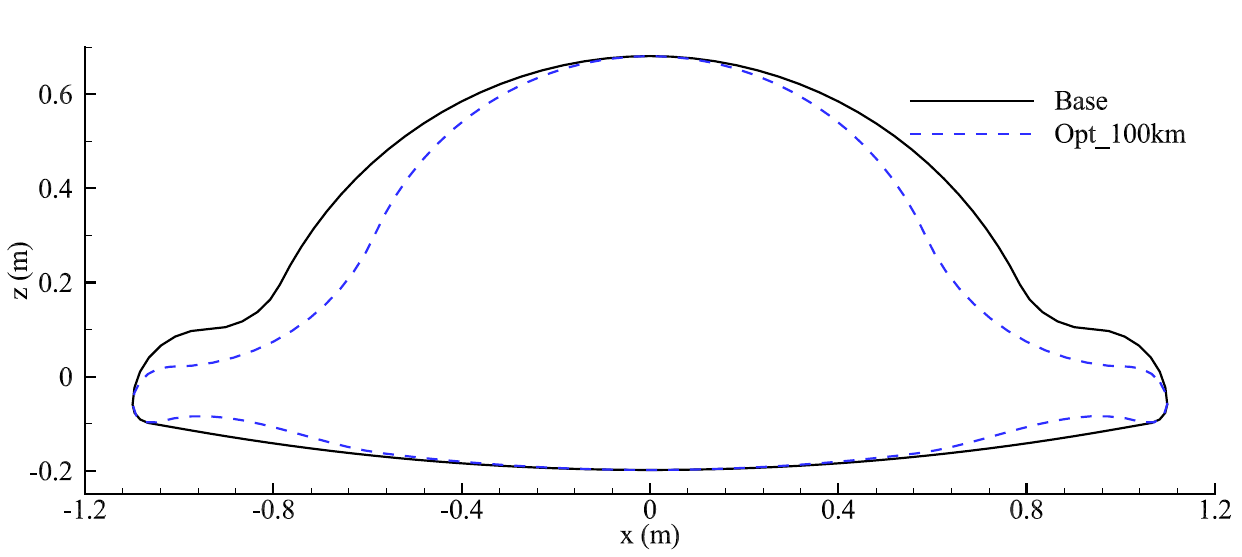}
    	}
	\caption{Geometric profile comparison between the baseline and optimized configurations at 70~km and 100 km.}
    \label{fig: geometric comparison_70 and 100}
\end{figure}

It is worth noting that a clear geometric trade-off is observed between the windward and leeward radii under the strict volume constraint. As detailed in the table, the $R_1$ is reduced by a substantial 52.62\% at 70~km, which is larger than the 19.78\% reduction at 100 km. Visually corroborated by Fig.~\ref{fig: geometric comparison_70 and 100}, the optimized windward surface at 70~km features a noticeably larger geometric curvature, whereas the 100 km configuration remains relatively flatter. Conversely, the $R_3$ undergoes a larger reduction at 100 km (-27.24\%) compared to 70~km (-17.16\%). The visual comparison reveals that the 70~km optimum retains more spatial volume on the upper body. This physical phenomenon suggests that at 70~km, modifying $R_1$ yields a more dominant aerodynamic benefit; therefore, $R_3$ is kept relatively larger to strictly satisfy the minimum volume requirement.

Regarding the aerodynamic performance, the volumetric efficiency ($\eta$) is strictly maintained at the allowable lower bound for both cases, verifying the effectiveness of the penalty function. The primary objective $L/D$ is effectively improved. Notably, the $L/D$ at 100 km achieves a remarkable enhancement of 9.00\%, while the 70~km case sees a solid 5.37\% increase. A detailed breakdown of the aerodynamic forces reveals that the $C_D$ is consistently reduced by approximately 5\% at both altitudes. However, the $C_L$ exhibits converse trends: similar to the 85 km design point, the $C_L$ at 100 km increases by 2.99\%. In contrast, the $C_L$ at 70~km experiences a marginal decrease of 0.46\%. This tiny decline indicates that enhancing $C_L$ becomes more challenging at lower altitudes with relatively denser freestream conditions.

To verify the reliability of the SBO framework at the extended altitudes, the optimized configurations are recalculated using multiscale UGKS solver, with the comparative results presented in Tab.~\ref{tab: CFD validation 70 and 100}. The Kriging surrogate model predictions demonstrate pretty good agreement with the numerical simulations. Specifically, the maximum relative errors across all parameters are merely 0.94\% at 70~km and 0.13\% at 100 km. These negligible discrepancies conclusively validate the high fidelity and robustness of the constructed surrogate models under varying highly rarefied conditions.

\begin{table}
\begin{center}
\begin{tabular*}{0.9\textwidth}{@{\extracolsep{\fill}}c cccccc}
\multirow{1}{*}{Parameters} & \multicolumn{3}{c}{70~km} & \multicolumn{3}{c}{100 km} \\
 & \makecell{Kriging \\ Predictions} & \makecell{Multiscale \\ UGKS results} & \makecell{Relative \\ error} & \makecell{Kriging \\ Predictions} & \makecell{Multiscale \\ UGKS results} & \makecell{Relative \\ error} \\
\midrule
$C_L$     & 62.103 & 61.889 & 0.35\%  & 56.446 & 56.428 & 0.033\%   \\
$C_D$     & 33.554 & 33.241 & 0.94\%  & 93.422 & 93.424 & -0.0020\% \\
$L/D$     & 1.857  & 1.862  & -0.26\% & 0.604  & 0.604  & 0.00099\% \\
$\eta$    & 0.1176 & 0.1178 & -0.14\% & 0.1176 & 0.1175 & 0.13\%    \\
\end{tabular*}
\end{center}
\caption{Validation of the optimized aerodynamic performance at 70~km and 100 km.}
\label{tab: CFD validation 70 and 100}
\end{table}

\subsubsection{Flow Field Evolution across Rarefied Regimes}

\begin{figure}
    \centering
    \subfigure[$x = 0\, \rm{m}$]
        {
            \includegraphics[width=0.42 \textwidth]{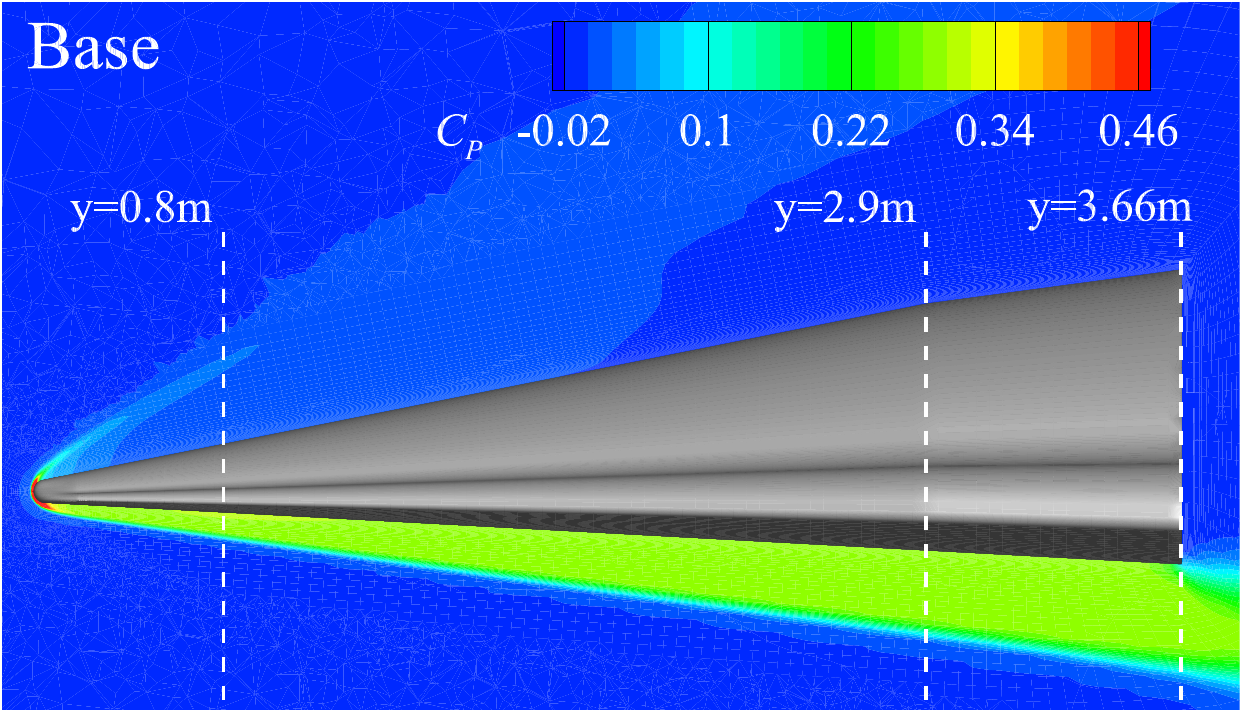}
        }
    \subfigure[$x = 0\, \rm{m}$]
        {
            \includegraphics[width=0.42 \textwidth]{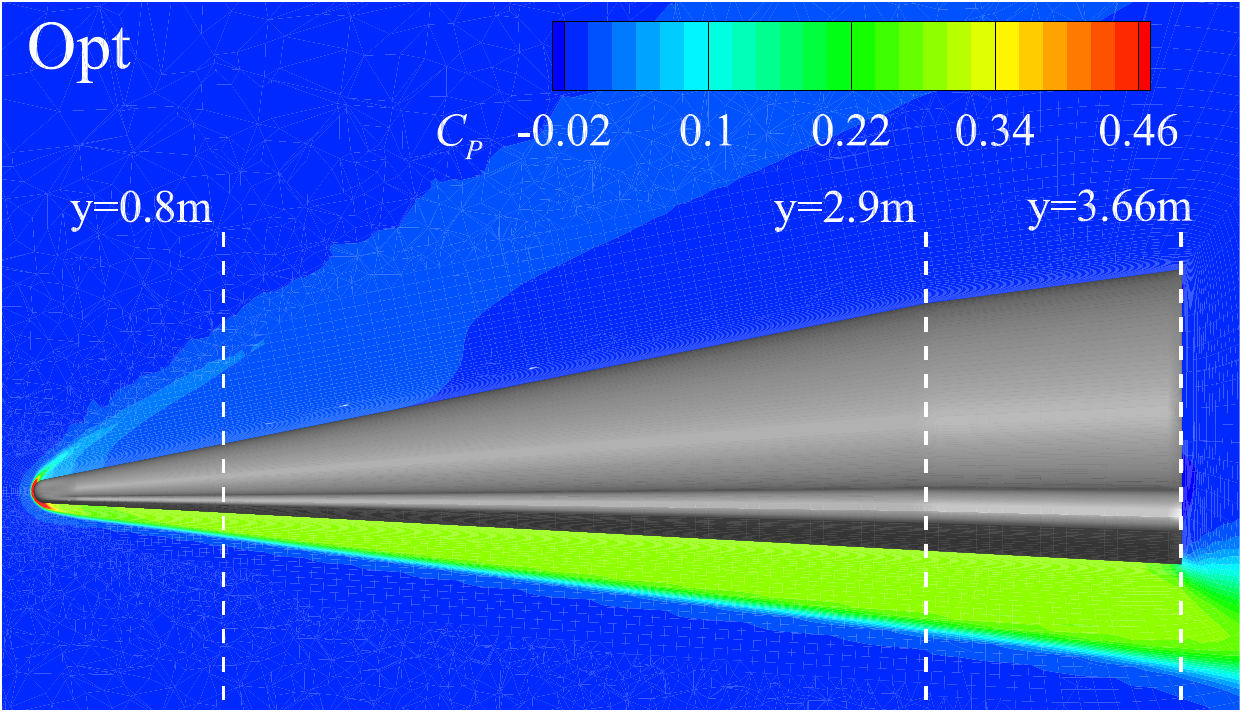}
        }
    \subfigure[$y = 0.8\,\rm{m}$]
        {
            \includegraphics[height=0.15 \textwidth]{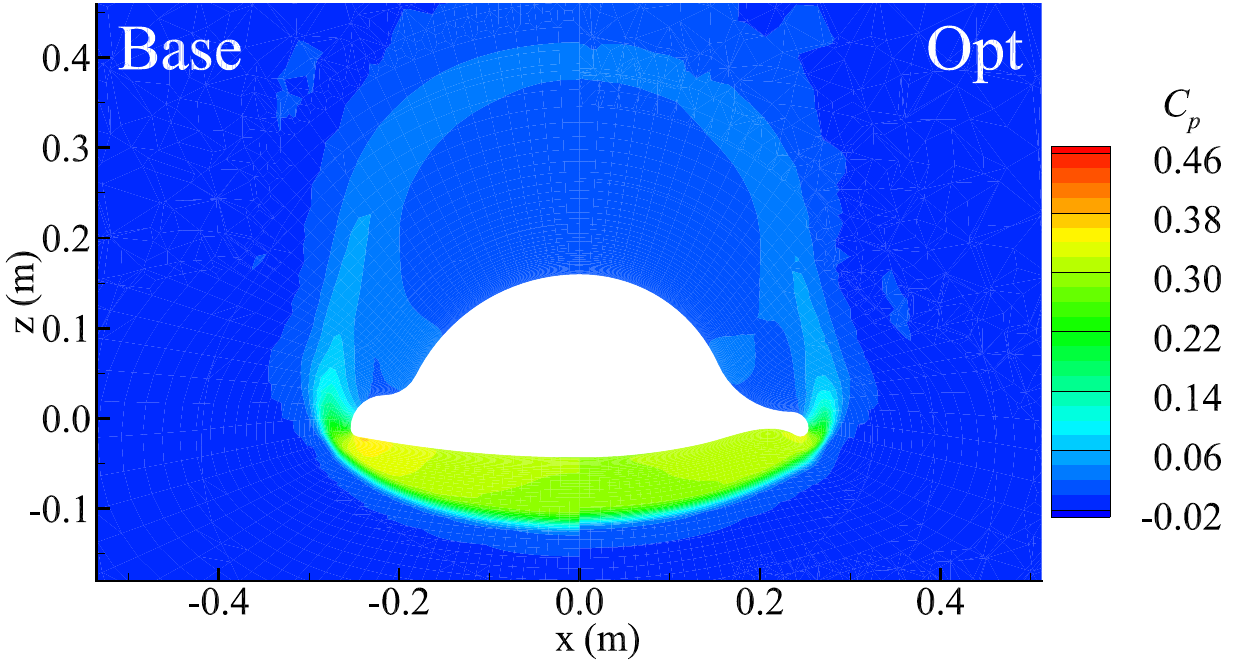}
        }
    \subfigure[$y = 2.9\,\rm{m}$]
        {
            \includegraphics[height=0.15 \textwidth]{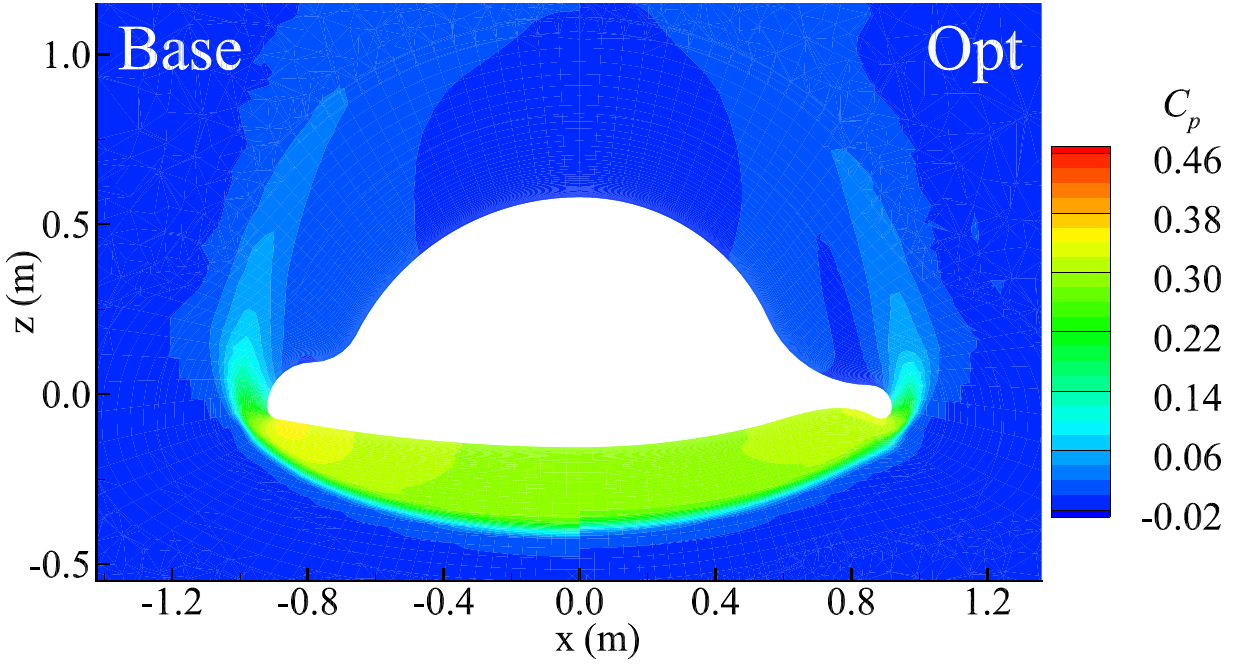}
        }
    \subfigure[$y = 3.66\,\rm{m}$]
        {
            \includegraphics[height=0.15 \textwidth]{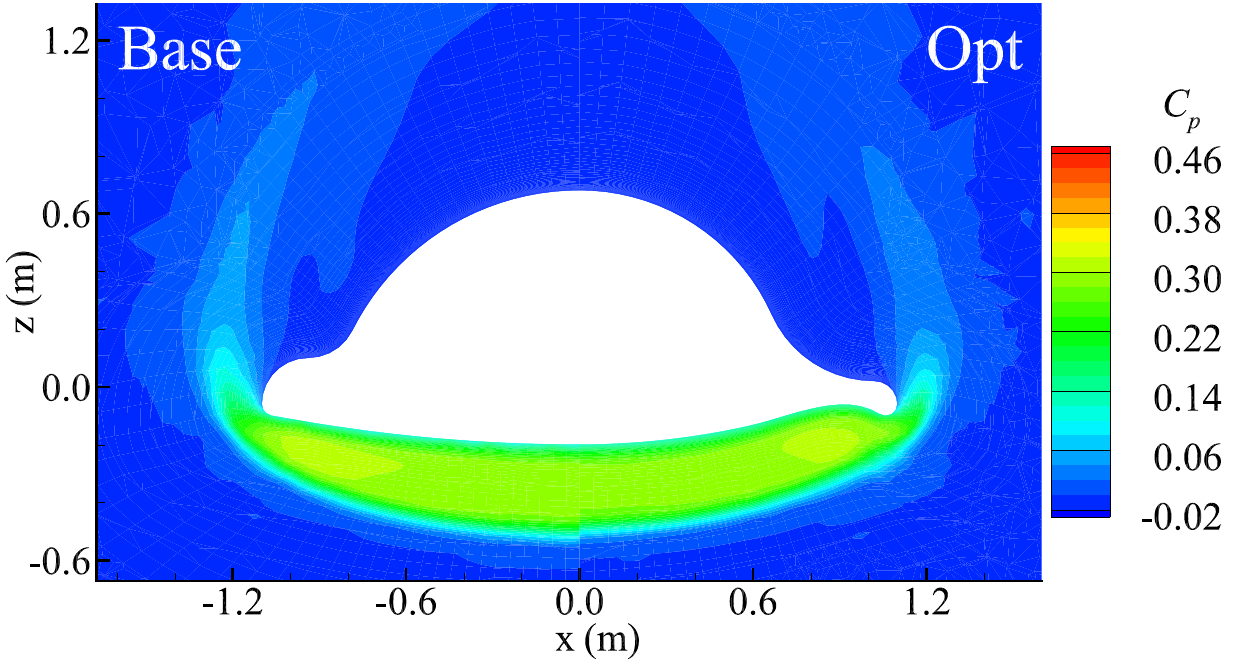}
        }
    \caption{Comparison of the pressure distribution of the HTV-2 type configurations along the longitudinal and axial cross-sections at 70~km.}
    \label{fig: cp counter at 70km}
\end{figure}

\begin{figure}
    \centering
    \subfigure[$C_p$]
        {\label{fig: cp slice at 70km}
            \includegraphics[width=0.45 \textwidth]{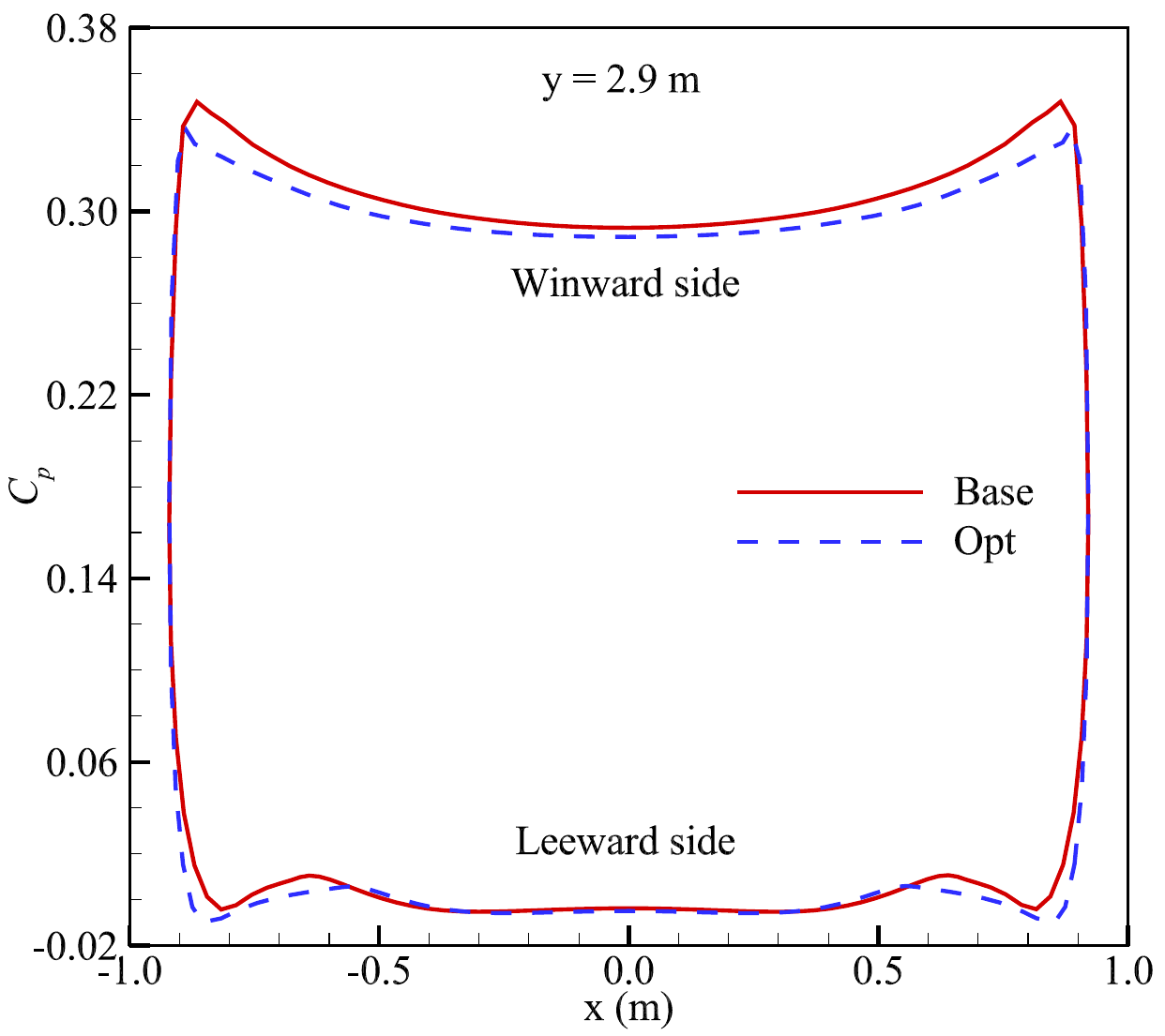}
        }
    \subfigure[$C_f$]
        {\label{fig: cf slice at 70km}
            \includegraphics[width=0.45 \textwidth]{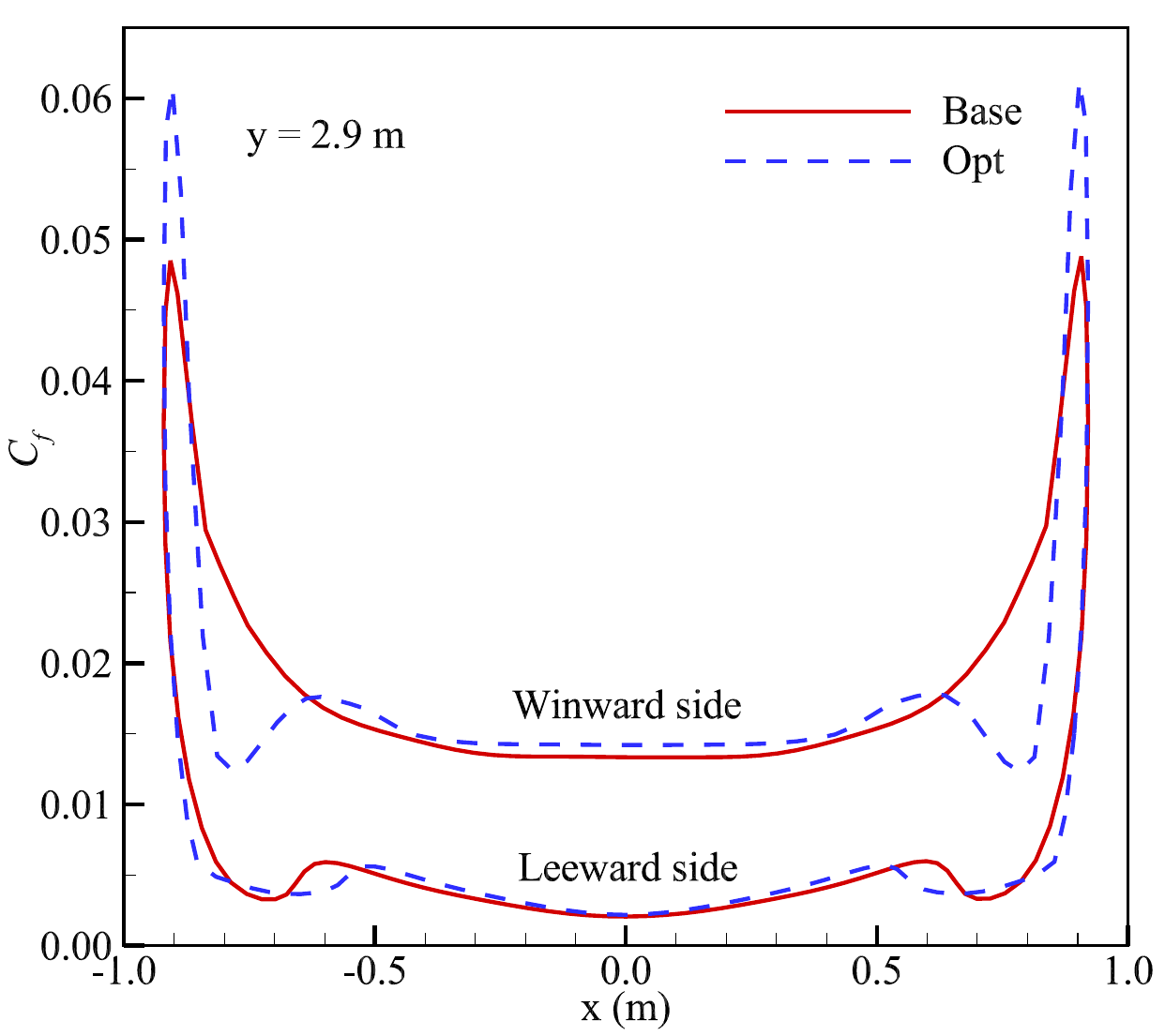}
        }
    \caption{Comparison of the $C_p$ and $C_f$ at $y = 2.9\,\rm{m}$ at 70~km.}
    \label{fig: clices at 70}
\end{figure}

To explore the reasons driving the aerodynamic modifications at 70~km, the flow field evolution is analyzed in detail. As illustrated in Fig.~\ref{fig: cp counter at 70km}, the longitudinal and axial pressure coefficient ($C_p$) distributions reveal distinct changes between the baseline and optimized configurations. On the windward side, the substantial reduction of $R_1$ weakens the flow compression, thereby reducing the intensity of the oblique shock wave and lowering the overall windward pressure coefficient.
On the leeward side, the reductions in the $R_3$ and $R_2$ enhance the flow expansion. As observed in the axial cross-sections, the decrease in $R_2$ shifts the expansion waves downward, lowering the local pressure near the wing surfaces. However, as illustrated in Fig.~\ref{fig: cp slice at 70km}, the $C_p$ drop on the windward surface is considerably larger than that on the leeward surface. This relatively modest pressure reduction on the leeward side is primarily because $R_3$ was not significantly decreased at 70~km due to the strict volumetric constraints. This asymmetric pressure variation—where the windward pressure loss outweighs the leeward pressure drop—physically explains the slight decrease in total lift observed at this altitude.

Fig.~\ref{fig: cf slice at 70km} presents the $C_f$ distribution at $y = 2.9\,\rm{m}$. The significant increase in $\theta_3$ amplifies the drooped characteristic of the wings, which reduces the friction on the lower wing surface. Physically, the modified wing shape blocks the cross-flow near the wing edges, which reduces the local velocity gradient and lowers the shear stress. The significant decrease in $C_D$ improves the overall $L/D$.

\begin{figure}
    \centering
    \subfigure[$x = 0\, \rm{m}$]
        {
            \includegraphics[width=0.42 \textwidth]{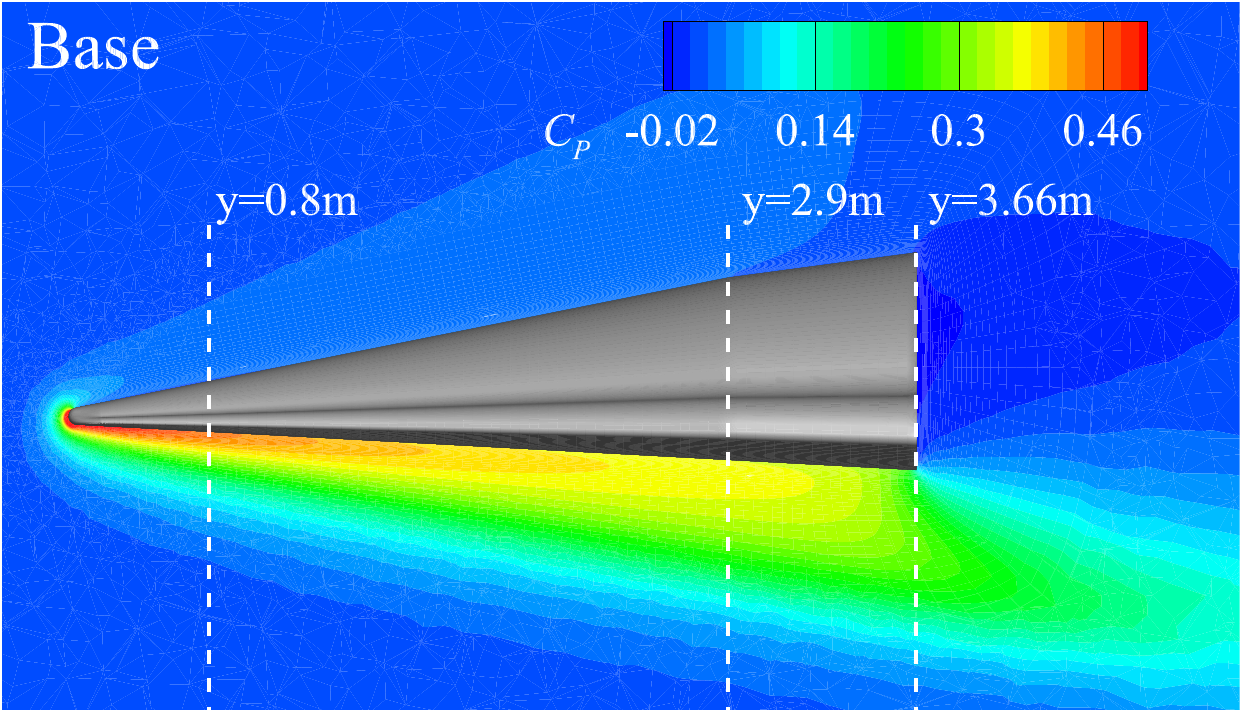}
        }
    \subfigure[$x = 0\, \rm{m}$]
        {
            \includegraphics[width=0.42 \textwidth]{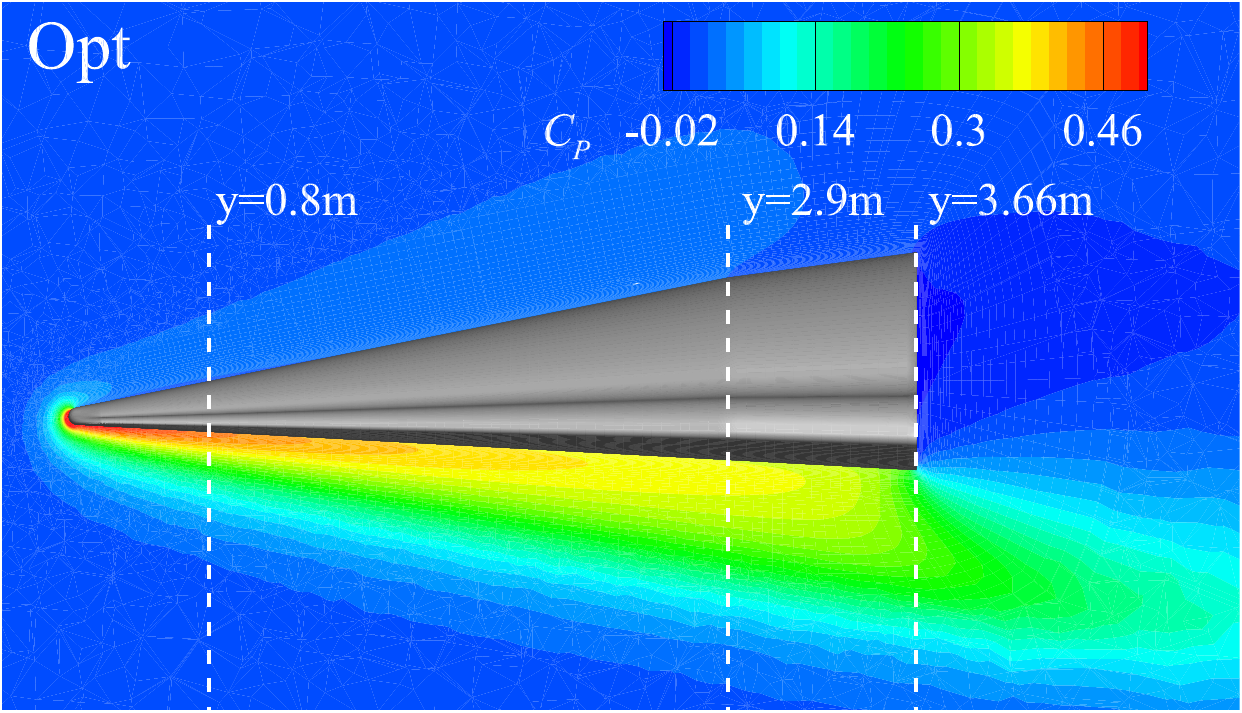}
        }
    \subfigure[$y = 0.8\,\rm{m}$]
        {
            \includegraphics[height=0.15 \textwidth]{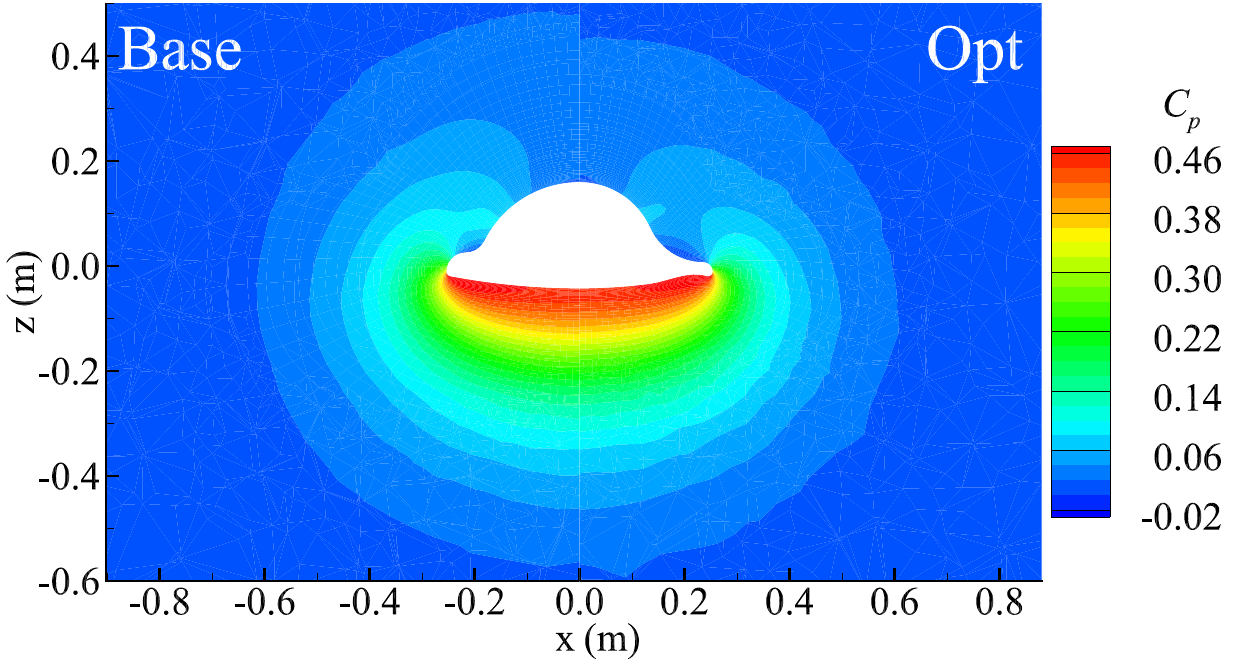}
        }
    \subfigure[$y = 2.9\,\rm{m}$]
        {
            \includegraphics[height=0.15 \textwidth]{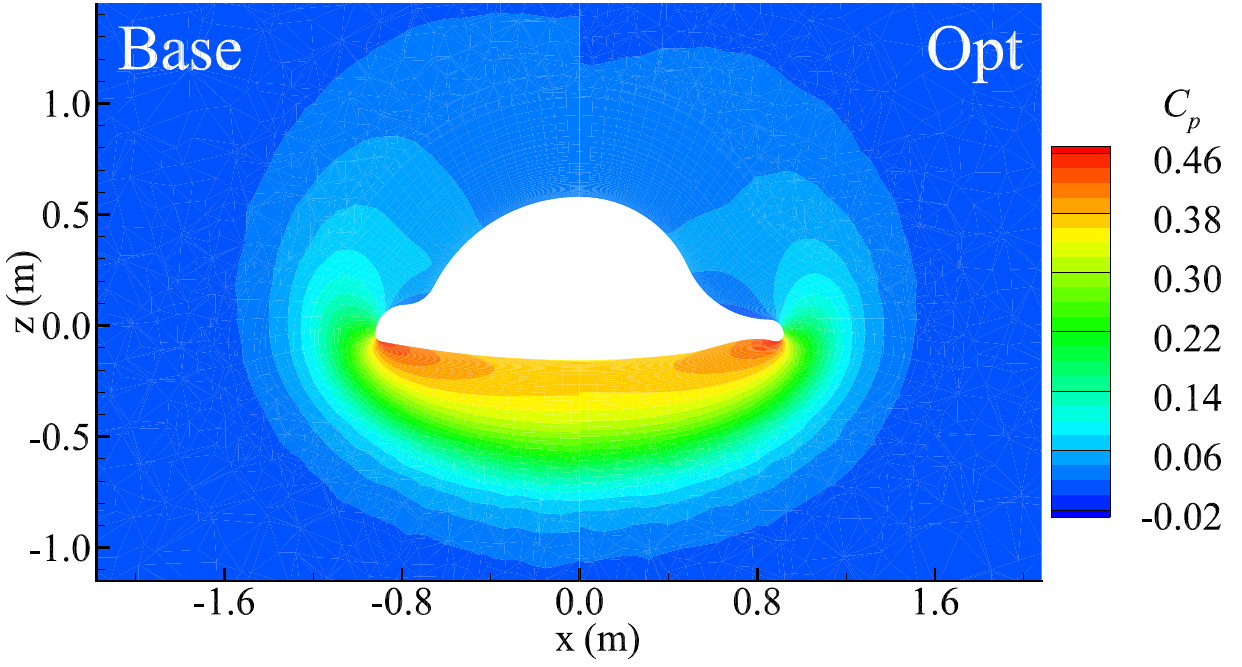}
        }
    \subfigure[$y = 3.66\,\rm{m}$]
        {
            \includegraphics[height=0.15 \textwidth]{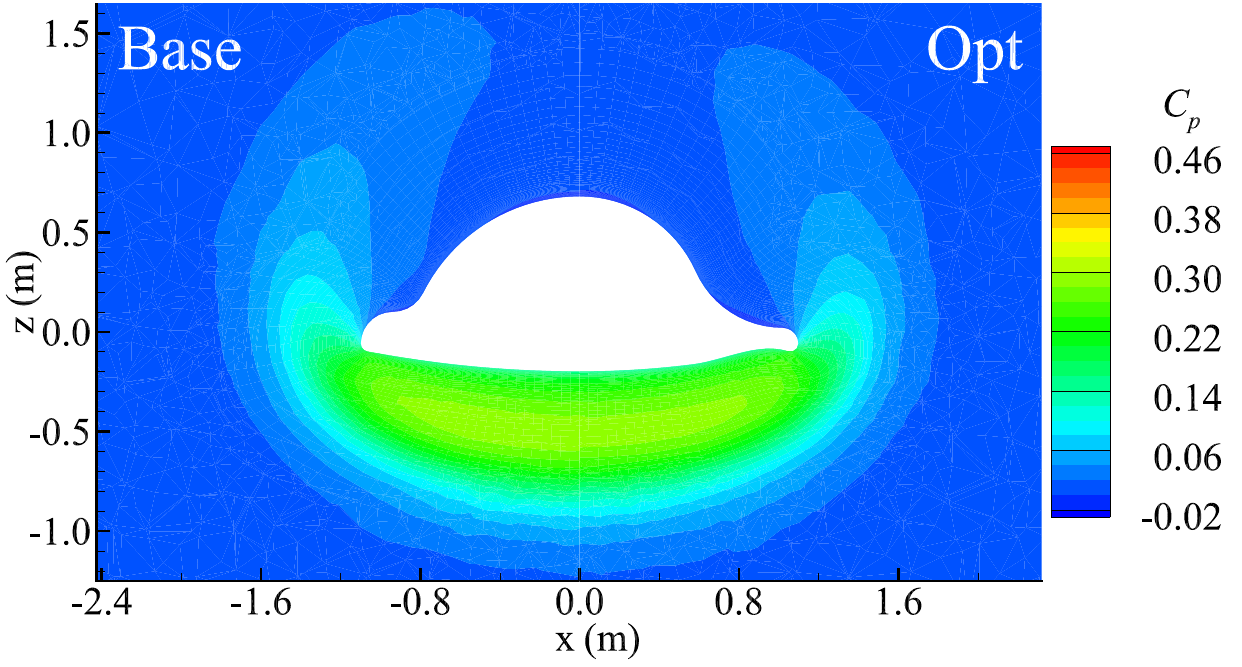}
        }
    \caption{Comparison of the pressure distribution of the HTV-2 type configurations along the longitudinal and axial cross-sections at 100 km.}
    \label{fig: cp counter at 100km}
\end{figure}

\begin{figure}
    \centering
    \subfigure[$C_p$]
        {\label{fig: cp slice at 100km}
            \includegraphics[width=0.45 \textwidth]{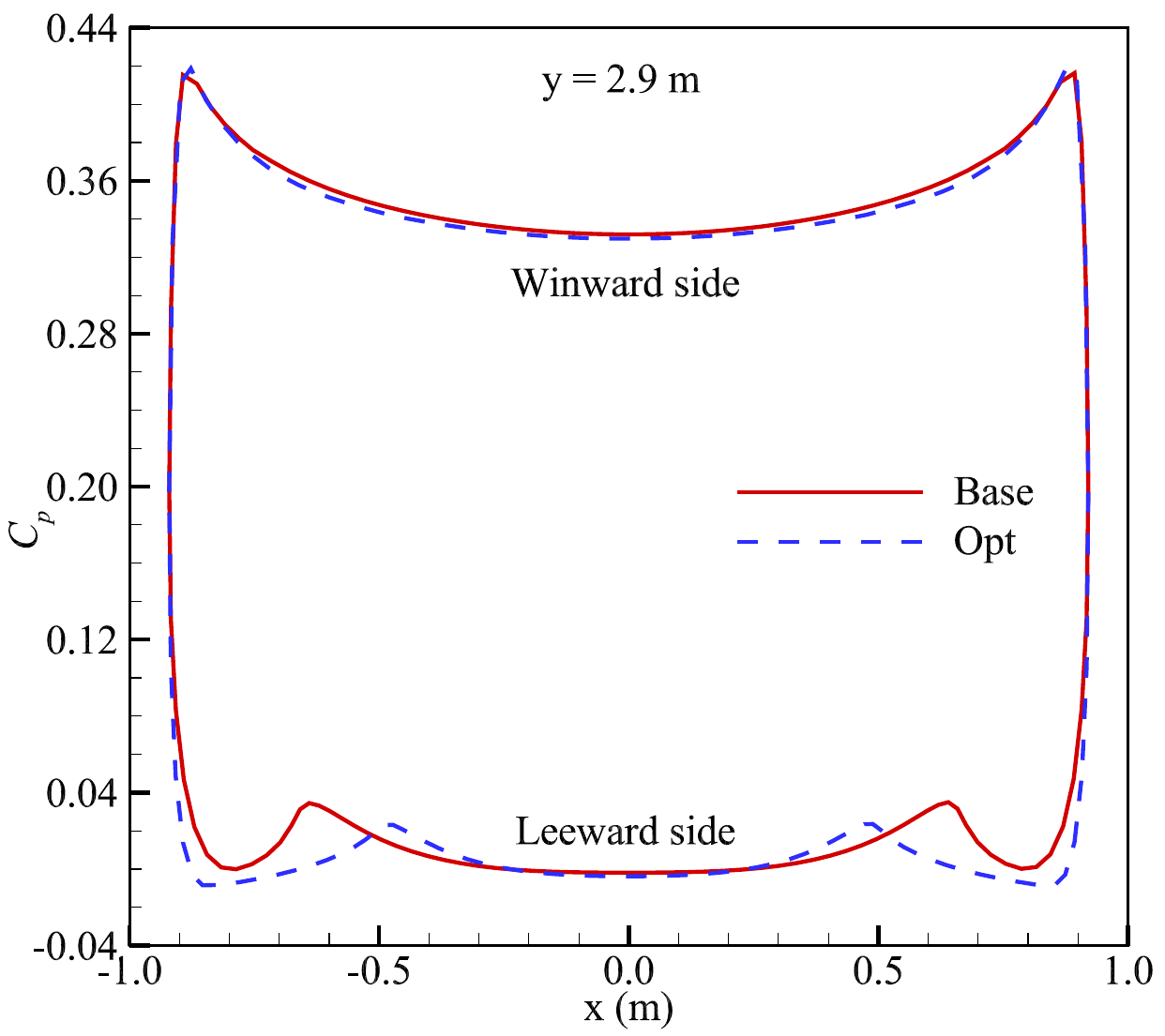}
        }
    \subfigure[$C_f$]
        {\label{fig: cf slice at 100km}
            \includegraphics[width=0.45 \textwidth]{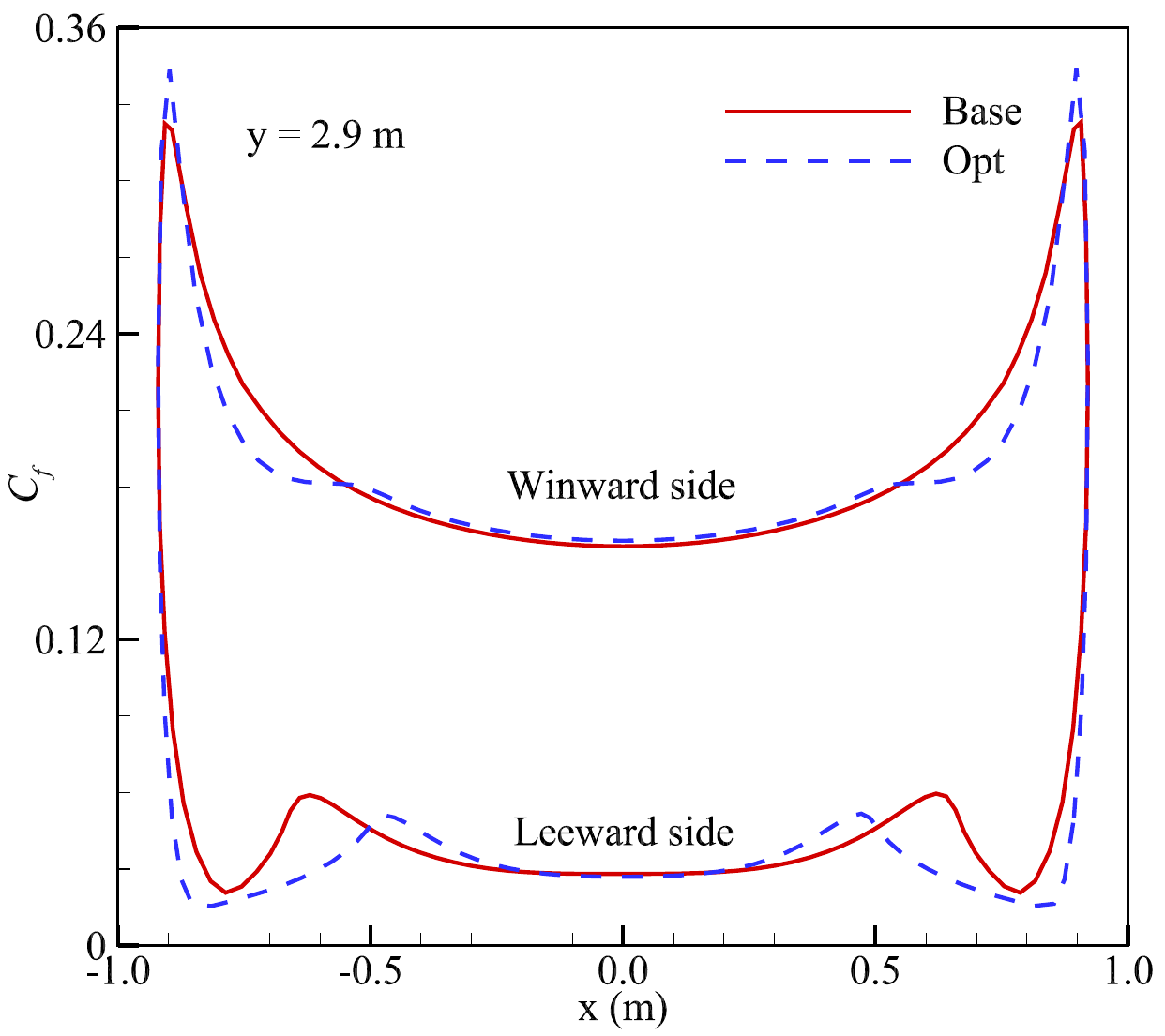}
        }
    \caption{Comparison of the $C_p$ and $C_f$ at $y = 2.9\,\rm{m}$ at 100 km.}
    \label{fig: clices at 100}
\end{figure}

Similarly, the flow field at 100 km is analyzed to understand the performance changes. As shown in Fig.~\ref{fig: cp counter at 100km}, on the leeward side, the reduction of $R_3$ and $R_2$ increases the curvature radius where the top body meets the wing. This geometry change enhances the flow expansion and lowers the local pressure on the upper surface. Unlike the 70~km case, the oblique shock wave at the nose does not change much. As seen in the surface pressure plots in Fig.~\ref{fig: cp slice at 100km}, the pressure coefficient on the windward side stays almost exactly the same before and after optimization. Because the windward high pressure is maintained while the leeward pressure drops, the $C_l$ at 100 km increases.
For the $C_f$ shown in Fig.~\ref{fig: cf slice at 100km}, the friction near the wing surface also drops. However, because the air at 100 km is much thinner, this decrease in $C_f$ is not as obvious as it is at 70~km. Even so, the combined effect on both drag reduction and lift increase leads to a higher $L/D$ at this highly rarefied altitude.

\begin{figure}
    \centering
    \subfigure[$x = 0\, \rm{m}$]
        {
            \includegraphics[width=0.42 \textwidth]{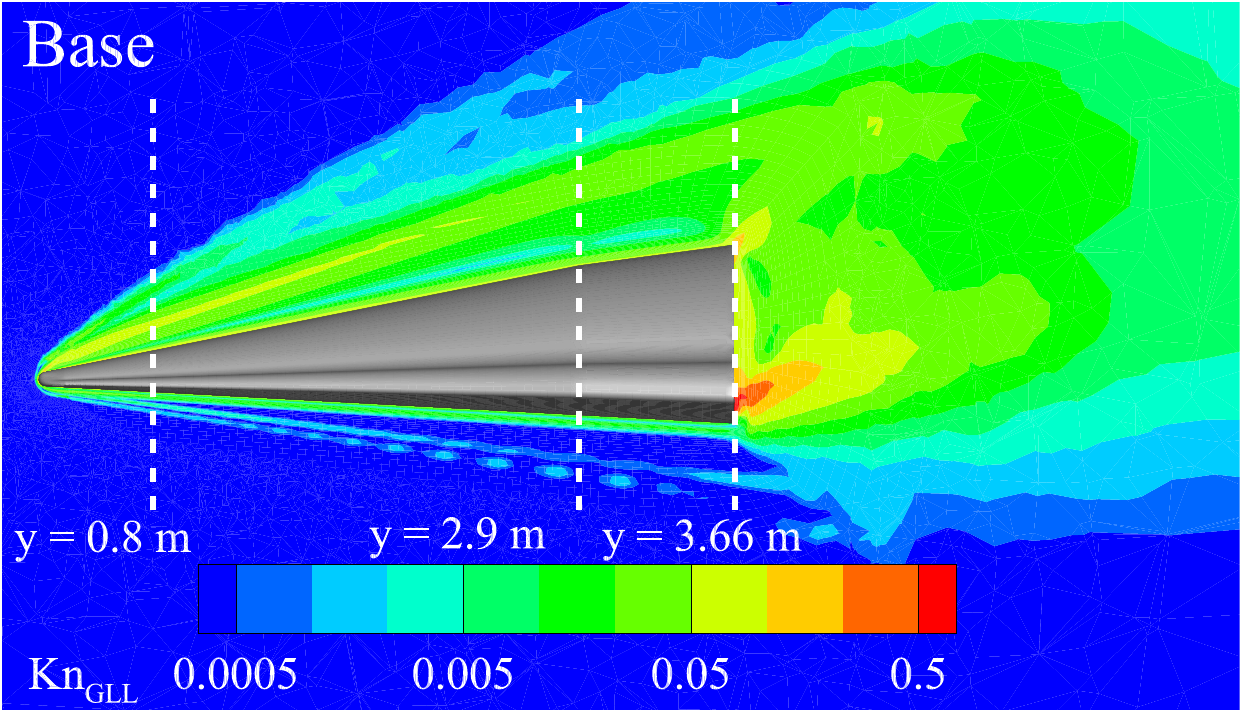}
        }
    \subfigure[$x = 0\, \rm{m}$]
        {
            \includegraphics[width=0.42 \textwidth]{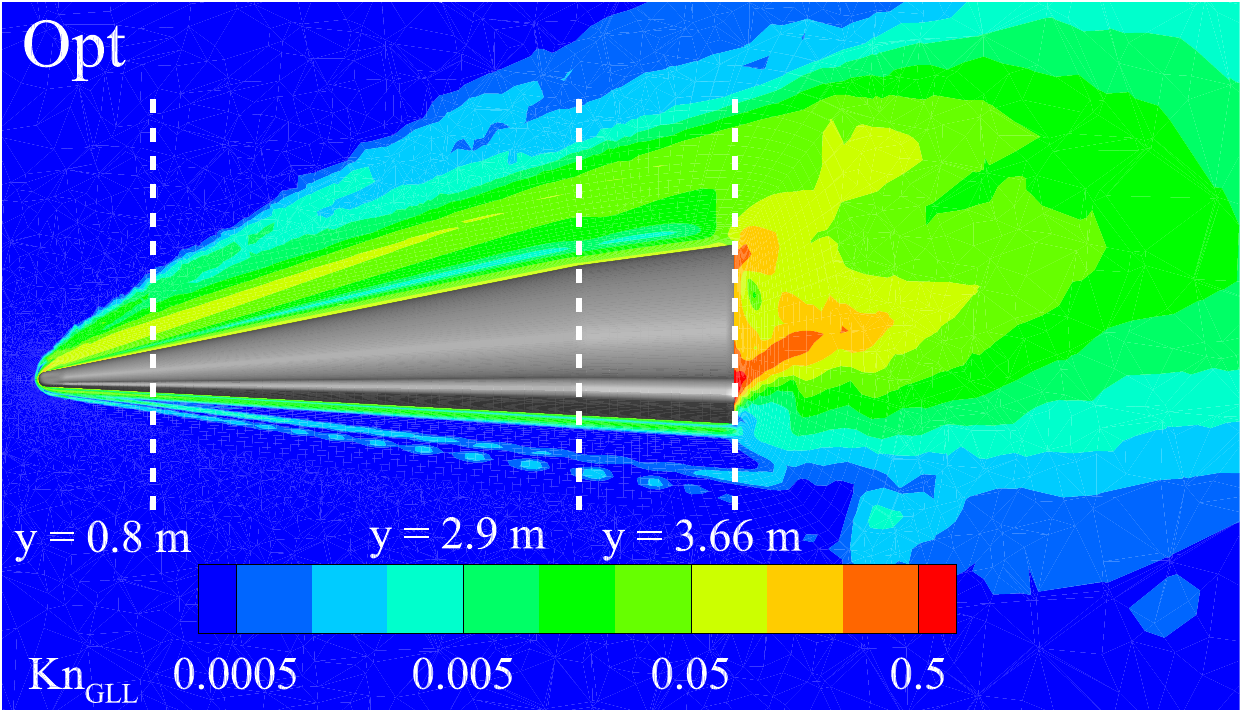}
        }
    \subfigure[$y = 0.8\,\rm{m}$]
        {
            \includegraphics[height=0.17 \textwidth]{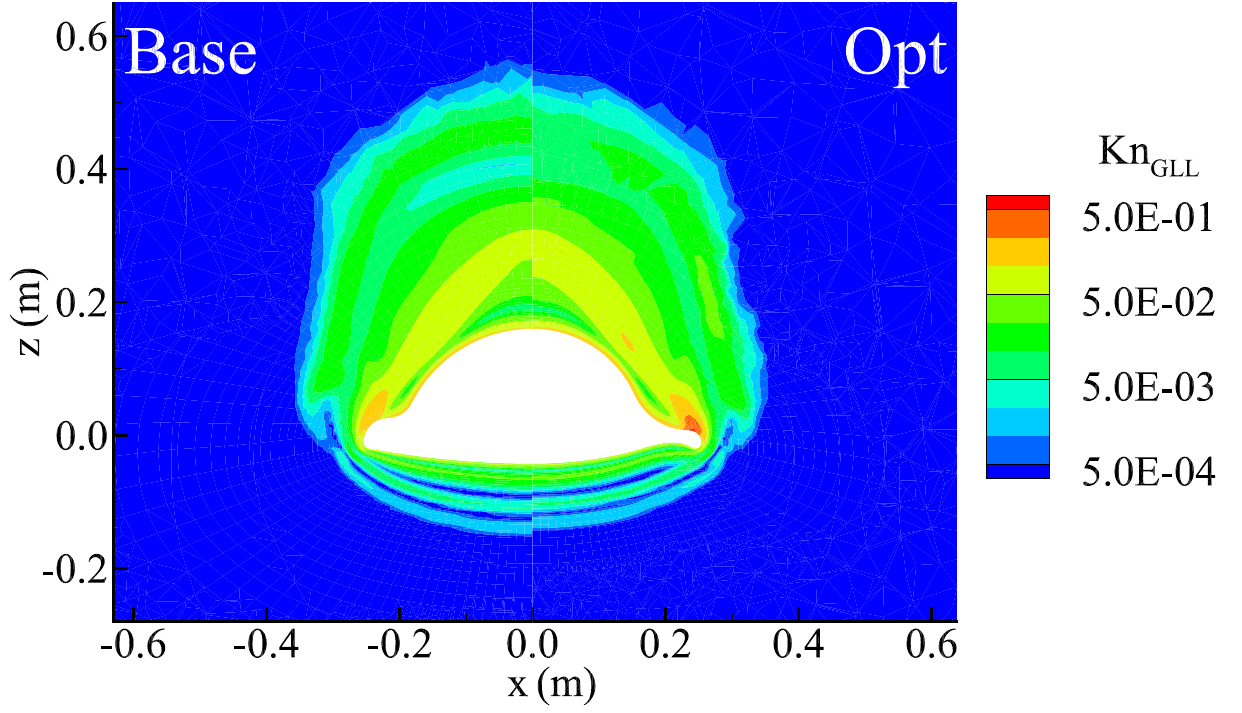}
        }
    \subfigure[$y = 2.9\,\rm{m}$]
        {
            \includegraphics[height=0.17 \textwidth]{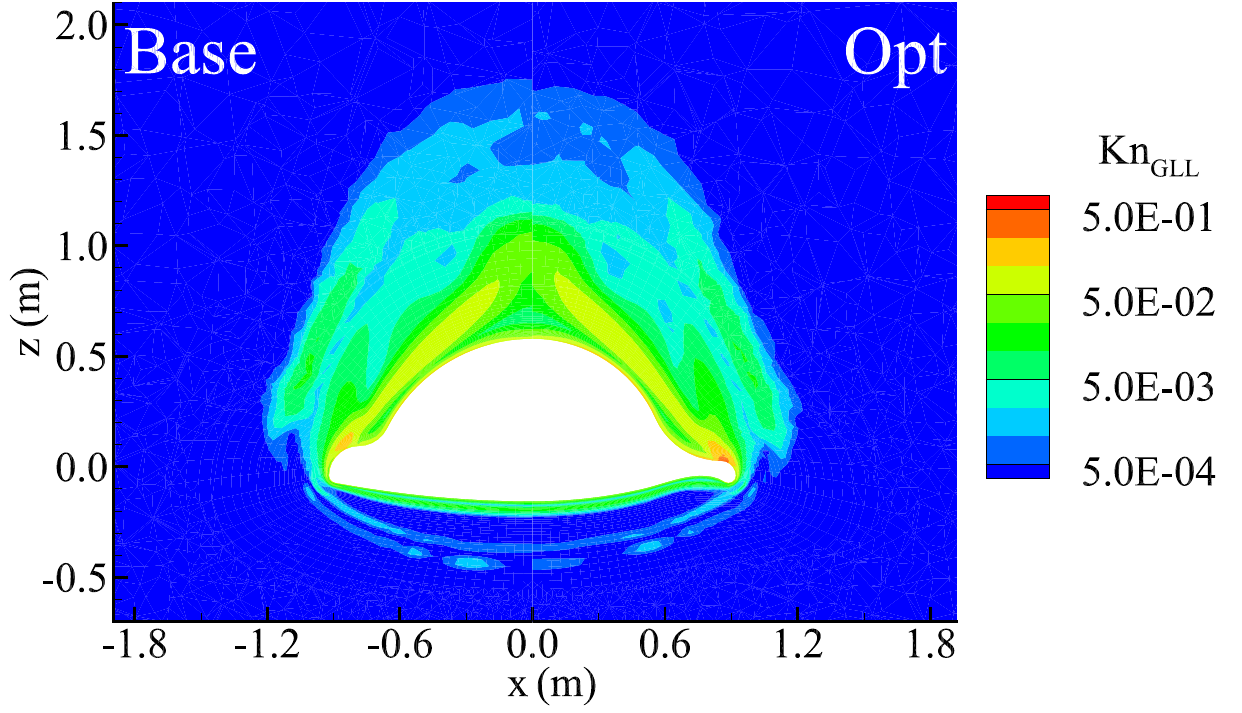}
        }
    \subfigure[$y = 3.66\,\rm{m}$]
        {
            \includegraphics[height=0.17 \textwidth]{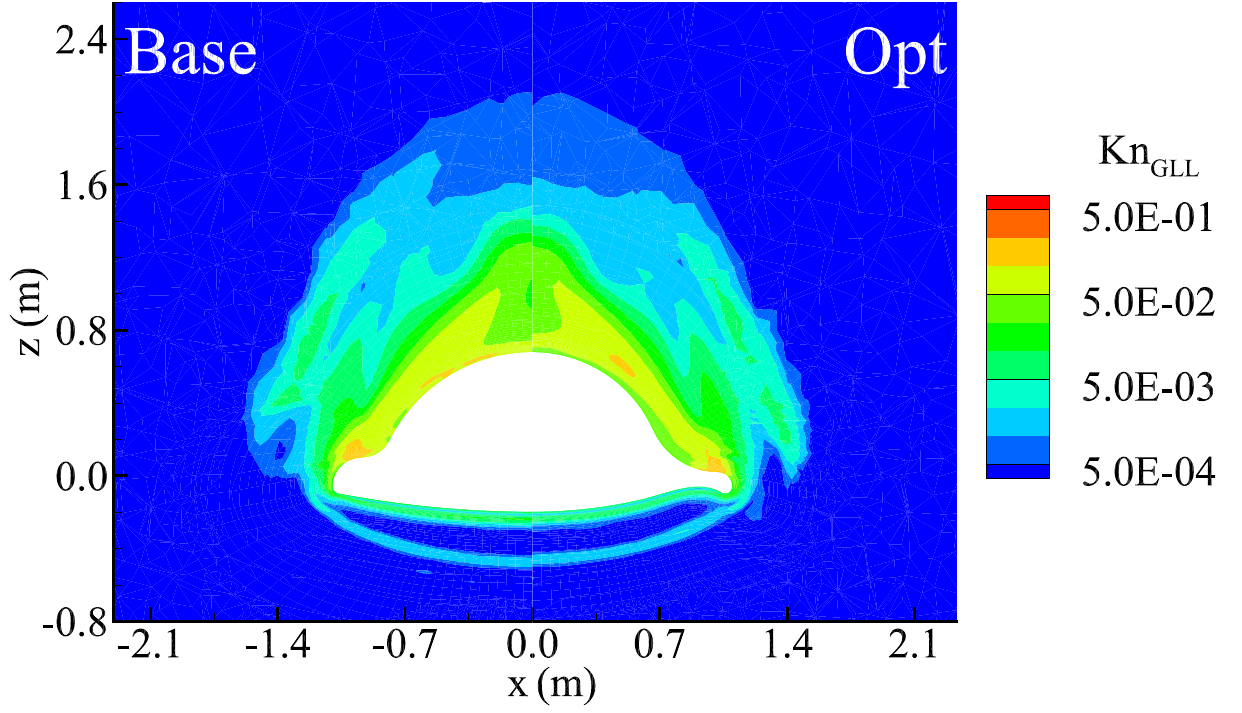}
        }
    \caption{Distributions of $\rm{Kn}_{\rm{GLL}}$ for the baseline and optimized configurations at 70~km.}
    \label{fig: Kn_GLL at 70km}
\end{figure}

\begin{figure}
    \centering
    \subfigure[$x = 0\, \rm{m}$]
        {
            \includegraphics[width=0.42 \textwidth]{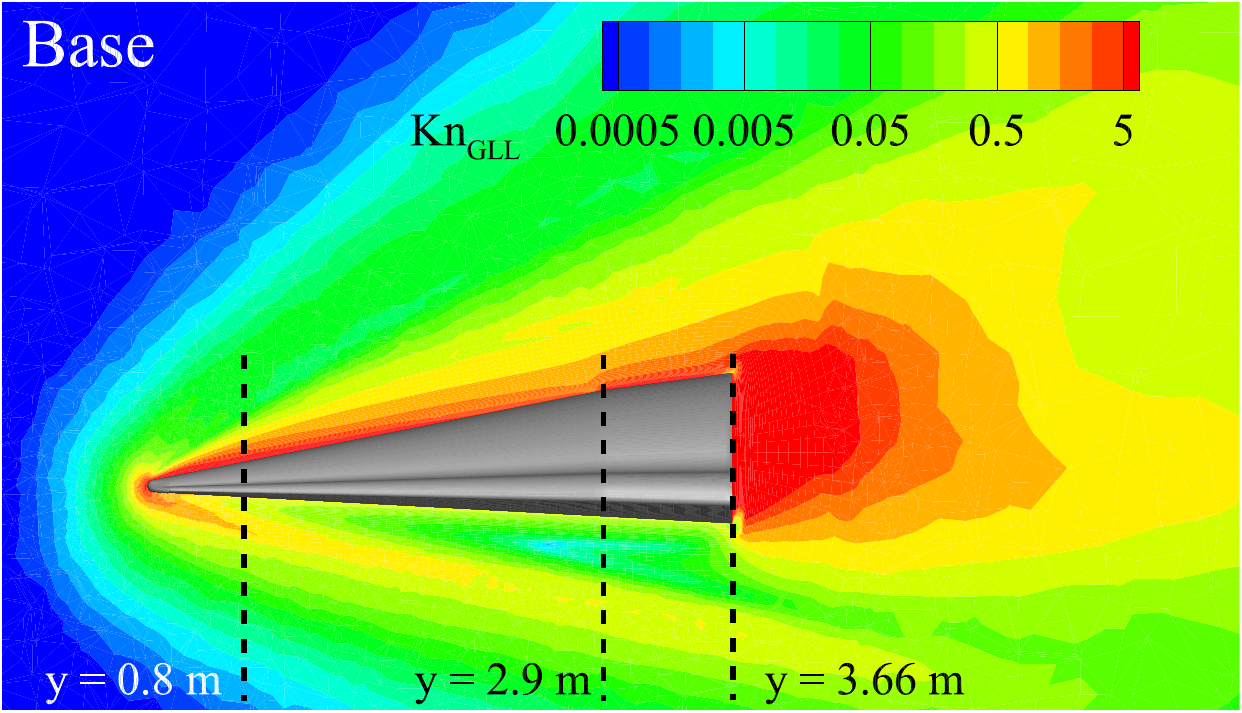}
        }
    \subfigure[$x = 0\, \rm{m}$]
        {
            \includegraphics[width=0.42 \textwidth]{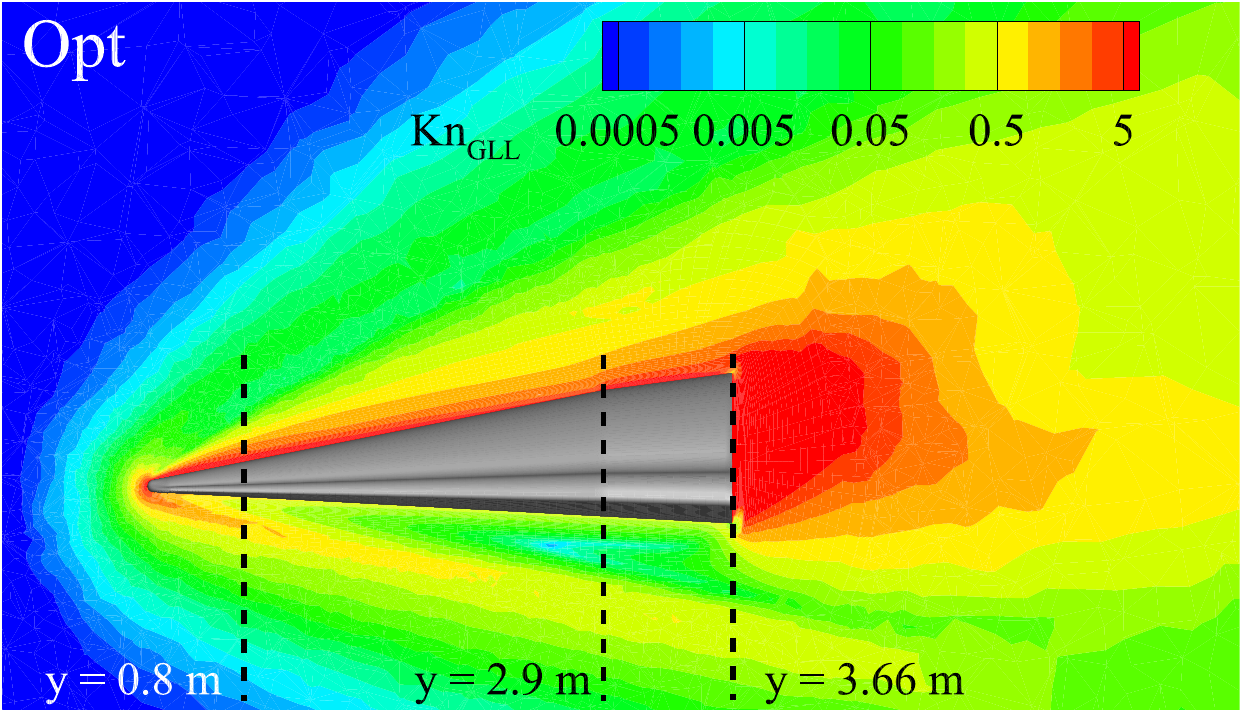}
        }
    \subfigure[$y = 0.8\,\rm{m}$]
        {
            \includegraphics[height=0.17 \textwidth]{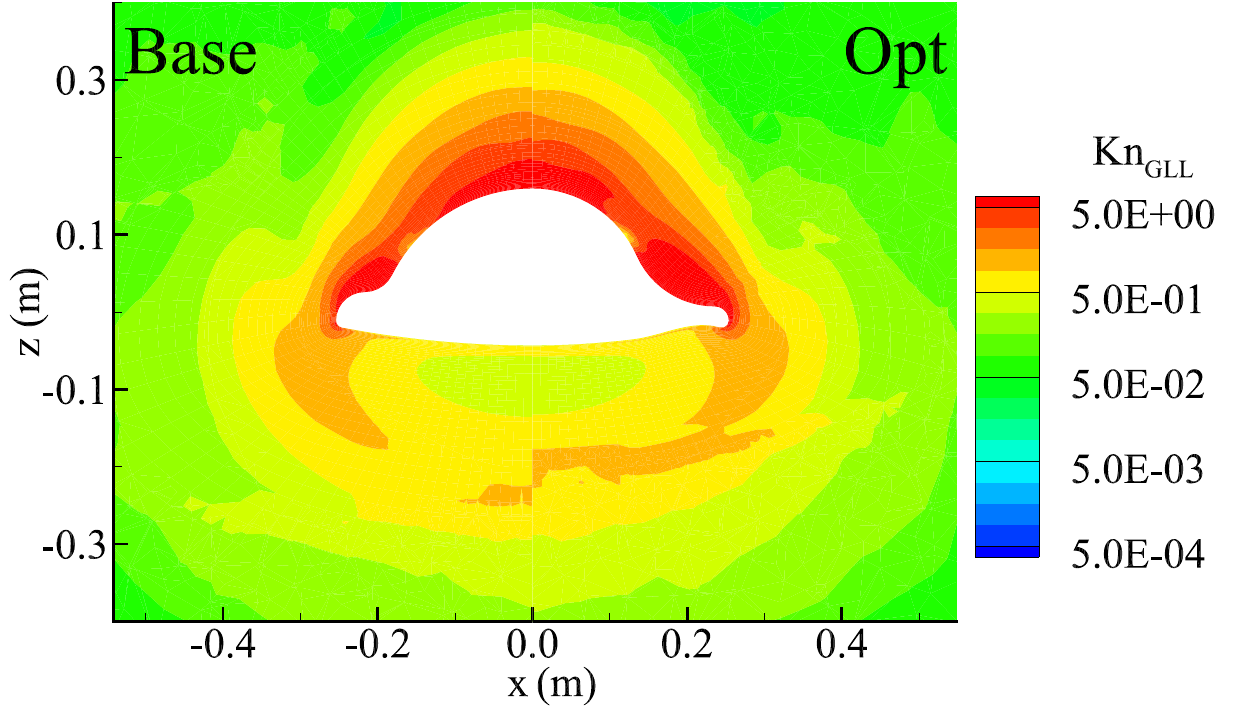}
        }
    \subfigure[$y = 2.9\,\rm{m}$]
        {
            \includegraphics[height=0.17 \textwidth]{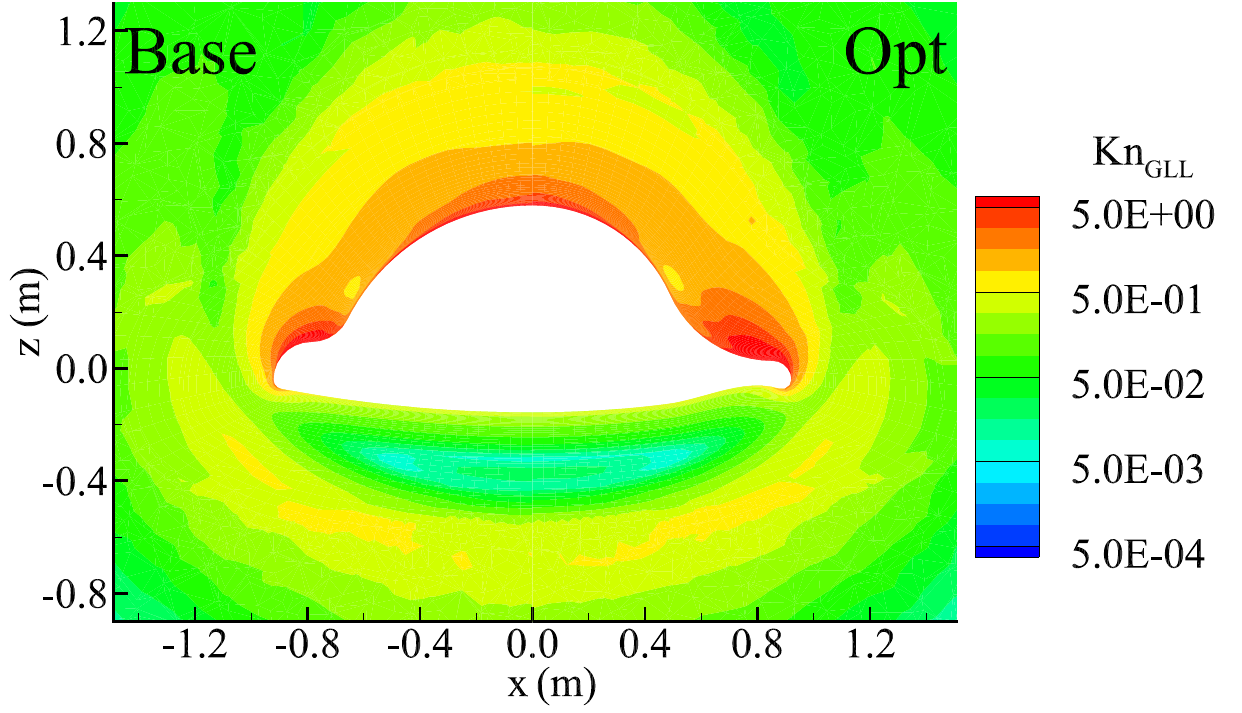}
        }
    \subfigure[$y = 3.66\,\rm{m}$]
        {
            \includegraphics[height=0.17 \textwidth]{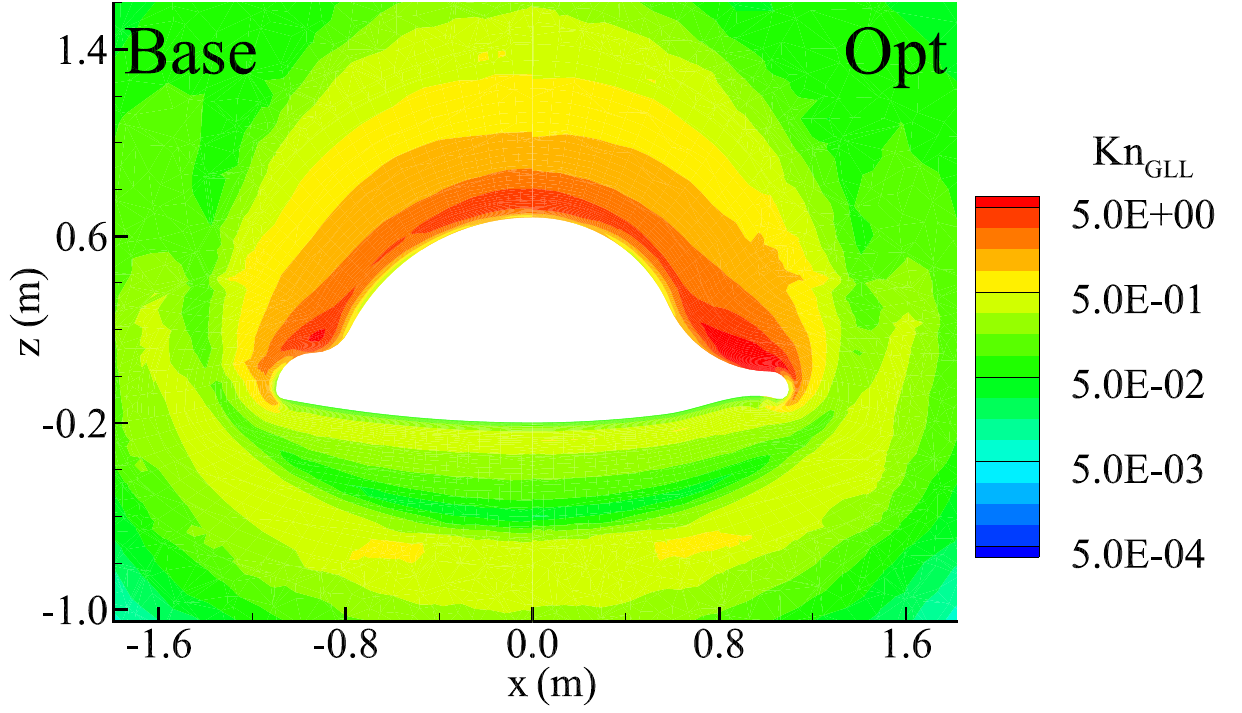}
        }
    \caption{Distributions of $\rm{Kn}_{\rm{GLL}}$ for the baseline and optimized configurations at 100 km.}
    \label{fig: Kn_GLL at 100km}
\end{figure}

Finally, Fig.~\ref{fig: Kn_GLL at 70km} and Fig.~\ref{fig: Kn_GLL at 100km} present the $\rm{Kn}_{\rm{GLL}}$ distributions at 70~km and 100 km. Similar to the 85 km case, the local Knudsen number above the wing increases for both optimized configurations, making the local flow more rarefied. Specifically at 70~km, the $\rm{Kn}_{\rm{GLL}}$ at the fuselage's bottom also increases, while the rest of the flow field shows no obvious changes.

\subsection{Comparative Discussion: Impact of Rarefaction on Optimization}

To further investigate the reasons behind these aerodynamic improvements, this section compares the variations in geometric parameters, aerodynamic forces, and first-order global sensitivity indices as the flow across the three altitudes. This comparison aims to identify the dominant geometric features governing the aerodynamic performance.

\begin{figure}
    \centering
    \subfigure[Geometric profiles]
        {\label{fig: baseall}
            \includegraphics[height=0.3 \textwidth]{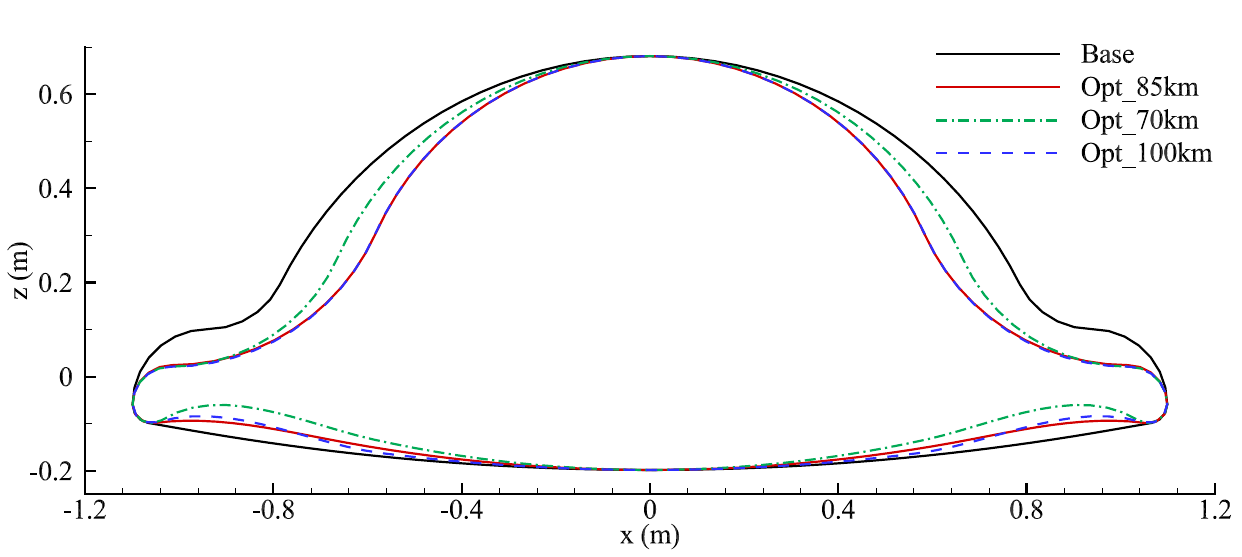}
        }
    \subfigure[Relative variations of variables]
        {\label{fig: relative_var}
            \includegraphics[height=0.3 \textwidth]{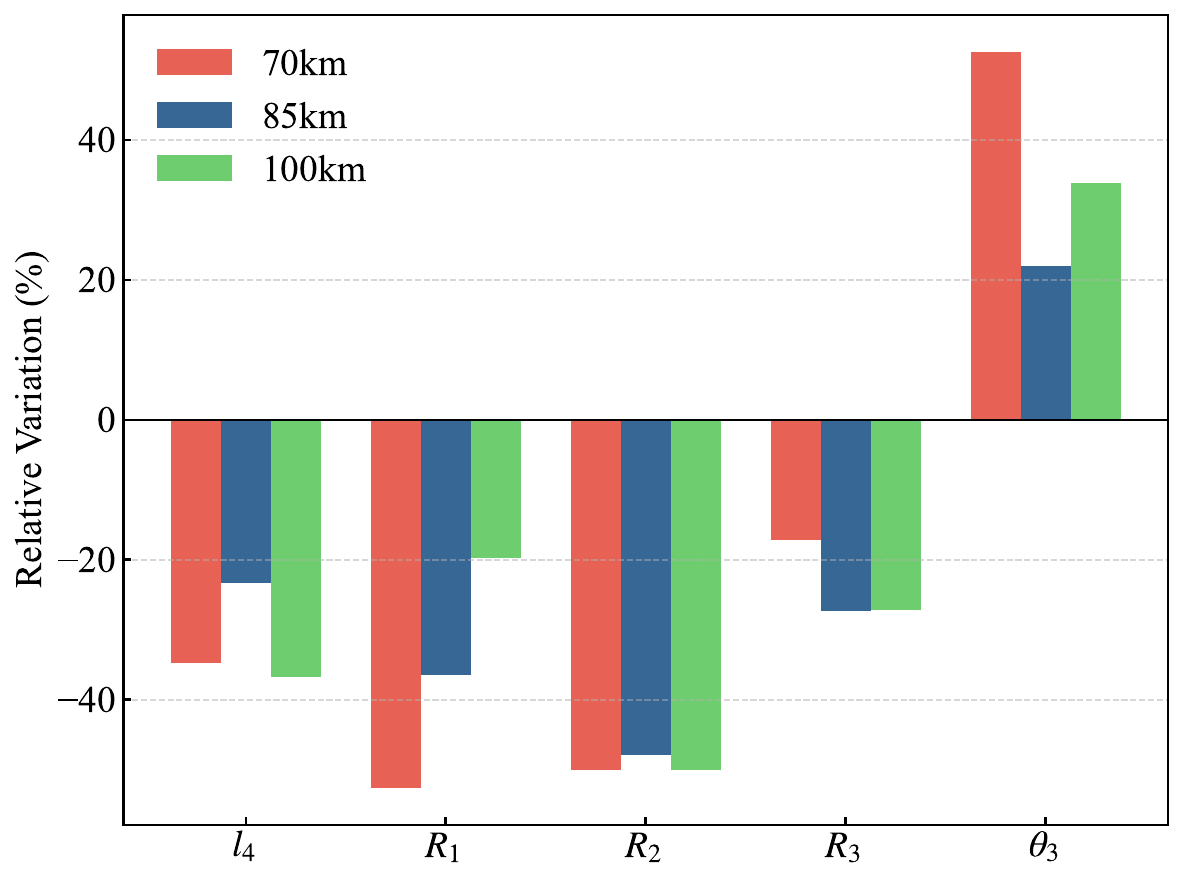}
        }
    \subfigure[Aerodynamic improvements]
        {\label{fig: aero_improve}
            \includegraphics[height=0.3 \textwidth]{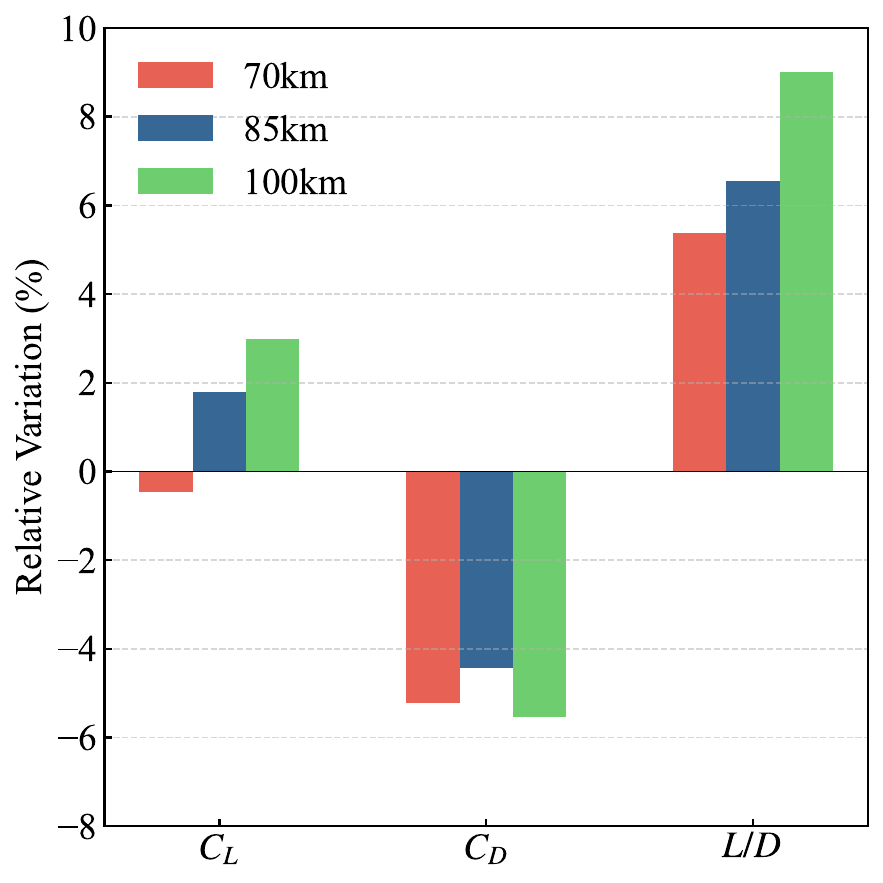}
        }
    \caption{Comparison of the geometric profiles and optimization results across different altitudes.}
    \label{fig: geometric_and_aero_comparison}
\end{figure}

As shown in Fig.~\ref{fig: relative_var}, the geometric parameters change in the same direction across all three altitudes. This indicates that the basic strategy to improve aerodynamic performance remains consistent within this flight envelope. Overall, the geometric changes are the smallest at 85 km. This is because the freestream condition for the HTV-2 type baseline configuration was chosen at this altitude, making it already closer to an optimal shape.

\begin{figure}
    \centering
    \subfigure[]
        {\label{fig: s1_l}
            \includegraphics[width=0.3 \textwidth]{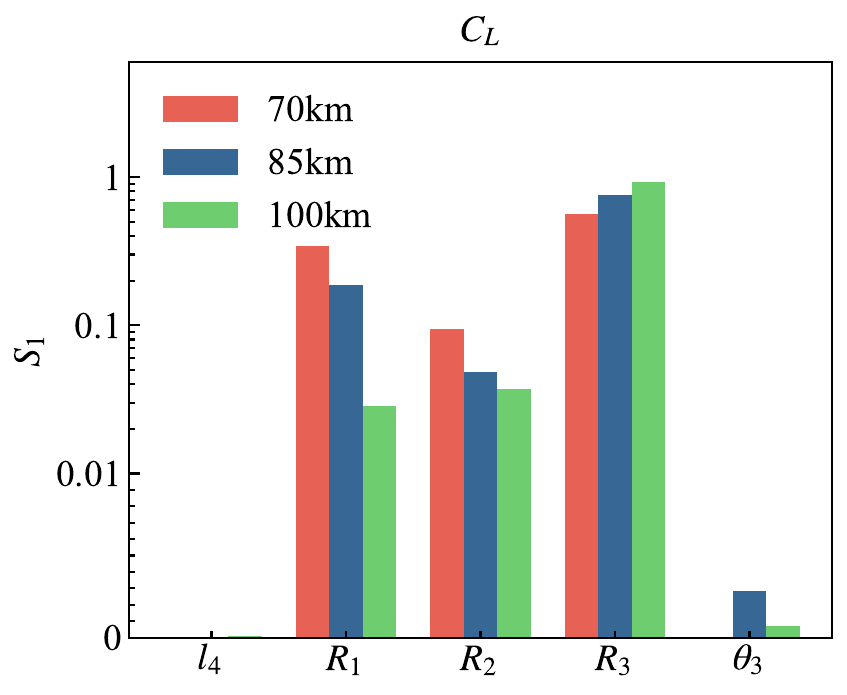}
        }
    \subfigure[]
        {\label{fig: s1_d}
            \includegraphics[width=0.3 \textwidth]{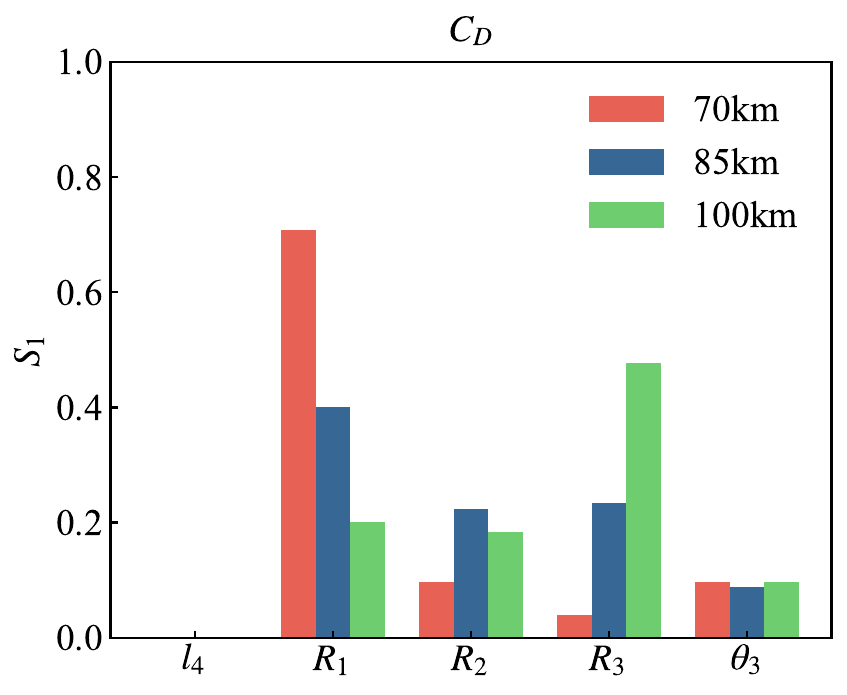}
        }
    \subfigure[]
        {\label{fig: s1_ltd}
            \includegraphics[width=0.3 \textwidth]{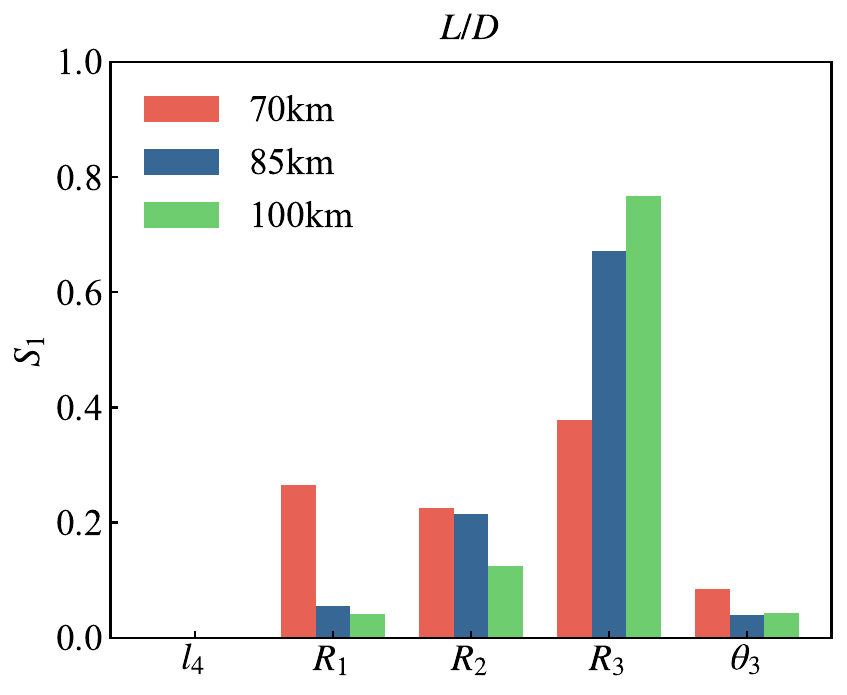}
        }
    \caption{Comparison of the Sobol first-order sensitivity indices across different altitudes.}
    \label{fig: sobol_indices_comparison}
\end{figure}

Fig.~\ref{fig: baseall} illustrates the base profiles of the three optimized configurations. Evidently, all optimized shapes trend toward a flatter design, emphasizing the role of the wing surfaces. Among the design variables, the reduction in the $R_2$ is almost identical across all cases. This shows that making the wing tips sharper consistently improves performance regardless of the altitude. The sensitivity analysis in Fig.~\ref{fig: sobol_indices_comparison} also confirms that $R_2$ maintains a steady and important influence on all aerodynamic forces.

In contrast to $R_2$, the variations of $R_1$ and $R_3$ exhibit a strong altitude dependence. Specifically, the 70~km optimized configuration retains a noticeably larger upper volume and a higher windward curvature compared to those at 85 and 100~km. While the upper profiles of the 85 and 100~km configurations are morphologically similar, the 100~km optimum features a flatter windward surface and a much more pronounced curvature transition at the wing-body junction.

Fig.~\ref{fig: aero_improve} quantifies the aerodynamic performance shifts of the optimized configurations relative to the baseline. With the exception of a marginal $C_L$ decrement at 70~km, the optimized geometries consistently yield elevated $C_L$ and reduced $C_D$. Crucially, the magnitude of $C_D$ reduction substantially outpaces the $C_L$ variations, thereby serving as the primary driver for the overall $L/D$ enhancement.

To further describe these improvements, Tab.~\ref{tab: cp cf to l and d} decomposes the aerodynamic forces into their pressure and skin friction components. As the data illustrates, the $C_L$ increment is predominantly governed by pressure contributions, although a slight contribution in friction lift is also observed at 100~km.
Conversely, the composition of the drag reduction changes fundamentally with altitude, reflecting the increasing dominance of skin friction in highly rarefied environments. At 70~km, the reduction in $C_D$ is almost entirely attributed to a decrease in pressure drag. At 85~km, the reductions in pressure and friction drag contribute equally. By 100~km, however, the overall decrease in $C_D$ is dominated by the reduction in friction drag.

\begin{table}
\begin{center}
\begin{tabular*}{0.9\textwidth}{@{\extracolsep{\fill}}c c cccccc}
Altitude & Case & $C_p-C_L$ & $C_f-C_L$ & $C_L$ & $C_p-C_D$ & $C_f-C_D$ & $C_D$ \\
\multirow{3}{*}{70~km}
 & Base      & 63.76 & -1.47 & 62.39 & 28.41 & 7.23  & 35.40 \\
 & Opt       & 63.44 & -1.59 & 62.10 & 26.62 & 6.96  & 33.55 \\
 & Variation & -0.32 & -0.12 & -0.29 & \textbf{-1.79} & -0.27 & -1.85 \\
\multirow{3}{*}{85~km}
 & Base      & 63.81 & -4.85 & 58.98 & 29.45 & 21.13 & 50.59 \\
 & Opt       & 65.00 & -5.08 & 60.03 & 28.29 & 19.95 & 48.35 \\
 & Variation & 1.19  & -0.23 & 1.06  & \textbf{-1.16} & \textbf{-1.18} & -2.24 \\
\multirow{3}{*}{100~km}
 & Base      & 73.49 & -18.64 & 54.81 & 33.58 & 65.47 & 98.91 \\
 & Opt       & 74.82 & -18.34 & 56.45 & 32.28 & 61.33 & 93.42 \\
 & Variation & 1.33  & 0.30   & 1.64  & -1.30 & \textbf{-4.14} & -5.49 \\
\end{tabular*}
\end{center}
\caption{Comparison of pressure and skin friction contributions to the $C_L$ and $C_D$ at different altitudes.}
\label{tab: cp cf to l and d}
\end{table}

At 70~km, where the gas density is relatively high and the $\mathrm{Kn}$ is small, pressure drag is the dominant component. Since the windward shock constitutes the primary source of this pressure drag, reducing $R_1$ effectively weakens the shock strength, thereby lowering $C_D$. The Sobol indices verify this physics: as shown in Fig.~\ref{fig: sobol_indices_comparison}, $R_1$ exhibits the highest sensitivity for both $C_L$ and $C_D$ at 70~km, although its influence clearly diminishes at 100~km.
On the upper surface, reducing $R_3$ enhances the expansion wave, which contributes to an increase in $C_L$. However, this geometric modification concurrently reduces the internal volume. At 70~km, the reduction of $R_3$ is relatively restricted to ensure the volumetric requirements are strictly satisfied. In contrast, when the altitude reaches 100~km and the flow becomes highly rarefied, the sensitivity analysis reveals that $R_3$ surpasses $R_1$ to emerge as the dominant geometric parameter governing the overall $L/D$.

\section{Conclusions}\label{sec: conclusion}

In this study, an automated optimization framework was constructed by coupling surrogate-based optimization (SBO) with the implicit unified gas-kinetic scheme (IUGKS). This framework was applied to the aerodynamic shape optimization of an HTV-2-type aircraft across multiple flow regimes (spanning altitudes of 70 to 100~km). Based on detailed optimization results, flow field evaluations, and Sobol sensitivity analyses, the main conclusions are summarized as follows:

\begin{itemize}
    \item  Under strict volumetric constraints, the lift-to-drag ratio $L/D$ of the optimized configurations was notably improved across all tested altitudes, achieving increases of 5.37\%, 6.55\%, and 9.00\% at 70km, 85km, and 100~km, respectively. Cross-validation using the multiscale UGKS method confirms that the SBO framework maintains high predictive accuracy across all flow regimes.

    \item The varying aerodynamic characteristics at different altitudes necessitate distinct optimization strategies. At 70~km, where pressure drag dominates, the optimized shape drastically reduces the windward surface radius to weaken the oblique shock wave. At the highly rarefied 100~km altitude, where skin friction becomes significantly more pronounced, the optimal configuration reduces the leeward surface radius to enhance upper-surface expansion and recover lift. Furthermore, across all altitudes, an enhanced drooped wingtip design effectively blocks spanwise cross-flow, thereby reducing local shear stress and friction drag.

    \item   The Sobol sensitivity analysis confirms a clear shift in the dominant geometric parameters across different flow regimes. As altitude increases and the gas becomes more rarefied, the sensitivity of the overall aerodynamic forces to $R_1$  (which controls the oblique shock wave) drops significantly. In contrast, the influence of $R_3$ (which controls the expansion wave) rises to become the most dominant factor. Meanwhile, reducing $R_2$ provides a consistent aerodynamic benefit throughout the entire flight envelope.
        
\end{itemize}

Overall, an optimal shape designed for a single altitude cannot fully adapt to the multiscale physical phenomena encountered across different flight regimes. To optimize aerodynamic performance across the entire trajectory, various physical effects must be considered and appropriate trade-offs must be established. Future work will focus on multi-objective optimization across the full flight envelope and the integration of machine learning techniques with larger sample sizes to further enhance the design framework.





\section*{Declaration of AI Usage}

During the preparation of this manuscript, the author used the Gemini large language model (version [Gemini free], accessible at https://gemini.google.com) solely for language polishing and improving text readability. The tool was used in its standard, unmodified form. The AI was not involved in data analysis, idea generation, or drawing academic conclusions. The author takes full responsibility for the final content and integrity of this paper.


\section*{Declaration of competing interest}

The authors declare that they have no known competing financial interests or personal relationships that could have appeared to influence the work reported in this paper.


\section*{Acknowledgements}

The current research is supported by National Natural Key R$\&$D Program of China (Grant Nos. 2022YFA1004500), National Natural Science Foundation of China (92371107), and Hong Kong research grant council (16208324).


\bibliographystyle{jfm}
\bibliography{sboref}

\end{document}